\documentclass[fleqn,usenatbib]{mnras}

\usepackage{newtxtext,newtxmath}

\usepackage[T1]{fontenc}

\DeclareRobustCommand{\VAN}[3]{#2}
\let\VANthebibliography\thebibliography
\def\thebibliography{\DeclareRobustCommand{\VAN}[3]{##3}\VANthebibliography}

\def\orcid#1{\kern .08em\href{https://orcid.org/#1}{\includegraphics[keepaspectratio,width=0.7em]{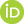}}}

\usepackage{graphicx}	
\usepackage{amsmath}	

\usepackage{mathtools}

\usepackage{tabularx}

\usepackage[dvipsnames]{xcolor}

\def\psj{PSJ}

\usepackage{grfext}

\title[NHD vs QHD equations in THOR]{Examining NHD vs QHD in the GCM THOR with non-grey radiative transfer for the hot Jupiter regime}

\author[Noti et al.]{
Pascal A. Noti,$^{1,2}$\thanks{E-mail: pascal-andreas.noti@unibe.ch}\orcid{0000-0002-8012-3400}
Elspeth K. H. Lee,$^{1}$\orcid{0000-0002-3052-7116}
Russell Deitrick $^{3}$\orcid{0000-0001-9423-8121}
and Mark Hammond $^{4}$\orcid{0000-0002-6893-522X}\\
$^{1}$Center for Space and Habitability, Universität Bern, Gesellschaftsstrasse 6, CH-3012 Bern, Switzerland\\
$^{2}$Physikalisches Institut, Universität Bern, Sidlerstrasse 5, CH-3012 Bern, Switzerland\\
$^{3}$School of Earth and Ocean Sciences, University of Victoria, Victoria, British Columbia, Canada\\
$^{4}$Atmospheric, Oceanic and Planetary Physics, University of Oxford, Oxford, United Kingdom
}

\date{Accepted XXX. Received YYY; in original form ZZZ}

\pubyear{2023}

\begin{document}
\label{firstpage}
\pagerange{\pageref{firstpage}--\pageref{lastpage}}
\maketitle

\begin{abstract}
Global circulation models (GCMs) play an important role in contemporary investigations of exoplanet atmospheres. Different GCMs evolve various sets of dynamical equations which can result in obtaining different atmospheric properties between models. In this study, we investigate the effect of different dynamical equation sets on the atmospheres of hot Jupiter exoplanets. We compare GCM simulations using the quasi-primitive dynamical equations (QHD) and the deep Navier-Stokes equations (NHD) in the GCM THOR. We utilise a two-stream non-grey "picket-fence" scheme to increase the realism of the radiative transfer calculations. We perform GCM simulations covering a wide parameter range grid of system parameters in the population of exoplanets. Our results show significant differences between simulations with the NHD and QHD equation sets at lower gravity, higher rotation rates or at higher irradiation temperatures. The chosen parameter range shows the relevance of choosing dynamical equation sets dependent on system and planetary properties. Our results show the climate states of hot Jupiters seem to be very diverse, where exceptions to prograde superrotation can often occur. Overall, our study shows the evolution of different climate states which arise just due to different selections of Navier-Stokes equations and approximations. We show the divergent behaviour of approximations used in GCMs for Earth, but applied for non Earth-like planets.

\end{abstract}




\section{Introduction}

Numerical weather and climate predictions provide useful information for our daily lives, naval and aviation safety, national policy, strategy development and for research in atmospheric science. Running numerical simulations can be computationally expensive, therefore, approximations of the Navier-Stokes equations \citep{navier1823,stokes1845,stokes1846} have been proposed for global scale simulations. \citet{bjerknes1904} proposed the basis of the hydrostatic primitive equations (HPEs). \citet{richardson1922} derived a variation from Bjerknes's primitive equations to perform the first attempt at a numerical weather forecast by hand. \citet{charney1949} produced the first numerical weather model on ENIAC in 1950. Already at the dawn of numerical forecasting, \citet{charney1955} identified those approximations as an important obstacle to overcome.

The limits of the HPEs are still assessed to this day; e.g. the energy conservation in global circulation models (GCMs) for Earth \citep{tort2015}, for short-period waves at small scales \citep[e.g.][]{alvarez2019}, as well as for global simulations of exoplanetary atmospheres \citep[e.g.][]{mayne2019,deitrick2020}.
While numerical models utilizing the primitive equations have been relatively successfully applied to Earth's atmosphere, the applicability of the primitive equation set has been questioned for exoplanet atmospheres. For example, \citet{mayne2019} discovered important differences in the zonal advection between simulations using the "primitive" equations and the "full" Navier-Stokes equations (according to the nomenclature of \citet{mayne2014a}). Those differences in the zonal advection lead, for example, to significant differences in the atmospheric redistribution of heat in simulations of the warm and tidally-locked small Neptune GJ 1214b. For hot Jupiters, \cite{deitrick2020} see changes of 15 to 20 \% in the peak zonal winds in simulations with the non-hydrostatic, deep atmospheres (NHD) and quasi-hydrostatic, deep atmosphere (QHD) equation sets.

Atmospheric simulations can be in the interest for observations of exoplanets; the era of JWST will bring us several phase curves observations of exoplanet atmospheres, ranging from hot giants to temperate terrestrials, at higher resolutions than ever before. Continuous and long duration observations combined with a larger spectral resolution, collecting area, and a wider spectral coverage ranging from $0.6\:\mu m$ to $20\:\mu m$ will lead the studies of exoplanets and their habitability to quantum leap forward in evolution \citep{stevenson2016, bean2018}. At the same time \citet{feng2016, dobbs_dixon2017, blecic2017, caldas2019, flowers2019, irwin2020, parmentier2020, taylor2020, beltz2021} highlight the importance of multidimensionality in interpreting observations. Therefore, simulations of the dynamics and the 3D structure of exoplanetary atmospheres are essential tools for helping to understand and interpret the new observation data from JWST. Moreover, phase curve data of hot Jupiters in the optical and infrared wavelength regimes can benefit from the findings of 3D simulations of exoplanetary atmospheres: the Transiting Exoplanet Survey Satellite  \citep[TESS,][]{ricker2014}, CHaracterising ExOPlanet Satellite \citep[CHEOPS,][]{broeg2013}, the Atmospheric Remote-sensing Infrared Exoplanet Large survey \citep[ARIEL,][]{tinetti2016}, and the high altitude ballon mission EXoplanet Climate Infrared TElescope  \citep[EXCITE,][]{nagler2019}. Since the 3D simulations of the exoplanetary atmospheres are necessary tools for the understanding of exoplanets, identifying significant differences between simulations with different dynamical equation sets is important. 

\citet{white2005} and \citet{mayne2014a} reviewed the shallow, deep, hydrostatic, quasi-hydrostatic and non-hydrostatic equations in GCMs. For a complete overview on the NHD and QHD equation sets, see \citet{deitrick2020}. Other conventions of dynamical equation sets can also be used e.g. \citet{mendonca2016,deitrick2020}.

Simulations with HPEs can represent gravity-waves and nearly-geostropic motions \citep{white2005}. For representing nearly-geostrophic or `balanced' motion much attention has been put into deriving approximations \citep[see reviews in][]{norbury2002a,norbury2002b}. Several approximations can be found in the HPEs: the `hydrostatic' assumption, `shallow atmosphere', `spherical geopotential approximation' and the `traditional approximation' \citep{eckart1960}.

The traditional approximation was first introduced to study the oceanic and atmospheric dynamics of Earth considering the negligible Coriolis terms in shallowness of the Earth \citep[e.g.][]{eckart1960,gerkema2008,zeitlin2018}. In the momentum equation, several terms go to zero  \citep[see][]{mayne2014a}: for longitudinal wind $u$ the terms $2\Omega\omega\cos{\phi}$ (traditional approximation) and $\frac{-u\omega}{r}$ (shallow approximation), for latitudinal wind $v$ the term $\frac{-v\omega}{r}$ (shallow approximation) and for the vertical wind $\omega$ the terms $2\Omega u\cos{\phi}$ (traditional approximation) and $\frac{u^2 + v^2}{r}$ (shallow approximation). In astrophysics, the traditional approximation of rotation (TAR) might describe the dynamics of gravito-inertial waves on stars \citep[e.g.][]{mathis2019} well, but it is problematic for some exoplanets such as dynamics of the warm and tidally-locked small Neptune GJ 1214b, as \citet{mayne2019} showed. The discussion of the $\cos{\phi}$ terms have been in contention for many years \citep{white2005}. Studies by \citet{phillips1990,thuburn2002a}, using linearized and adiabatic analysis, showed those $\cos{\phi}$ terms are minor given the parameters of Earth if the ratio of planetary rotation frequency to buoyancy frequency is very small $(\ll 1)$. \citet{white2005} regarded the terms to be unsettling, because buoyancy frequency differs across the globe and diabatic processes drive the global circulation. Furthermore, they find that the $\cos{\phi}$ terms are problematic if the buoyancy frequency increases through climate change. \citet{bretherton1964} an \citet{deverdiere1994} showed the importance of the $\cos{\phi}$ terms near the equator. Moreover, the $\cos{\phi}$ terms become relevant for the mesoscale motion \citep{draghici1989}. The traditional approximation to models simulating exoplanets varies widely in their climate regimes. Therefore, we could assume that the traditional approximation might be not valid for many exoplanets.

Models with non-hydrostatic equations (NHEs) for global simulations are used for 3 reasons \citep{white2005}; models with HPEs cannot resolve effectively at high resolution, so \citet{daley1988} suggested to apply a single equation set for all scales. Secondly, \citet{tanguay1990} saw that semi-implicit methods treat acoustic waves efficiently and that more accurate NHEs should be developed. Thirdly, \citet{white2005} judged the mathematically evolutionary derivations of HPEs as less mature compared to NHEs which are designed for classical compressible fluid dynamics. Already outside the original discipline, the meteorology, some approximations perform already less well on Earth; For the dynamics of deep oceans, the $\cos{\phi}$ terms become more important \citep{white2005} because of the larger ratio of the planetary rotation frequency to the buoyancy frequency. The larger ratio is due to the smaller buoyancy frequency in the ocean, by one order of magnitude \citep[see p. 52 of][]{gill1982}.

For understanding the observational data better, \citet{yamazaki2004}, \citet{mullerWodarg2006}, \citet{hollingsworth&kahre2010} and \citet{lebonnois2010} implemented GCMs for Jupiter, Saturn, Mars and Venus. Since first discovered \citep{mayor1995}, several hundreds of exoplanets have been observed. Exoplanets and their central stars vary widely in their parameters which makes modelling challenging \citep[see for review][]{showman2010}. Hot Jupiters are of prime interest, since they represent easier targets for observation due to their large radius and the stronger thermal emitted radiation. \citet{showman2009}, \citet{dobbs_dixon2008} and  \citet{dobbs_dixon2009} adapted some of the first GCMs to hot Jupiters.

Several groups have used GCMs or Radiative-Hydrodynamic models (RHD) to study atmospheres of (ultra) hot Jupiters and warm Neptunes \citep[e.g.][]{showman&guillot2002,showman2009,rauscher2010,heng2011a,dobbs_dixon2010,dobbs_dixon2013,mayne2014b,mayne2019,charnay2015b,kataria2015,amundsen2016,mendonca2016,zhang2017,deitrick2020,lee2021,carone2020,deitrick2022,lee2022}. Several physical processes have been added to GCMs. Regarding radiative transfer (RT), GCMs contain the Newtonian relaxation \citep[e.g.][]{showman2008, rauscher2010,heng2011a, mayne2014b,carone2020} and multi-band grey or non-grey schemes in various adaptations \citep[e.g.][]{heng2011b,rauscher2012, dobbs_dixon2013,mendonca2018} in studies for hot Jupiters. Such simplified RT schemes run in GCMs efficiently. The computational efficiency enables easier benchmarking between GCMs \citep[e.g.][]{heng2015} and to explore parameters \citep[e.g.][]{komacek2016,komacek2017,tan2019,tan2020} for investigations of dynamical regimes. \citet{showman2009}, \citet{charnay2015b} and \citet{amundsen2016} combined detailed real gas, correlated-k RT schemes to GCMs which led to more computational expensive operations. Studies such as \citet{kataria2014}, \citet{kataria2016}, \citet{amundsen2016}, \citet{parmentier2016}, \citet{schneider2009} and \citet{deitrick2022} perform GCM simulations including real gas RT schemes. In \citet{lee2021}, they compared semi-grey, non-grey picket-fence and correlated-k RT schemes and suggested to use the picket-fence scheme as simple and computationally efficient, but realistic solution.

 Regarding the validity, \citet{tokano2013} raises doubts about the primitive equations in relatively thick atmospheres. In such thick atmospheres, the ratio of scale height to the planetary radius gets sufficiently large so that the traditional approximation becomes inappropriate. Similarly, \citet{tort2015} and \citet{gerkema2008} analysed the limits of the primitive equations for Earth respectively, the traditional approximation in particular. In the past decade, a few models with the full or deep Navier-Stokes equations have been developed for exoplanets: the 3D radiation-hydrodynamics model of \citet{dobbs_dixon2013}, the dynamical core of THOR \citep{mendonca2016,deitrick2020}, and the modified UM ENDGame of \citep{mayne2014b}. However, only a few studies \citep[e.g.][]{mayne2014b,mayne2019,deitrick2020} have investigated differences between simulations with different dynamical equations for exoplanets. While two studies uses two-stream, double-grey RT respectively, only \citet{mayne2019} applied detailed real gas, correlated-k RT scheme for the comparison of the dynamical equations. They suggested to study differences emerging out of different dynamical equations by implementing a full radiative transfer solution as used in \citet{amundsen2016}.

In this study, we investigate the differing effects of simplified Navier-Stokes equations in a GCM. We use THOR GCM because of its computational efficiency, and update the RT using the picket fence scheme of \citet{lee2021}. THOR allows us to simulate atmospheres with different dynamical equations, as shown by \citep{deitrick2020} with NHD and QHD equation sets. We will focus on the NHD and QHD equation sets in our investigation similarly.

For investigating the effects between the NHD and QHD equation sets, we analyse effects in a parameter grid space appropriate for the hot exoplanet regime. We alter the gravity, rotation period and irradiation temperature at the top of the atmosphere separately to see the differences among the equations and their dependence of those parameters.

\section{THOR Model}

\citet{mendonca2016} developed the open-source GCM THOR for the purpose to study exoplanet atmosphere dynamics. Further model developments were published by \citet{mendonca2018model}, \citet{mendonca2018}, \citet{mendonca2018chemical}, \citet{deitrick2020}, and \citet{deitrick2022}. THOR simulates the global atmospheres in a full 3D icosahedral grid with a given horizontal resolution (customizable by the $g_{l\!e\!v\!e\!l\!s}$ settings). Consequently, singularities and resolution crowding at the poles do not occur like in latitude-longitude grids.

\subsection{Hydrodynamics}
THOR evolves the general non-hydrostatic Euler equations \citep{mendonca2016}. The integration schemes are horizontally explicit and vertically implicit. \citet{mendonca2018} and \citet{mendonca2018chemical} added a dry convective adjustment and a `sponge layer', as a form of drag for numerical stability similar to most contemporary GCMs. Furthermore, the model offers hydrostatic shallow (HSS), quasi-hydrostatic deep (QHD), and non-hydrostatic deep (NHD) equation sets \citep{deitrick2020}. In summary, the vertical momentum flux differs between both equation sets.

NHD and QHD vary mainly in 3 terms: $\frac{Dv_{r}}{Dt}$ the Langrangian derivative of the vertical velocity, $\mathcal{F}_{r}$ the hyperdiffusive flux and $\mathcal{A}_{r}$ the vertical component of the advection term. The terms $\frac{Dv_{r}}{Dt}$ and $\mathcal{F}_{r}$ turn to zero in the QHD case. $\mathcal{A}_{r} = \nabla(\rho \vec{v} \otimes \vec{v})$ becomes
\begin{equation}
\mathcal{A}^{QH}_{r} = \frac{\rho \vec{v}_{h} \cdot \vec{v}_{h}}{r},
	\label{eq:advection_term_QHD}
\end{equation}
where $\rho$ is the density of the air, $\vec{v}_{h}$ the horizontal momentum vector and $r$ the radial distance from the center of the planet.
For a more complete review on the NHD and QHD equation sets, see \citet{deitrick2020}.

\subsection{Picket-fence RT scheme}

A two-stream, double-grey RT scheme is available in THOR since the update made by \citet{deitrick2020}.
However, to increase the realism of the RT scheme, we use the non-grey "picket-fence" \citep{chandrasekhar1935} translated from \citet{lee2021} which refers to the approaches of \citet{parmentier2014} and \citet{parmentier2015}. The picket-fence approach of \citet{lee2021} simulates the radiation propagating in 5 bands (3 visible, 2 infrared) through the atmospheric layers. The picket fence scheme uses two representative opacities: the molecular and atomic line opacity, and the general continuum opacity. The values of these opacities are derived from the Rosseland mean opacity computed through fitting functions \citep[analytically derived by][]{parmentier2014,parmentier2015}.

Ignoring the effects of multiple scattering, the net flux, $F_{n\!e\!t,i} [Wm^{-2}]$, at each level $i$ is given by the difference of the outgoing longwave flux, $F_{I\!R\uparrow,i}$, to the downwards longwave flux, $ F_{I\!R\downarrow,i}$, and shortwave fluxes, $ F_{V\downarrow, i}$,
\begin{equation}
F_{n\!e\!t,i} = F_{I\!R\uparrow,i} - F_{I\!R\downarrow,i} - F_{V\downarrow, i}.
	\label{eq:Fnet}
\end{equation}
Assuming hydrostatic equilibrium, the partial optical depth, $\Delta \tau_i$,  \citep{parmentier2014} is given by
\begin{equation}
\Delta \tau_{i,b}=\kappa_{R,i,b}(p_i,T_i) \Delta h_i \rho_i,
	\label{eq:tau}
\end{equation}
where the opacity, $\kappa_{R,i,b}[m^2 kg^{-1}]$, for the level $i$ and for the band $b$, the height difference between levels $\Delta h_i$ and the density $\rho_i$ determines the partial optical depth. We implemented a B\'ezier interpolation to compute $p_i$ and $T_i$ from the pressure and temperature at the layers of the model from the altitude levels. We consider the atmosphere above the model grid using a ghost layer with optical depth
\begin{equation}
\Delta \tau_{g\!h\!o\!s\!t}=\frac{\kappa_{R,t\!o\!p}(p_{t\!o\!p},T_{t\!o\!p}) p_{t\!o\!p}}{g}.
	\label{eq:tau_ghost}
\end{equation}
where $p [Pa]$ stands for the pressure and $g [ms^{-2}]$ for the gravity. The Rosseland mean opacity is calculated \citep{parmentier2015} as
\begin{equation}
\frac{1}{\kappa_R} \equiv \frac{ \int_{0}^{\infty} \frac{1}{\kappa_\lambda} \frac{d B_\lambda}{d T} \,d\lambda }{\int_{0}^{\infty}  \frac{d B_\lambda}{d T} \,d\lambda},
	\label{eq:rosseland_mean}
\end{equation}
where $\kappa_\lambda [m^2g^{-1}]$ is the wavelength dependent opacity and $dB-{\lambda}/dT$ the temperature derivative of the Planck function. In order to quantify the non-greyness of the atmosphere, $\kappa_{i,b}$ is computed for each level as well as for each V and IR band through the relation
\begin{equation}
\kappa_{P,i,b} \equiv \gamma_b\kappa_{R,i,b}(p_i,T_i),
	\label{eq:kappa}
\end{equation}
where $\gamma_b$ is the opacity ratio coefficient \citep[]{parmentier2014,parmentier2015} for each band, $b$,  and $\kappa_R(p_i,T_i) [m^2 kg^{-1}]$ the Rosseland mean opacity for each band $b$. Adding the opacity ratio coefficient to the Equations \ref{eq:tau} and \ref{eq:tau_ghost}, the equations become
\begin{subequations}
\begin{align}
\Delta \tau_{i,b}=\gamma_b\kappa_{R,i,b}(p_i,T_i) \Delta h_i \rho_i,\\
\Delta \tau_{g\!h\!o\!s\!t}=\frac{\gamma_b\kappa_{R,t\!o\!p,b}(p_{t\!o\!p},T_{t\!o\!p}) p_{t\!o\!p}}{g},
\label{eq:tau_new}
\end{align}
\end{subequations}
where $\gamma_b = 1$ accounts for a grey atmosphere and $\gamma_b >1$ for a non-grey
atmosphere in the band $b$ \citep{king1956}. Applying the formation definition in Equation \ref{eq:rosseland_mean}, the Rosseland mean opacity is computed from fitting function and tables in \citet{freedman2014}.

The $\gamma_b$, $\beta$, and the Bond albedo ,$A_B$, depend on the effective temperature, $T_{\rm eff} [K]$. Therefore, $T_{\rm eff} [K]$ is computed in advance according to \citet{parmentier2015} for each column as
\begin{equation}
T_{\rm eff} = \sqrt[4]{T_{i\!n\!t}^{4} + (1 - A_B)\mu_\star T_{i\!r\!r}^{4}},
	\label{eq:Teff}
\end{equation}
where $T_{i\!n\!t} [K]$ is the internal temperature, $\mu_\star = \cos{\phi}\cos{\theta}$  the cosine angle from the sub-stellar point, $A_B$ the Bond albedo and $T_{i\!r\!r}$ the irradiation temperature at the substellar point. Equation \ref{eq:Teff} simplifies to $T_{\rm eff} = T_{i\!n\!t}$ for nightside profiles. We use the fit of \citet{parmentier2015} to the Bond albedo, $A_B$, which depends on $g$, the gravity, and $T_{\rm eff}$.

The RT scheme operates for each column as follows:
\begin{enumerate}
\item Computation of the Bond albedo according to \citet{parmentier2015}, with $T_{\rm eff}$ assuming $\mu_{\star} = 1/\sqrt{3}$.
\item Computation of all $\gamma_b$ and $\beta$ with $T_{\rm eff}$ calculated according to Equation \ref{eq:Teff} for each column and according to the fitting coefficient tables in \citet{parmentier2015}
and definitions in \citet[]{parmentier2014}.
\item Compute the IR band Rosseland mean opacity, $\kappa_R(p_i,T_i)$, in each layer from the fits and tables of \citet[]{freedman2014}.
\item Compute the V band opacities in each layer using the $\gamma_b$ and $\kappa_R$ relationships as in  the Equation \ref{eq:kappa} .
\item Compute the IR band opacities in each layer using the $\gamma_b$ and $\kappa_R$ relationships as in the Equations \ref{eq:kappa}.
\item Compute the optical depth as in the Equation \ref{eq:tau_new}.
\item Compute the two-stream calculations for each V and IR band.
\end{enumerate}

\subsubsection{Shortwave radiation}
For the stellar flux at the top of the atmosphere, $F_0$ $[Wm^{-2}]$, is given by the irradiation temperature, $T_{i\!r\!r} [K]$,  \citep{guillot2010} as
\begin{equation}
F_0 = \sigma T_{irr}^{4} = \bigg(\frac{R_\star}{a}\bigg)^{2} \sigma T_{\star}^{4},
	\label{eq:f0_guillot}
\end{equation}
where $\sigma [Wm^{-2}K^{-4}]$ is the Stefan-Bolzmann constant, $R_\star [m]$ the stellar radius, $a [m]$ the semi-major axis and $T_\star [K]$ the effective temperature of the star.

The downward shortwave flux at each layer $i$ is summed over the short-wave bands with the optical depth to layer $i$, $\tau_{i,b}$
\begin{equation}
F_{V\downarrow, i} = (1 - A_B)F_{0}\mu_\star \sum \limits_{b=1}^{N_b}\beta_{V,i}\exp \Bigg(- \frac{\tau_{i,b}}{\mu_\star} \Bigg),
	\label{eq:V_flux_down}
\end{equation}
where $N_b$ stays for the number of V bands (3 in this study), and $\beta_{V,i}$ the fraction of stellar flux in band $b$ (1/3 in this study). 

\subsubsection{Longwave radiation}
We implement a two-stream solution using the short characteristic method with linear interpolants introduced by \citet{olsen1987}. The downward intensity, the intensity of the ghost layer, the upward intensity and the upward intensity at the bottom $I_{IR, g,i}[Wm^{-2}sr^{-1}]$, at levels $i$ and in $IR$ bands for a Gaussian quadrature $g$ point are given by
\begin{subequations}
\begin{align}
I_{\downarrow,IR, g,i} &=  (\epsilon_{0i} -1)I_{\downarrow,IR, g,i+1} + \alpha_{i}^{-}B_{i+1,IR} +\beta_{i}^{-}B_{i,IR},\\
I_{\downarrow,IR, g,g\!h\!o\!s\!t\!} &= [1 -\exp{(\tau_{IR,top)} / \mu_g}] B_{top-1},\\
I_{\uparrow,IR, g,i} &=  (\epsilon_{0i}-1)I_{\uparrow,IR, g,i-1} + \beta_{i}^{+} B_{i,IR} + \gamma_{i}^{+} B_{i-1,IR},\\
I_{\uparrow,IR, g,b\!o\!t\!t\!o\!m} &=  B_{i\!n\!t\!} + I_{\downarrow,IR, g,b\!o\!t\!t\!o\!m},
\end{align}
	\label{eq:IR_intensities}
\end{subequations}
where 
\begin{subequations}
\begin{align}
\epsilon_{0i} &= 1 - exp(-\Delta \tau_{IR,i} / \mu_g), \\
\epsilon_{1i} &= \Delta \tau_{IR,i} / \mu_g - 1 + exp(-\Delta \tau_{IR,i} / \mu_g) = \Delta \tau_{IR,i} / \mu_g - \epsilon_{0i},
\end{align}
	\label{eq:epsilon_1i}
\end{subequations}
with the coefficients for linear interpolation
\begin{subequations}
\begin{align}
\alpha_{i}^{-} &=  \epsilon_{0i} - \epsilon_{1i} / \Delta \tau_{IR,i}, \\
\beta_{i}^{-} &= \epsilon_{1i} / \Delta \tau_{IR,i},  \\
\gamma_{i}^{-} &= 0, \\
\alpha_{i}^{+} &= 0, \\
\beta_{i}^{+} &= \epsilon_{1i} / \Delta \tau_{IR,i},  \\
\gamma_{i}^{+} &= \epsilon_{0i} - \epsilon_{1i} / \Delta \tau_{IR,i}, 
\end{align}
	\label{eq:parabolic_coefficients}
\end{subequations}
and for optical depth lower than $10^{-6}$ the coefficients are set to
\begin{subequations}
\begin{align}
\alpha_{i}^{-} &=  0.5\cdot \epsilon_{0i} (B_{IR,i+1} + B_{IR,i}) /B_{IR,i+1}, \\
\beta_{i}^{-} &= 0,  \\
\gamma_{i}^{-} &= 0, \\
\alpha_{i}^{+} &= 0, \\
\beta_{i}^{+} &= 0.5 \cdot \epsilon_{0i} (B_{IR,i} + B_{i-1,b}) /B_{IR,i},  \\
\gamma_{i}^{+} &= 0, 
\end{align}
	\label{eq:parabolic_coefficients_iso}
\end{subequations}
which reduces to the isothermal approximation to avoid numerical instability.  $\mu_g$ is the emission angle, and $B_{IR,i} [Wm^{-2}sr^{-1}]$ the wave-length integrated blackbody intensity defined as
\begin{equation}
B_{IR,i} = \beta_{I\!R} B_{i} = \beta_{I\!R} \sigma T_{i}^{4}/\pi,
	\label{eq:blackbody_intensity}
\end{equation}
where $\beta_{I\!R,b}$ is the fraction of flux in band $b$. This forces the RT scheme to return to the isothermal approximation at low optical depths where numerical stability would be an issue.
The upward and downward longwave fluxes $F_{I\!R,i} [Wm^{-2}]$ are given by
\begin{subequations}
\begin{align}
F_{I\!R\downarrow,i} &=  2\pi \sum \limits_{b}^{N_{I\!R}} \sum \limits_{g}^{N_g} w_g \mu_g I_{\downarrow,I\!R, g,i} \\
F_{I\!R\uparrow,i} &= 2\pi \sum \limits_{I\!R}^{2} \sum \limits_{g}^{5} w_g \mu_g I_{\uparrow,I\!R, g, i},
\end{align}
\label{eq:IR_flux}
\end{subequations}
where $N_{I\!R}$ is the number of IR bands (here 2), $N_g$ the number of Gauss quadrature points (here 2) and $w_g$ the quadrature weight.

\subsection{Altitude setup}

\begin{figure}
	\includegraphics[width=\columnwidth]{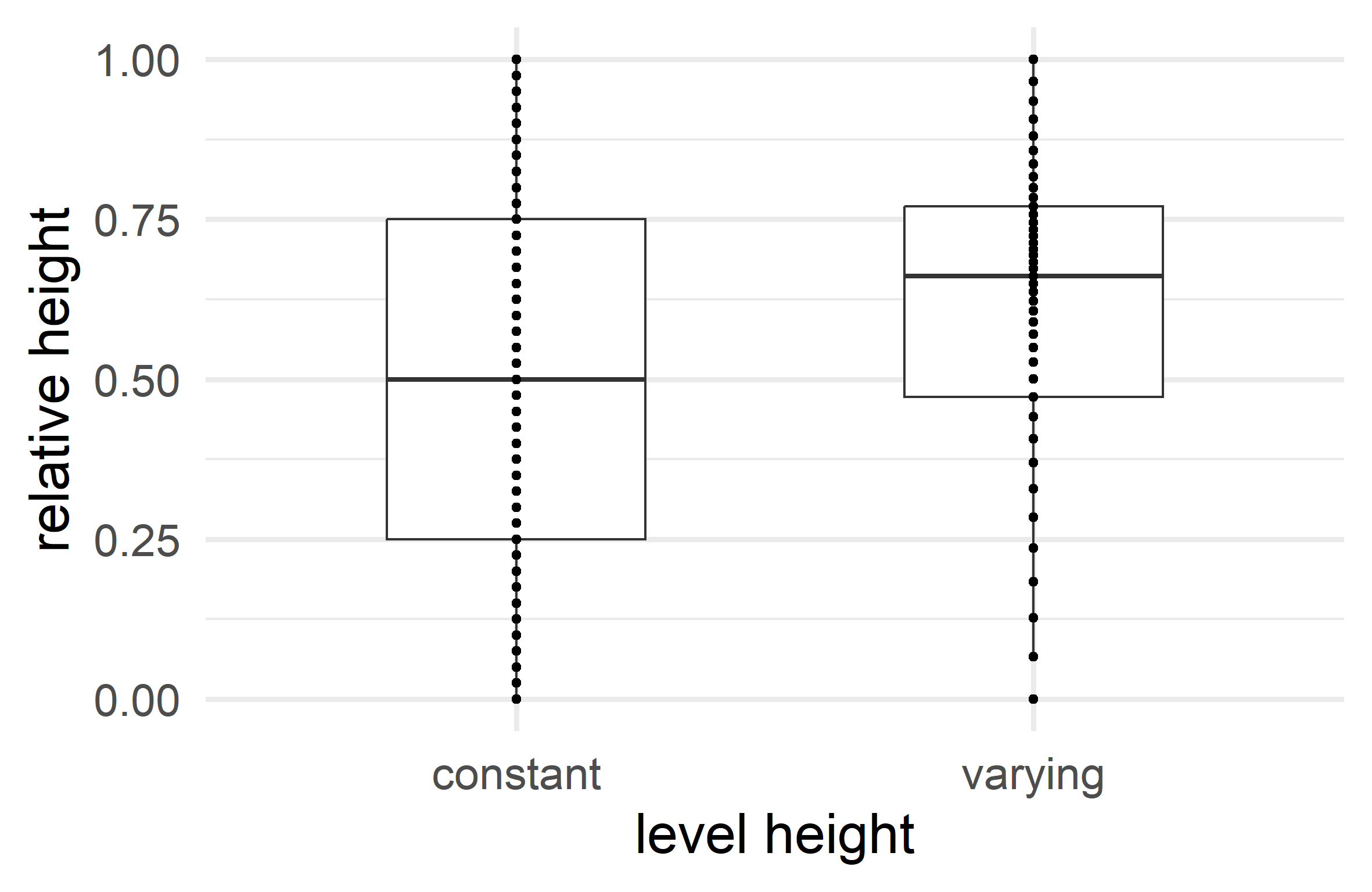}
    \caption{Distribution of the cumulative heights of the levels in the new varying height setting compared to the standard setting with constant level heights.}
    \label{fig:height_grids}
\end{figure}

Strong temperature gradients pose a problem in the simulations with a low vertical resolution. Instead of increasing vertical resolution, which would increase numerical cost, we instead alter the relative thickness of the atmosphere layers. Where the temperature gradient remains relatively constant (e.g. deeper atmosphere), a higher thickness can be tolerated. Therefore, we create a function, which increases the vertical resolution at a chosen relative height, $h_{rel}$, defined by
\begin{subequations}
\begin{align}
h_{lev}(i) &= z(i)h_{top}, \\
h_{lay}(i) &= [h_{lev}(i) + h_{lev}(i+1)]/2,
\end{align}
	\label{eq:alt_grid}
\end{subequations}
where $i$ stands for the height index, $h_{lev}$ for the altitude at the levels (interfaces), $h_{lay}$ for the altitude at the layers, $h_{top}$ is the chosen top altitude of the model, $z(i)$ gives the relative height and was defined by
\begin{subequations}
\begin{align}
y(i) &= a(i-c)^{3} + b(i-d)^{2}, \\
z(i) &= \frac{y(i)+y(0)}{y(N_{lev}-1)+y(0)}, 
\end{align}
	\label{eq:alt_top}
\end{subequations}
where $c$ and $d$ are parameterized as
\begin{subequations}
\begin{align}
c &= \frac{h_{rel}(N_{lev}-1)}{2} + \frac{b}{3a} + \frac{(N_{lev}-1)}{4}, \\
d &= \frac{1}{2},
\end{align}
	\label{eq:alt_coefficients}
\end{subequations}
where $a$ and $b$ are parameters which can be chosen.
In this study, we set $h_{rel}=0.7$, $a=1$ and $b=6$ for our simulations. Figure \ref{fig:height_grids} illustrates the different heights of the levels in the new setting compared the standard setting. The new scheme aims to have a slightly smother T-p profiles where temperature gradients are large like at pressures $p < 10^{5}\: Pa$.

\subsection{Initial condition setup}

We assume an initial T-p profile given by the picket-fence analytical solution at the substellar point. We implemented the suggestion of \citet{sainsbury2019} aiming for a hot adiabatic profile for the deep atmosphere of hot Jupiters. Furthermore, a hotter T-p profile can quickly cool down towards a realistic adiabatic gradient compared to a warming up from colder temperatures. The internal temperature, $T_{i\!n\!t} [K]$, was calculated in advance, using the expression of \citet{thorngren2019}. A pressure grid with $1'000$ grid points is generated by
\begin{equation}
p(x) = p_{r\!e\!f}e^{-\frac{20(x)}{10^{3}}},
	\label{eq:init_press_grid}
\end{equation}
where $p_{r\!e\!f}$ is the reference pressure.
The opacity at the layer $i$ is defined as
\begin{equation}
\tau_{i}=\tau_{i+1} + \kappa( p_{i+1},T_{i+1}) (p_i - p_{i+1})/g.
	\label{eq:tau_ghost_initial}
\end{equation}
The scheme of the initial conditions operates as follows:
\begin{enumerate}
\item Computation of the Bond albedo according to \citet{parmentier2015}, with $T_{\rm eff}$ assuming $\mu_{\star} = 1/\sqrt{3}$.
\item Computation of all $\gamma_b$, $\gamma_p$ and $\beta$ with $T_{\rm eff}$ calculated according to Equation \ref{eq:Teff} for each column and according to the fitting coefficient tables in \citet{parmentier2015}
and definitions in \citet{parmentier2014}.
\item Compute the IR band Rosseland mean opacity, $\kappa_R(p_i,T_i)$, in each layer from the fits and tables of \citet[]{freedman2014}.
\item Compute the temperature from the top to the bottom of the atmosphere with a first guess followed by a convergence loop.
\item Compute the adiabatic correction of the initial T-p profile according to \citet{parmentier2015}.
\item Compute an initial altitude grid in addition to the T-p profile with the hydrostatic equation in the bottom up approach.
\item Interpolate the temperature with both altitude grids and the initial temperature structure.
\item Compute the T-p profile with the hydrostatic equation and the reference pressure from bottom up.
\end{enumerate}

\section{Test cases}

\begin{figure}
	\includegraphics[width=\columnwidth]{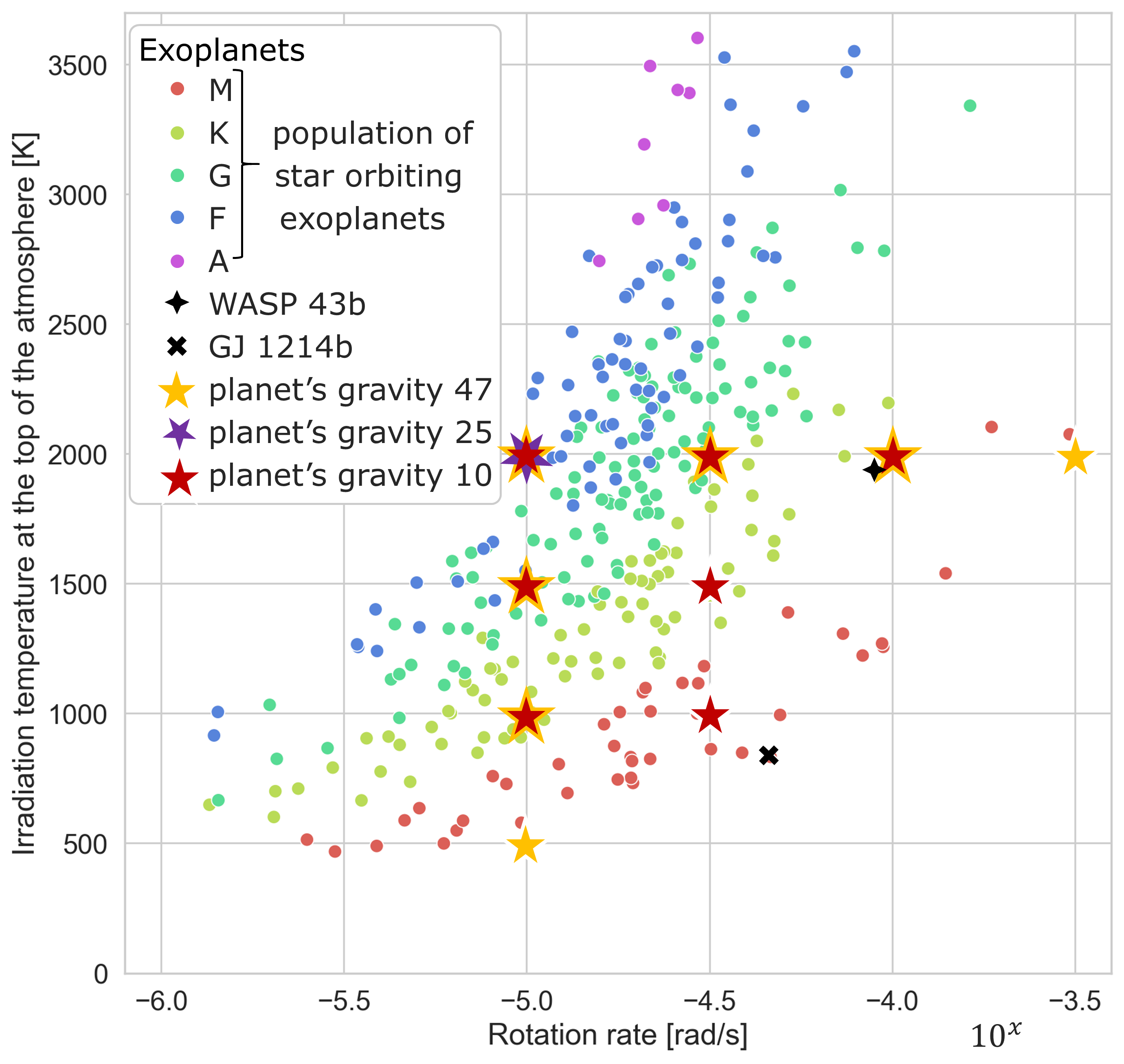}
    \caption{Grid of simulated exoplanetary parameters compared to the known exoplanets organized by host star type \citep[retrieved on the 22.11.2021 from the entries with sufficient information in the NASA exoplanet archive][]{nasa2020planetary}.}
    \label{fig:grid_exoplanets}
\end{figure}

For investigating the differences between the NHD and QHD equation sets, we run simulations across a parameter grid. In the JWST mission, WASP 43b will be among the first exoplanets to be observed with the MIRI/LRS instrument \citep{bean2018, venot2020} and many more exoplanets will follow in the coming years. Therefore, we used WASP 43b as role model planet and altered only the parameters for the rotation rate $\Omega$, $g$ and $T_{\rm eff}$. The $T_{\rm eff}$ in the Equation \ref{eq:Teff} was changed in the way that the $T_{irr}$ reaches our targeted values. Additionally, we analyse the effects rising from altering $\Omega$, $g$ and $T_{\rm eff}$ in regard to the differing terms $\frac{Dv_{r}}{Dt}$, $\mathcal{F}_{r}$ and $\mathcal{A}_{r}$ in the NHD and QHD case. Due to the lack of computational resources, we performed simulations across 9 parameter sets. Figure \ref{fig:grid_exoplanets} illustrates the grid values with the altering $\Omega$, $g$ and $T_{\rm eff}$ one by one. Table~\ref{tab:sim_paras} lists the other parameters for the simulations. For the divergence-damping and hyperdiffusion coefficients, we follow the suggestions by \citet{hammond2022}. The simulations are computed over $5'100$ days. We take the mean of the last $10$ outputs covering $100$ days. Each pair of NHD-QHD simulations share the same altitude grid. To compare the $18$ simulations, the outputs are interpolated and extrapolated to pressures ranging from $10^{8}\: Pa$ to $10^{3}\: Pa$. For the first 100 days, $D_{d\!i\!v}$ and $D_{h\!y\!p,v}$ was increased by a factor of $10$ to damp waves caused by initial instabilities.

In our results, we compare and contrast the NHD and QHD T-p profiles, maps showing the temperature and horizontal wind velocity at $10^{4}\: Pa$, mean zonal wind, vertical and horizontal momenta-pressure profiles, Outgoing Longwave Radiation (OLR), OLR phase curve, radiative and zonal wind timescales. Additionally, we generate further composites with NHD and QHD equation sets which we present in the \textbf{supplementary file}; temperature, horizontal and vertical wind at $10^{4}\: Pa$, the streamfunction $\Psi$, the tidally-locked streamfunction $\Psi '$, the components of the Helmholtz decomposition, vertical and horizontal density acceleration and the sign of the $\frac{vtan(\Phi)}{10w} - 1$ for quality assessment  \citep[like in][]{mayne2019}. The vertical and horizontal (zonal) density acceleration is computed as in \citet{hammond2020,hammond2021}. In the discussion, we classify the results into climate states based on the simulations with the NHD case and relate the results to the literature. Furthermore, we computed (large-scale flow) characteristic quantities and scales including the scale height $H$, Rossby number $Ro$, Rossby deformation radius $L_{D}$, Rhines scale and the Brunt–Väisälä frequency $N$. We relate these characteristic values to climate states in the discussion.
The sections \ref{tidally}, \ref{streamfuction}, \ref{helmholtz} and \ref{OLR}, \ref{timescales} and \ref{large_scale_flow_quantities} of the appendix describe how the tidally-locked coordinates and wind, the streamfunction $\Psi$, the tidally-locked streamfunction $\Psi '$, Helmholtz decomposition, the OLR phase curve, the radiative and zonal timescales and the large-scale flow quantities and scales are calculated.

\begin{table*}
	\centering
	\caption{Defined parameters for the all simulations}
	\label{tab:sim_paras}
	\begin{tabular}{lcccll} 
		\hline \hline
		Symbol & Model runs & Units & Description & Source\\
		\hline \hline
		$R_p$ & $72'427'000$ & $[m]$ & Planet radius & \citet{gillon2012}\\
		$g$ & $10$, $25$, $47.39$ & $[m\,s^{-2}]$ & Gravity & -\\
		$\Omega$ & $10^{-5}, 10^{-4.5}, 10^{-4}$ & $[rad \, s^{-1}] $ & Rotation rate &\\
		$R_d$ & $3714$ & $[JK^{-1}kg^{-1}]$ & Gas constant & \citet{deitrick2020}\\
		$C_p$ & $13'000$ & $[JK^{-1}kg^{-1}]$ & Atmospheric heat capacity & \citet{deitrick2020}\\
		$P_{r\!e\!f}$ & $1\cdot 10^{8}$ & $[Pa]$ & References pressure at the bottom & -\\
		$T_{i\!n\!t}$ & $535$ & $[K]$ & temperature of internal heat flux & according to \cite{thorngren2019}\\
		\hline
        $T_{i\!r\!r}$ & $500$, $1'000$, $1'500$, $2'000$ &$[K]$ & Irradiation temperature at TOA & -\\
		$T_{\star}$ & $1'108$, $2'217$, $3'325$, $4'434$ &$[K]$ & Stellar effective temperature & computed from $T_{irr}$ see \citet{guillot2010}\\
		$R_{\star}$ & $0.667$ & - & Stellar radius ratio relative to Earth & \citet{gillon2012}\\
		$a$ & $0.01525$ & $[au]$ & Orbital distance & \citet{gillon2012}\\
		$met$ & $0$ & - & stellar metalicity [Fe/H] & -\\
		
		\hline
		$\Delta t_M$ & $300$ & $[s]$  & Time step & -\\
		$t_{M,t\!o\!t}$ & $5'100$ & $[Earth days]$ & Run length & -\\
		$g_{l\!e\!v\!e\!l}$ & $5$ & - & Grid refinement level ($g_{l\!e\!v\!e\!l}=5 \sim 2^\circ$) & -\\
		$v_{l\!a\!y\!e\!r}$ & $40$ & - & Number of vertical layers & -\\
		$O_{h\!y\!p,v}$ & $6$ & - & Order of hyperdiffusion operator & \citet{hammond2022}\\
		$D_{d\!i\!v}$ & $0.01$ & - & Divergence damping coefficient & \citet{hammond2022}\\
		$D_{h\!y\!p,h}$ & $0.0025$ & - & Horizontal hyperdiffusion coefficient & \citet{hammond2022}\\
		$D_{h\!y\!p,v}$ & $0.001$ & - & Vertical hyperdiffusion coefficient & \citet{hammond2022}\\
		\hline
		$\eta_{s\!p}$ & $0.8$ & - & Bottom of sponge layer (fraction of $z_{t\!o\!p}$) & \citet{hammond2022}\\
        $k_{s\!p,h}^{R}$ &  $0.001$ &$[s^{-1}]$ & Horizontal rayleigh sponge strength & \citet{hammond2022}\\
        $k_{s\!p,v}^{R}$ &  $0.0001$ &$[s^{-1}]$ & Vertical rayleigh sponge strength & -\\
        $k_{s\!p}^{HD}$ &  $0.01$ &$[s^{-1}]$ & Hyperdiffusive sponge strength & \citet{hammond2022}\\
		$n_{l\!a\!t\!s}$ & $20$ & - & Number of sponge layer latitude bins &  \citet{deitrick2020}\\
		
		\hline
	\end{tabular}
\end{table*}

\section{Results}

\subsection{Altering Rotation Rate}

\begin{figure*}
	\includegraphics[width=\textwidth]{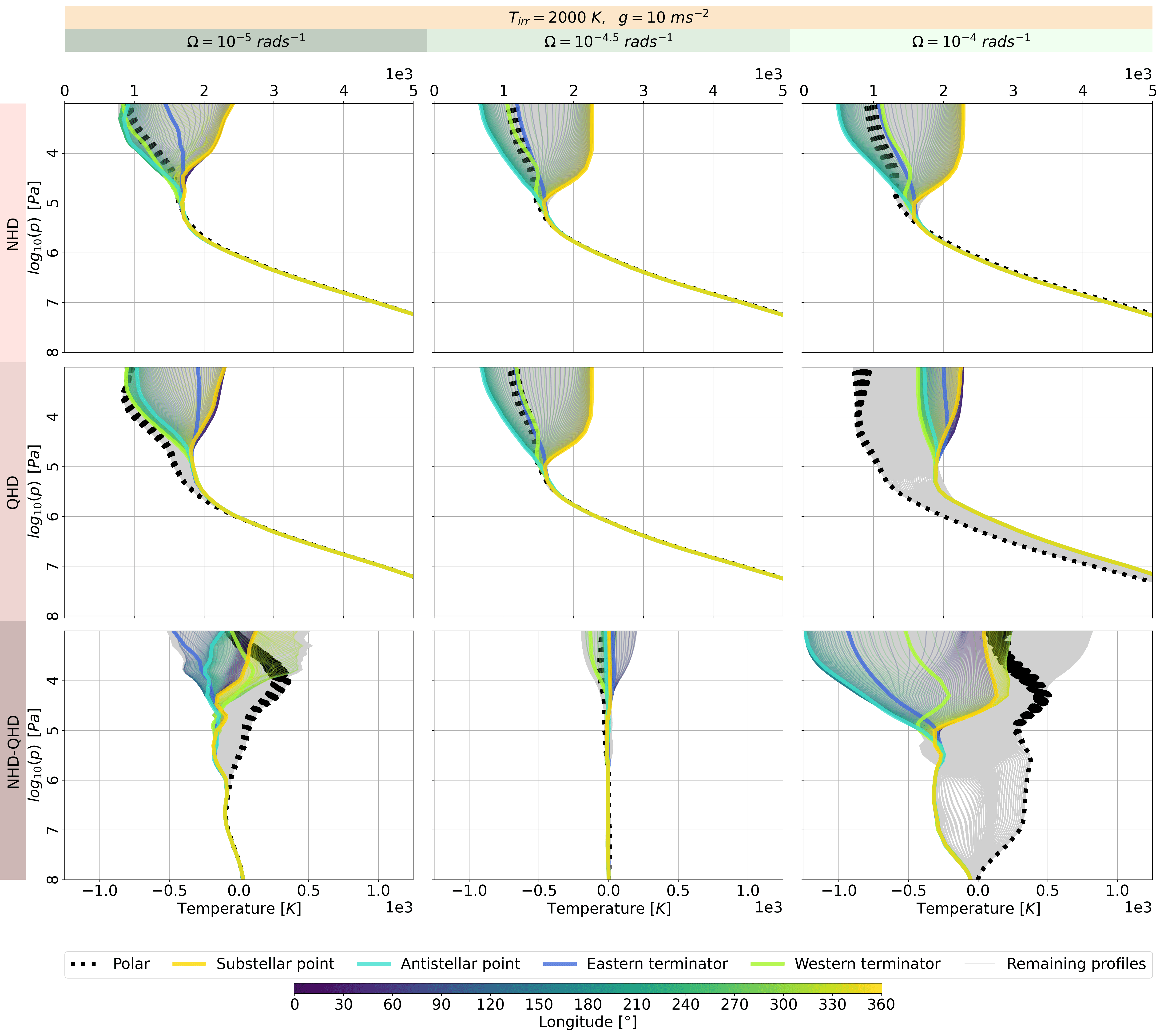}
    \caption{All font sizes and line widths have been increased. Shared scales and colour-bars T-p profiles of covering entire planet for the NHD and QHD equation sets with $g = 10 \:ms^{-2}$, $T_{irr} = 2'000 \:K$ and with altering $\Omega$. The coloured lines indicate T-p profiles along the equator and its coordinates by the colourbar. The dotted black thin line shows T-p profiles at the latitudes 87°N and 87°S. The bold coloured lines represent T-p profiles at the western, eastern terminators, sub- and antistellar point. The grey lines represents all the other T-p profiles.}
    \label{fig:TP_profile_composit_10G_2000}
\end{figure*}

Figure \ref{fig:TP_profile_composit_10G_2000} shows T-p profiles (vertical temperature-pressure profiles) for the NHD and QHD equation sets with $g = 10 \:ms^{-2}$, $T_{irr} = 2'000 \:K$ and altering $\Omega$. Looking at the differences between the NHD and QHD equation sets at the slow rotation rate, the regions around the eastern terminator and antistellar point reach much lower temperatures in the NHD case at pressures $< 50'000 \:Pa$. In contrast, the areas around the poles and western terminator are warmer in the NHD case. At the fast rotation rate, the temperature differences between the NHD and QHD cases increase by two times in many regions. The temperatures at antistellar point, eastern terminator and at the western terminator differ more than $1'000 \:K$, $800 \:K$ and $450 \:K$ at pressures $< 10^{5} \:Pa$. In general, the differences in temperatures diminish at higher pressures. In the lower atmosphere, the high rotation rate produces larger temperature differences. At the low rotation rate, temperature differences almost vanish in the deep atmosphere.

\begin{figure*}
	\includegraphics[width=\textwidth]{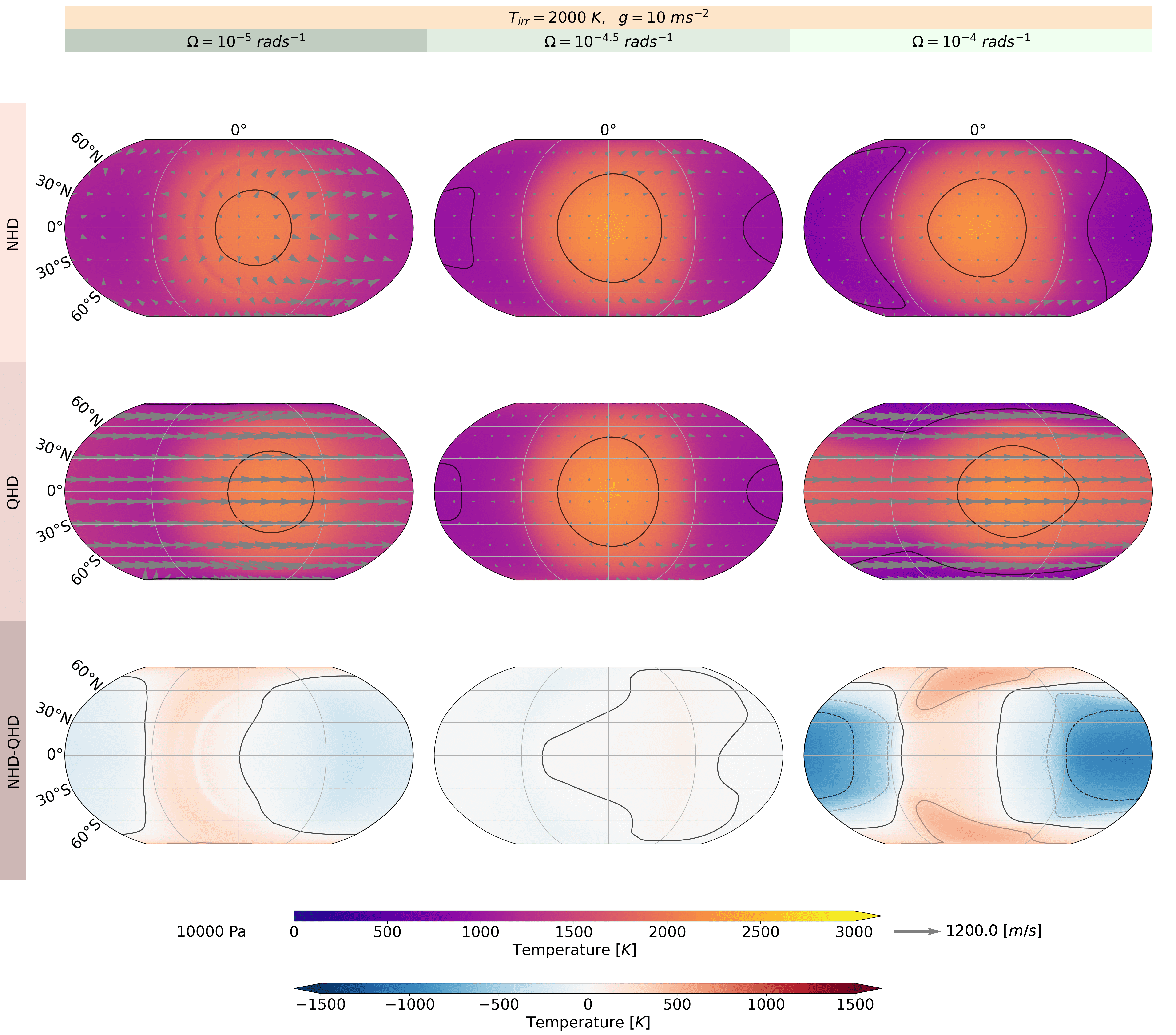}
    \caption{Temperature and wind speed at $10^{4} \:Pa$ for the the NHD and QHD equation sets with $g = 10 \:ms^{-2}$, $T_{irr}=2'000 \:K$ and with altering $\Omega$.}
    \label{fig:overview_composit_10G_2000}
\end{figure*}

Figure \ref{fig:overview_composit_10G_2000} shows the temperature and horizontal wind at $10^{4} \:Pa$ for the NHD and QHD equation sets with $g = 10 \:ms^{-2}$, $T_{irr} = 2'000 \:K$ and altering $\Omega$. The NHD case shows a hotspot shift to the east at low $\Omega$. Increasing $\Omega$ leads to smaller hotspot shifts to the east. The QHD case leads to the opposite effect with a larger shift to the east with higher $\Omega$. Regarding the horizontal wind, we see strong divergence at the substellar point at low $\Omega$ in the NHD case. Higher $\Omega$ cause more deflection by Coriolis forces. Furthermore, jets have evolved at high latitudes on the eastern hemisphere, while a retrograde equatorial jet occurs on the western hemisphere. The QHD case has evolved a large jet spanning from pole to pole at low and high $\Omega$, but a different wind field at moderate $\Omega$ interestingly. The wind field at moderate $\Omega$ looks similar to the NHD case, but varies at different pressures. The different wind field to the NHD case leads to different advection at low and high $\Omega$. Therefore, the NHD case has lower temperatures at the nightside and higher temperatures at the poles than the QHD case.

\begin{figure*}
	\includegraphics[width=\textwidth]{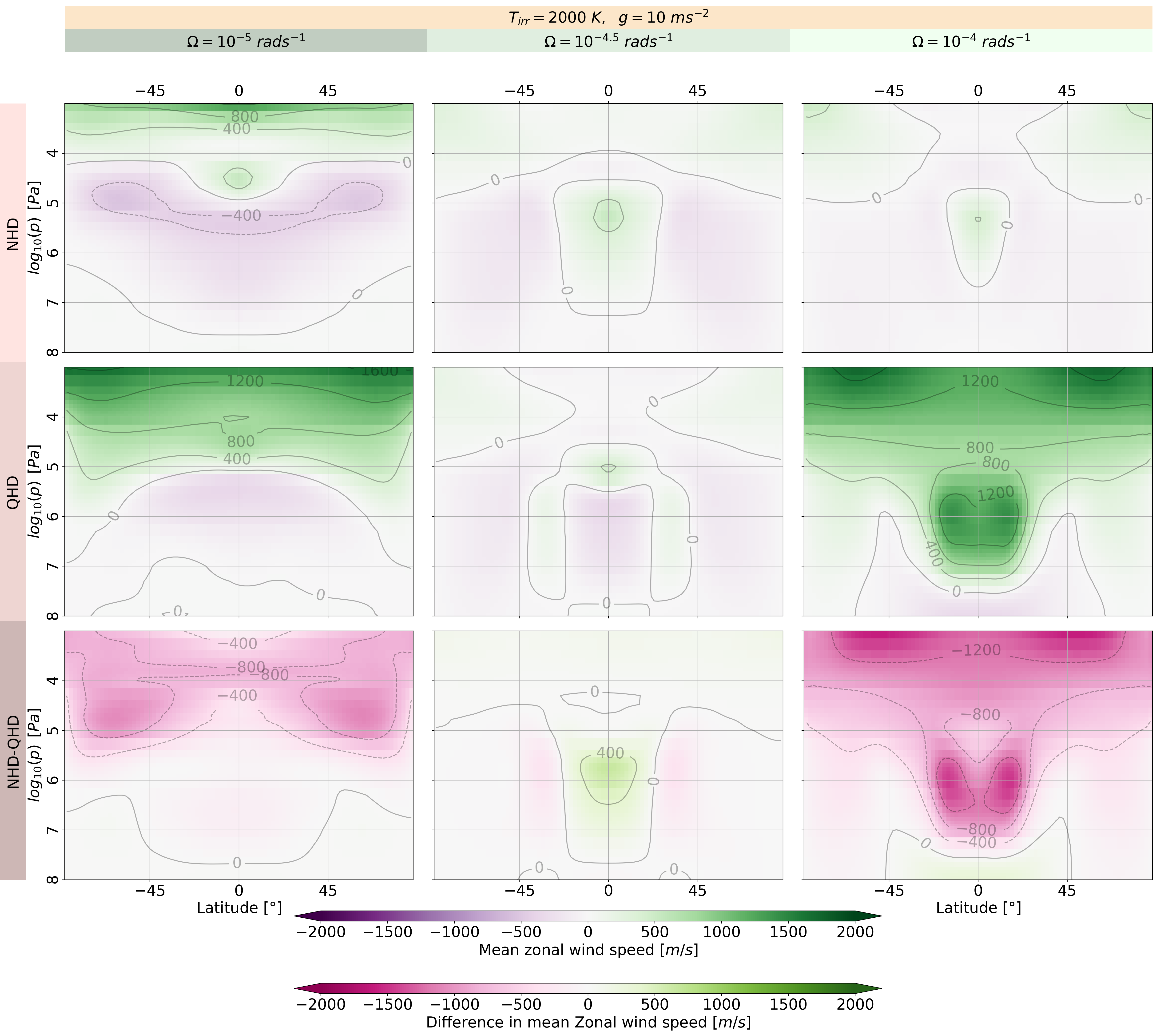}
    \caption{Zonal mean wind at each grid point for the NHD and QHD equation sets with $g = 10 \:ms^{-2}$, $T_{irr} = 2'000 \:K$ and with altering $\Omega$.}
    \label{fig:zonal_wind_composit_10G_2000}
\end{figure*}
Figure \ref{fig:zonal_wind_composit_10G_2000} shows the zonal mean wind for the NHD and QHD equation sets with $g = 10 \:ms^{-2}$, $T_{irr} = 2'000 \:K$ and altering $\Omega$. We see a 3 prograde jet system  at all $\Omega$ in the NHD case and at some $\Omega$ in the QHD case. The QHD case seems to be in transition to a 2 prograde jet system with superrotation at low $\Omega$. We ignore the very top layers because they might be affected by extrapolation and boundary conditions in some simulations. The QHD case has much higher horizontal wind speeds which increase with $\Omega$, except for the moderate $\Omega$. There is a deep retrograde jet at low $\Omega$ in both cases, but more pronounced in the NHD case. The height of the westerlies decreases the faster the rotation rate gets in the NHD case at pressure $p<10^{6} \: Pa$ (in the upper atmosphere) as observed in \citet{showman2015}.

\begin{figure*}
	\includegraphics[width=\textwidth]{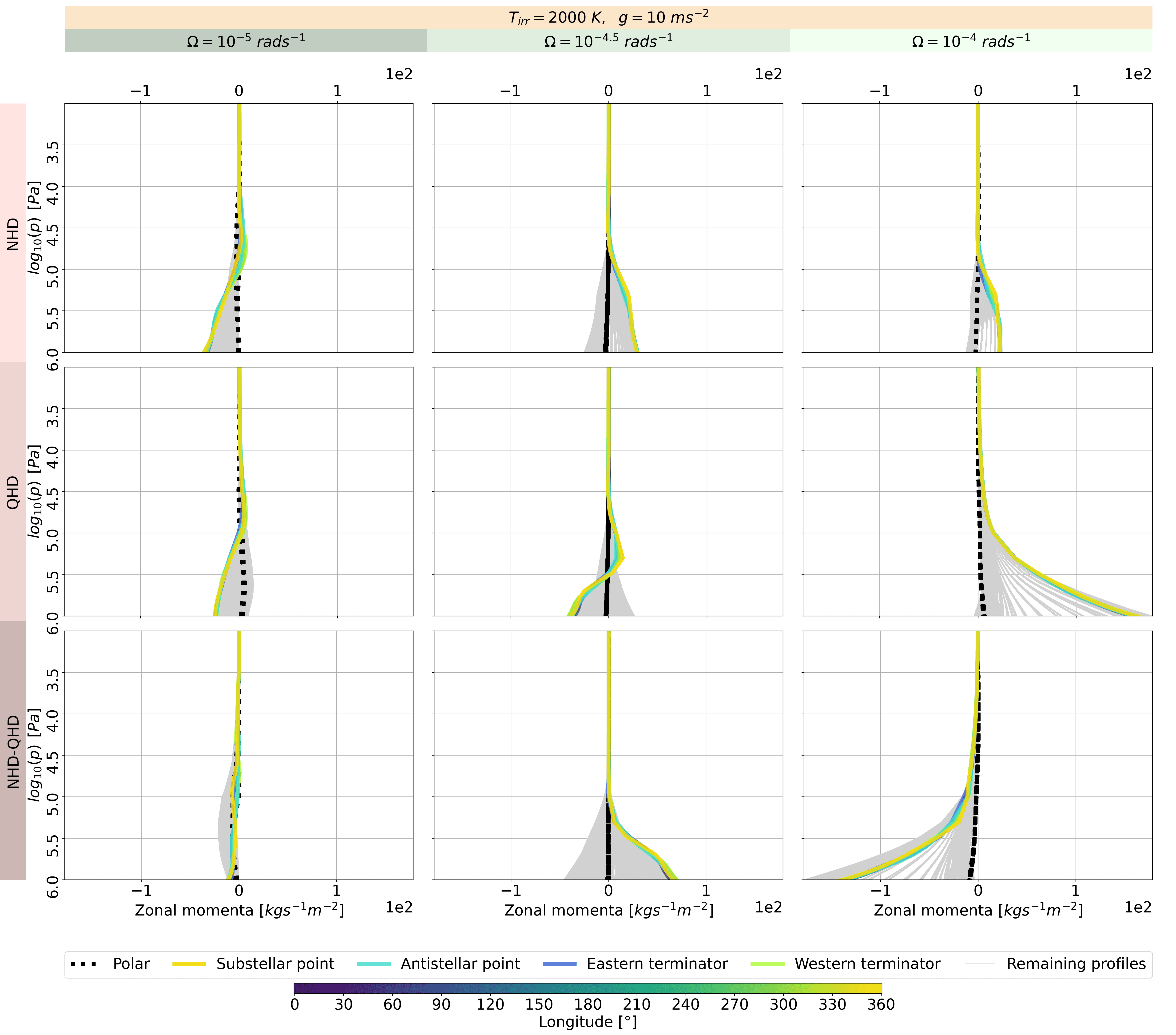}
    \caption{Zonal momenta at each grid point for the NHD and QHD equation sets with $g = 10 \:ms^{-2}$, $T_{irr} = 2'000 \:K$ and with altering $\Omega$. The profiles show only pressures $p\le 10^{6} \:Pa$ (without the pressure range $10^{6} \ge p \le 10^{8}$). The coloured lines indicate momenta profiles along the equator and its coordinates by the colourbar. The dotted black thin line shows momenta profiles at the latitudes 87°N and 87°S. The bold coloured lines represent momenta profiles at the western, eastern terminators, sub- and antistellar point. The grey lines represents all the other momenta profiles.}
    \label{fig:density_u_wind_composit_10G_2000}
\end{figure*}

Figure \ref{fig:density_u_wind_composit_10G_2000} shows the zonal momenta $[kg/m^3 m/s]$ along vertical profiles at each grid point for NHD and QHD equation set with $g = 10 \:ms^{-2}$, $T_{irr} = 2'000 \:K$ and altering $\Omega$  (without the deep atmosphere). Throughout all profiles and simulation cases, the range of the momenta get smaller with higher altitude mainly due to decreasing density.  The QHD case would follow the same trend at pressure $p<10^{6} \: Pa$, if the simulation of the moderate rotation rate did not resemble the NHD case. In the NHD case at the poles, the zonal momenta changes from a divergent to a more zonal field of momenta (see divergent component of the Helmholtz decomopostion in the \textbf{supplementary file}). The balance between eastward acceleration and vertical advection of westward momentum  \citep{showman2011superrotation} favour westward winds above major westerly jet at lower latitudes in the upper atmosphere at higher rotation rates.
The QHD simulations show two regime changes at pressure $p<10^{5} \: Pa$ with increasing rotation rate; At high rotation rates, high positive momenta dominates at pressure $p<10^{7} \: Pa$ and the flow pattern varies qualitatively to the NHD simulations. Interestingly, the flow pattern in the QHD case is qualitatively much more similar to that of the NHD case at moderate rotation rate at pressure $p<10^{5} \: Pa$ (in the upper atmosphere). But in the deep atmosphere (at pressure $p>10^{5} \: Pa$, the dynamical regime of the QHD case varies from that of the NHD case substantially.
Considering the entire simulated altitudes, the simulation with the QHD case has the smaller range of zonal momenta than the NHD case at low rotation rate. But at high rotation rate, range of the QHD case exceeds by around 5 times that of the NHD case at high rotation rates.

\begin{figure*}
	\includegraphics[width=\textwidth]{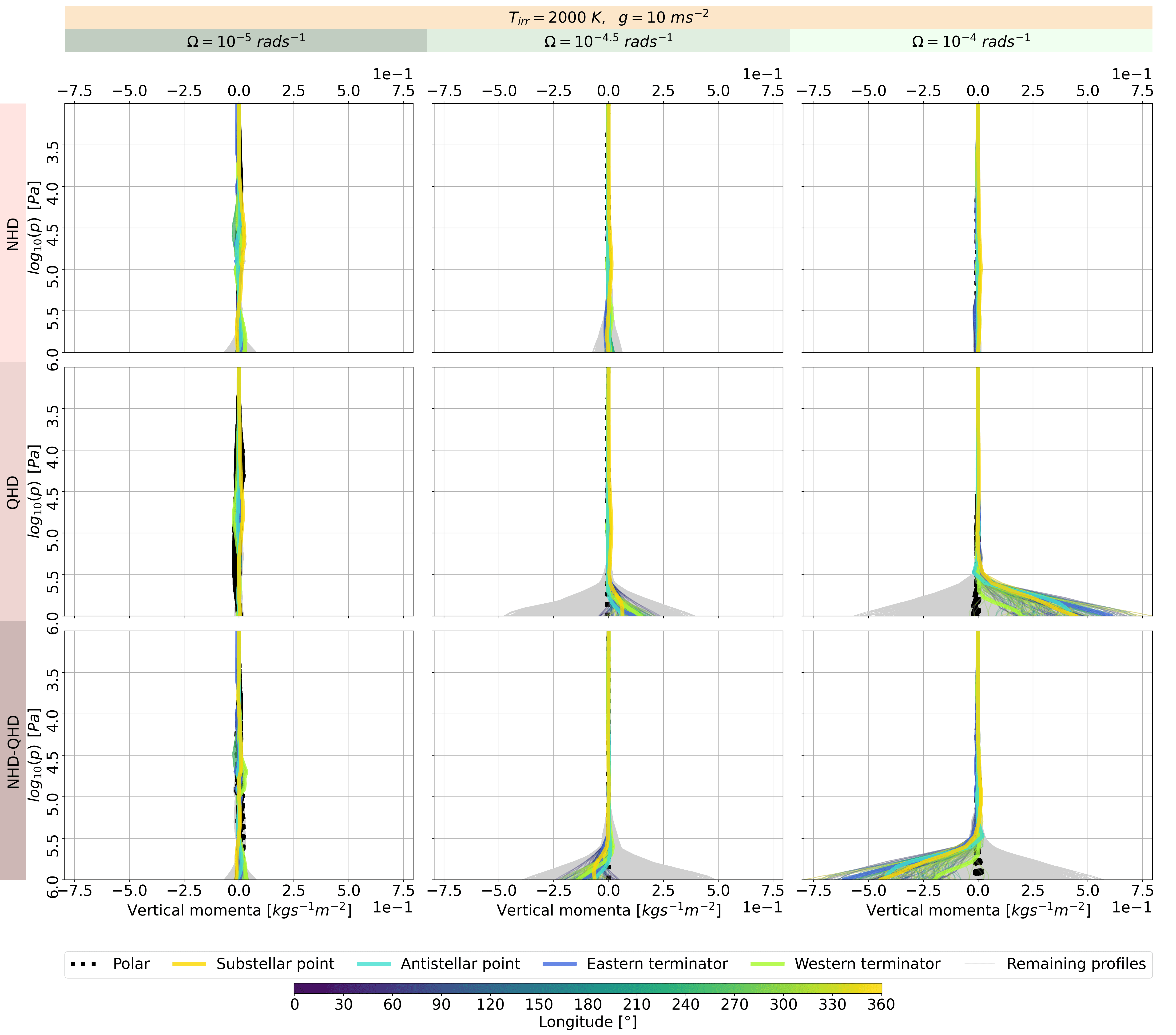}
    \caption{Vertical momenta at each grid point for the NHD and QHD equation sets with $g = 10 \:ms^{-2}$, $T_{irr} = 2'000 \:K$ and with altering $\Omega$. The profiles show only pressures $p\le 10^{6} \:Pa$ (without the pressure range $10^{6} \ge p \le 10^{8}$). The coloured lines indicate momenta profiles along the equator and its coordinates by the colourbar. The dotted black thin line shows momenta profiles at the latitudes 87°N and 87°S. The bold coloured lines represent momenta profiles at the western, eastern terminators, sub- and antistellar point. The grey lines represents all the other momenta profiles.}
    \label{fig:density_w_wind_composit_10G_2000}
\end{figure*}

Figure \ref{fig:density_w_wind_composit_10G_2000} shows the vertical momenta $[kgm^{-3} ms^{-1}]$ along vertical profiles at each grid point for the NHD and QHD equation sets with $g = 10 \:ms^{-2}$, $T_{irr} = 2'000 \:K$ and altering $\Omega$ (without the deep atmosphere). The maxima of the upward momenta sinks to higher pressure the faster the planet rotates as observed in \citet{showman2015}.

\begin{figure*}
	\includegraphics[width=\textwidth]{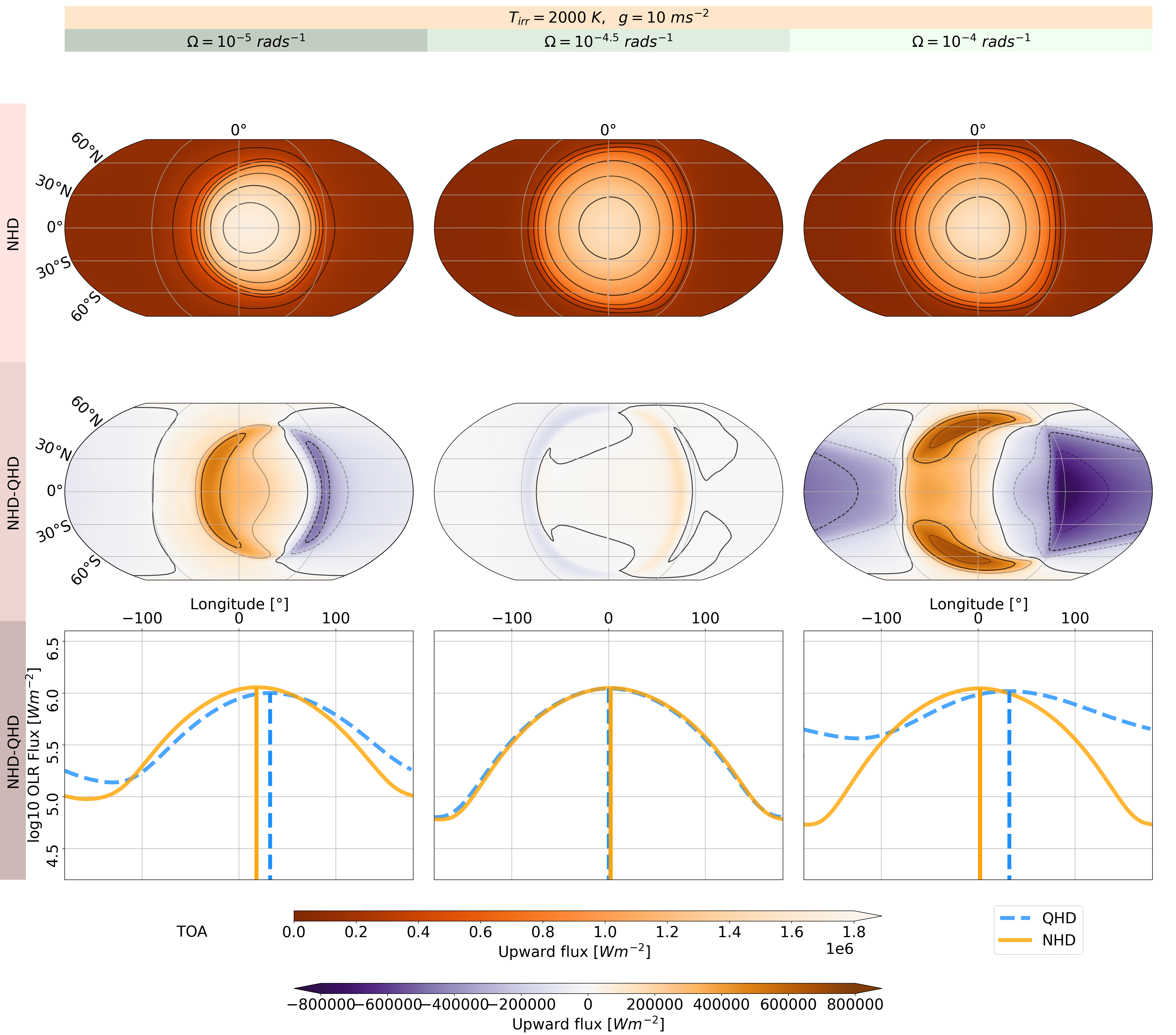}
    \caption{OLR fluxes at the top of the atmosphere for the NHD and QHD equation sets with $g = 10 \:ms^{-2}$, $T_{i\!r\!r} = 2'000 \:K$ and with altering $\Omega$. Third row: OLR phase curves.}
    \label{fig:flux_composit_10G_2000} 
\end{figure*}

Figure \ref{fig:flux_composit_10G_2000} shows the phase curves of the upward flux at the top of the atmosphere (Outgoing Long-wave Radiation - OLR) for the NHD and QHD equation sets with $g = 10 \:ms^{-2}$, $T_{irr} = 2'000 \:K$ and altering $\Omega$. The OLR reaches the highest values in the NHD case at the lowest rotation rate, whereas the QHD case does at moderate rotation rate. Furthermore, the hotspot is shifted more eastwards in the QHD case at low and high rotation rate (see the phase curves). At high rotation rate, the gap between the hotspot shifts in the simulations with the QHD and NHD equation sets developed the largest at high rotation rate. At moderate rotation, the difference in the hotspot shifts reaches the smallest value. In the region around the eastern terminator and on the night side, the NHD case remains cooler.

\begin{figure*}
	\includegraphics[width=\textwidth]{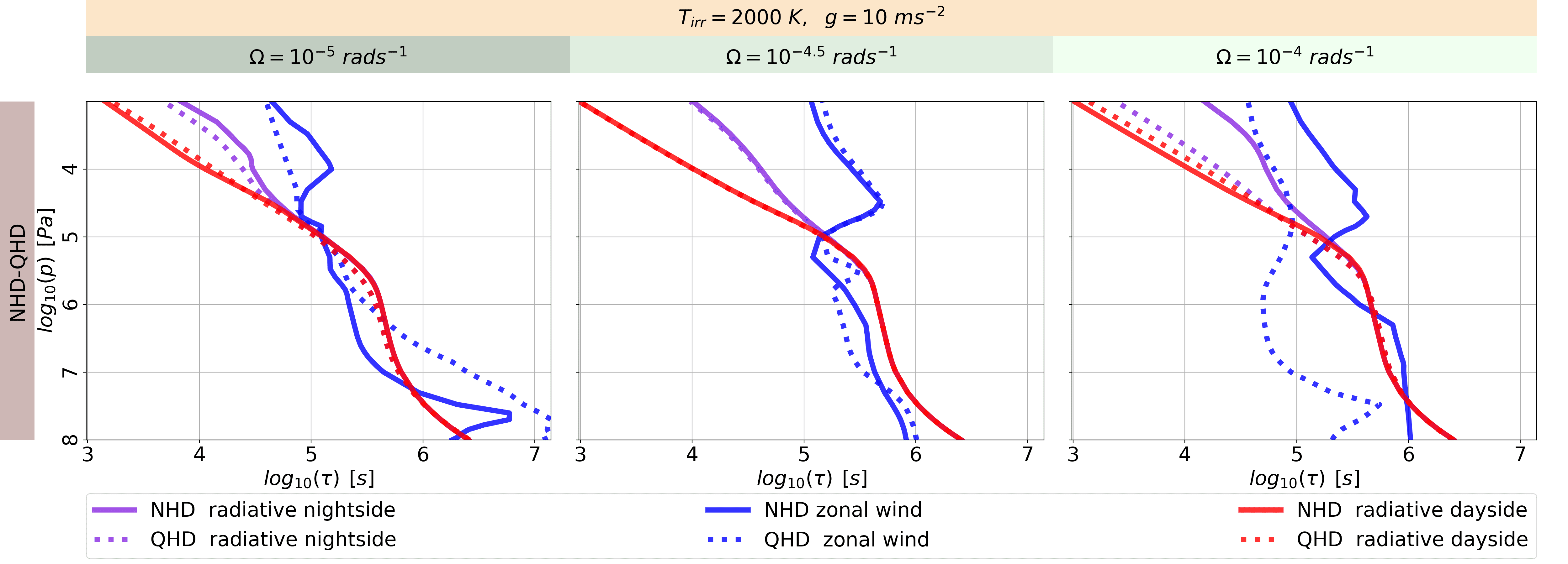}
    \caption{Radiative and zonal wind timescales for the NHD and QHD equation sets with $g = 10 \:ms^{-2}$, $T_{irr} = 2'000 \:K$ and with altering $\Omega$.}
    \label{fig:timescale_composit_10G_2000}
\end{figure*}

Figure \ref{fig:timescale_composit_10G_2000} shows the radiative and zonal wind timescales for the NHD and QHD equation sets with $g = 10 \:ms^{-2}$, $T_{irr} = 2'000 \:K$ and with altering $\Omega$. The radiative timescales varies less than the zonal wind timescales for the NHD and QHD equations sets. The radiative timescales on the dayside is conserved more than other timescales when the planet rotates faster. Above about $10^{5} \:Pa$, the radiative timescales on the day- and nightside remain the shortest for the NHD and QHD case. Furthermore, the radiative timescales on the day- and nightside fall together in the deep atmosphere. But at pressures $p<10^{5} \:Pa$, the radiative timescales on the day- and nightside start to divert more and more in both cases. Around $10^{5} \:Pa$, we see the timescales of the zonal wind become the shortest for both cases. For the slow and fast rotation rates, there may be a few switches between radiative and dynamical timescales to be the shortest.

Regarding differences between the NHD and QHD equation sets, we see most differences occurring in the timescale of the zonal wind. At the fast rotation rate, the QHD equation set shortens the timescale of the zonal wind throughout the atmospheres and especially in the deep atmosphere. We see a slightly higher radiative timescales on the dayside respectively lower on the nightside which speaks for an higher heat transport in the QHD case. There is an increase in the timescales of the zonal wind in the NHD case when the planet rotates faster. Whereas, the QHD equation set leads to an decrease of the zonal wind timescales in the deep atmosphere when the planet's rotation increases.

\subsection{Altering Gravity}

\begin{figure*}
	\includegraphics[width=\textwidth]{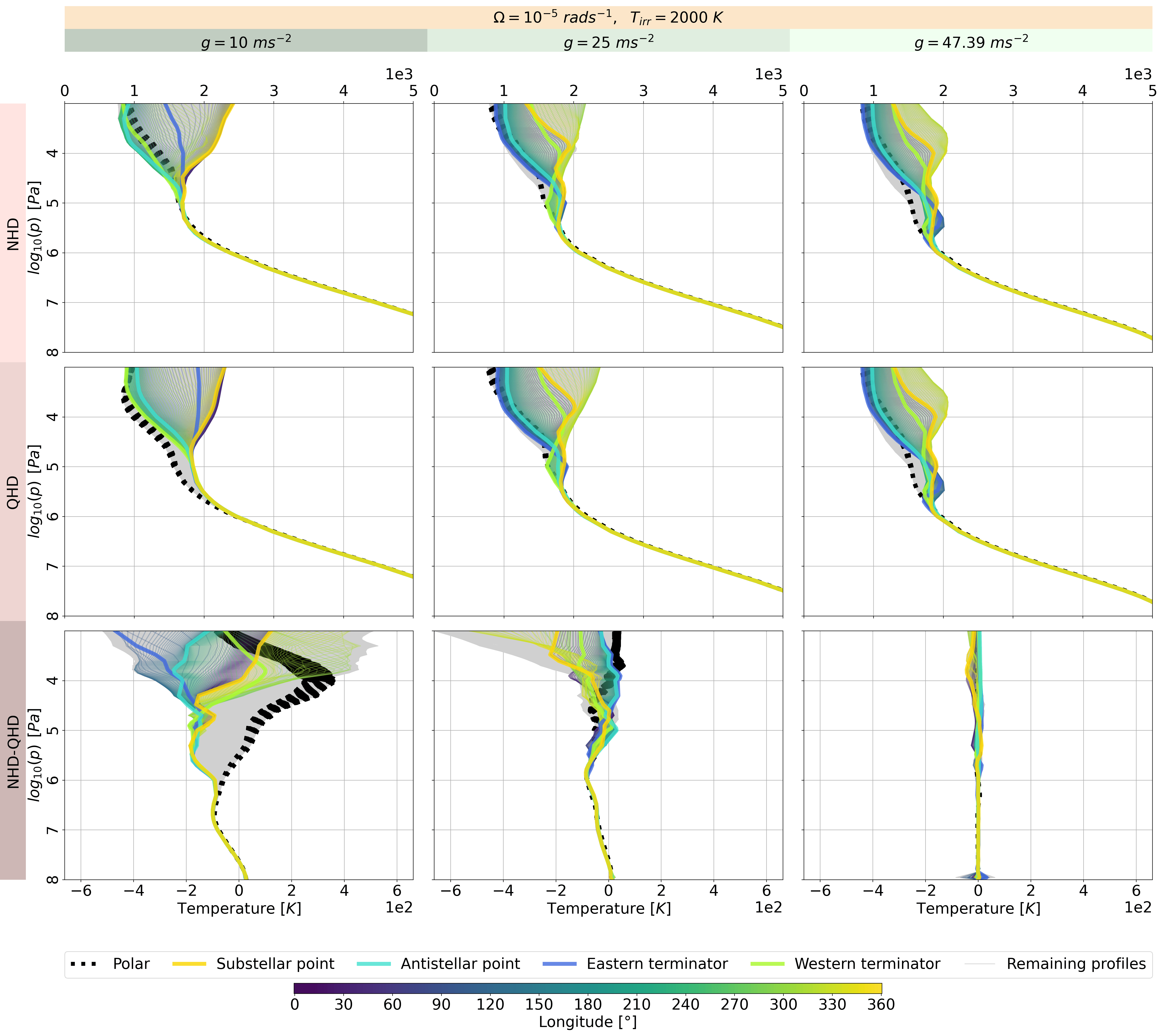}
    \caption{T-p profiles of covering entire planet for NHD and QHD equation sets with the same $\Omega = 1 \cdot 10^{-5} \:rad/s$, $T_{irr} = 2'000 \:K$ and with altering $g$. The coloured lines indicate T-p profiles along the equator and its coordinates by the colourbar. The dotted black thin line shows T-p profiles at the latitudes 87°N and 87°S. The bold coloured lines represent T-p profiles at the western, eastern terminators, sub- and antistellar point. The grey lines represents all the other T-p profiles.}
    \label{fig:TP_profile_composit_1e-5_2000_varying_G}
\end{figure*}

Figure \ref{fig:TP_profile_composit_1e-5_2000_varying_G} shows the T-p profiles for the NHD and QHD equation sets with the same $\Omega = 1 \cdot 10^{-5} \:rad/s$, $T_{irr} = 2'000 \:K$ and with altering $g$. Looking at similarities between NHD and QHD equation sets, the spread of temperatures shrinks the stronger the gravity becomes. The decreasing day-night contrast occurs together with additional inversions with increasing rotation rates. The number of inversions increases in the T-p profiles around the equator with higher gravity. Furthermore, the base of the lowest inversions reach higher pressures the larger the gravity gets. Therefore, the temperatures are substantially lower in the deep atmosphere with higher gravity.

Looking at pressures $p \sim 10^{5} \:Pa$, differences between simulations with NHD and QHD equation sets, we see a decrease in the differences the stronger the gravity gets.

\begin{figure*}
	\includegraphics[width=\textwidth]{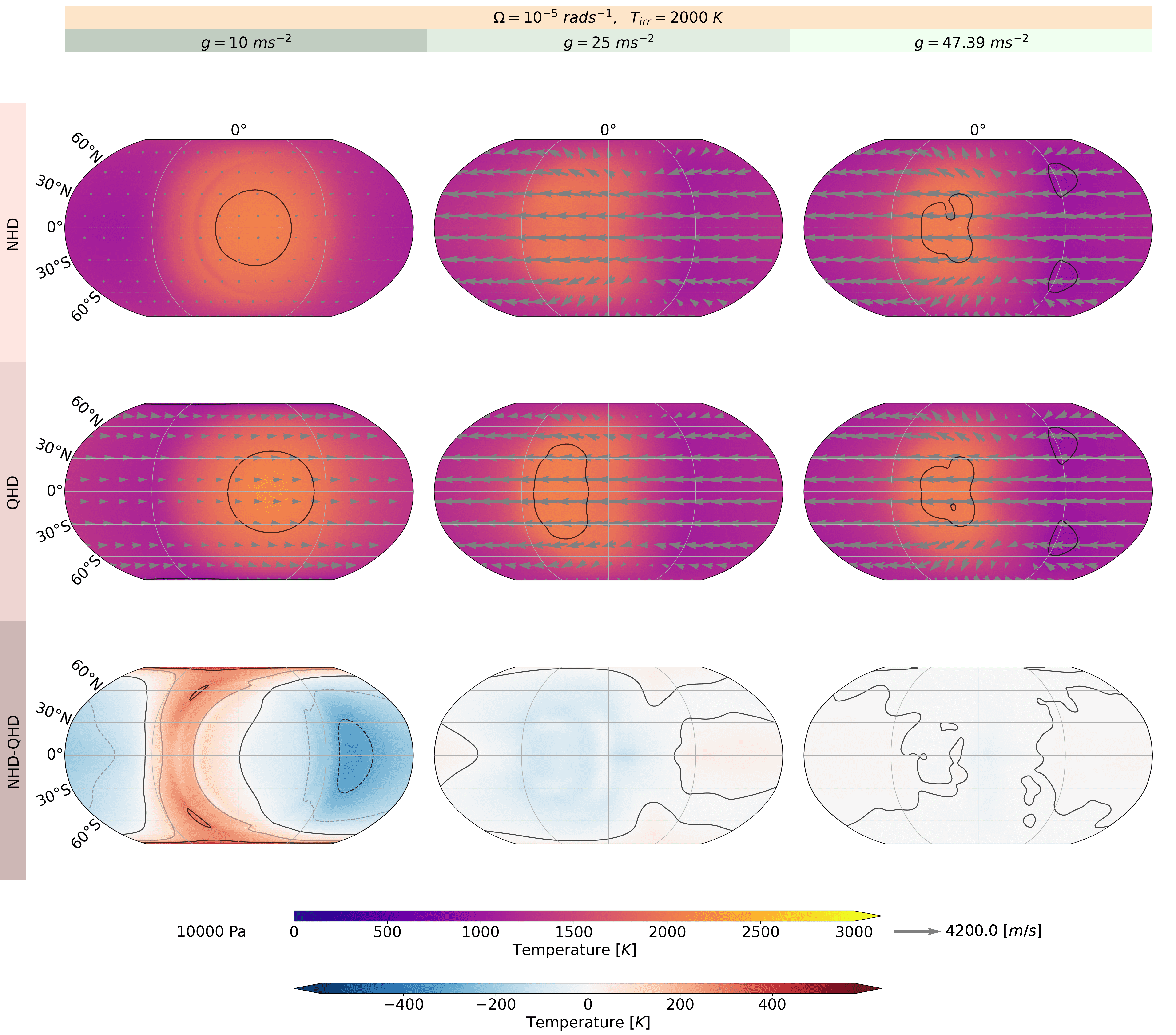}
    \caption{Temperature and wind speed at $10^{4} \:Pa$ for the NHD and QHD equation sets with $\Omega = 1 \cdot 10^{-5} \:rad/s$, $T_{irr} = 2'000 \:K$ and with altering $g$.}
    \label{fig:overview_composit_1e-5_2000_varying_G}
\end{figure*}

Figure \ref{fig:overview_composit_1e-5_2000_varying_G} shows the temperature and horizontal wind at $10^{4} \:Pa$ for the NHD and QHD equation sets with the same $\Omega = 1 \cdot 10^{-5} \:rad/s$, $T_{irr} = 2'000 \:K$ and with altering $g$. While the hotspot shift has an eastern offset at low $g$, it gets a western offset at higher $g$. The hotspot shift comes along with retrograde jet ranging to high latitudes with much higher wind speeds. The offset got larger with the high wind speeds, but decreases with higher $g$. Differences between the NHD and QHD case decreases with higher $g$.

\begin{figure*}
	\includegraphics[width=\textwidth]{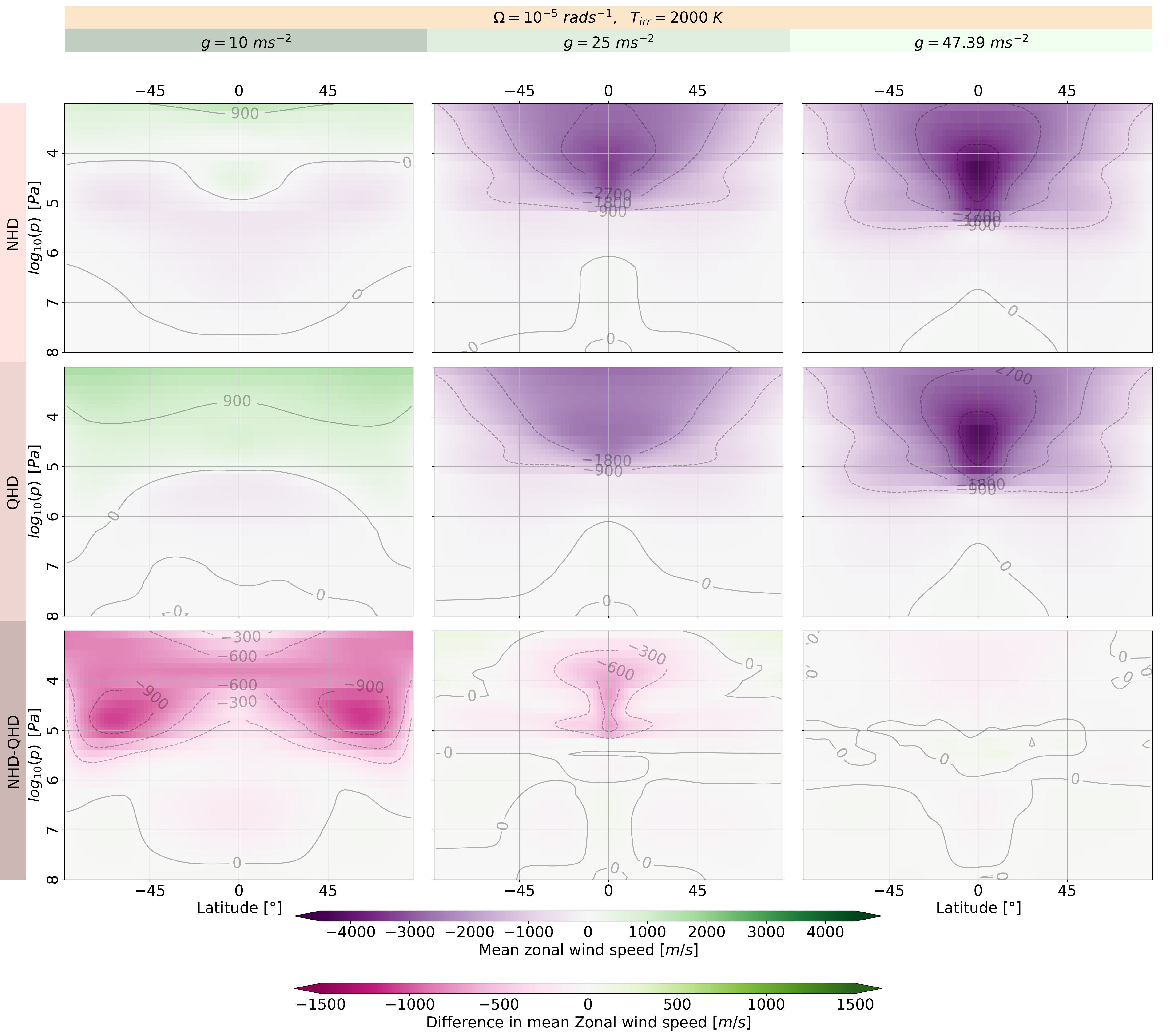}
    \caption{Zonal mean wind for the NHD and QHD equation sets with $\Omega = 1 \cdot 10^{-5} \:rad/s$, $T_{irr} = 2'000 \:K$ and with altering $g$.}
    \label{fig:zonal_wind_composit_1e-5_2000_varying_G}
\end{figure*}

Figure \ref{fig:zonal_wind_composit_1e-5_2000_varying_G} shows the zonal mean wind for the NHD and QHD equation sets with the same $\Omega = 1 \cdot 10^{-5} \:rad/s$, $T_{irr} = 2'000 \:K$ and with altering $g$. The higher $g$ leads to a change from the 3 prograde jet system to a one retrograde jet system. The system and climate state change brings higher wind speeds for the jet along. Furthermore, the wind flows in the deep atmosphere become weaker at higher $g$.

\begin{figure*}
	\includegraphics[width=\textwidth]{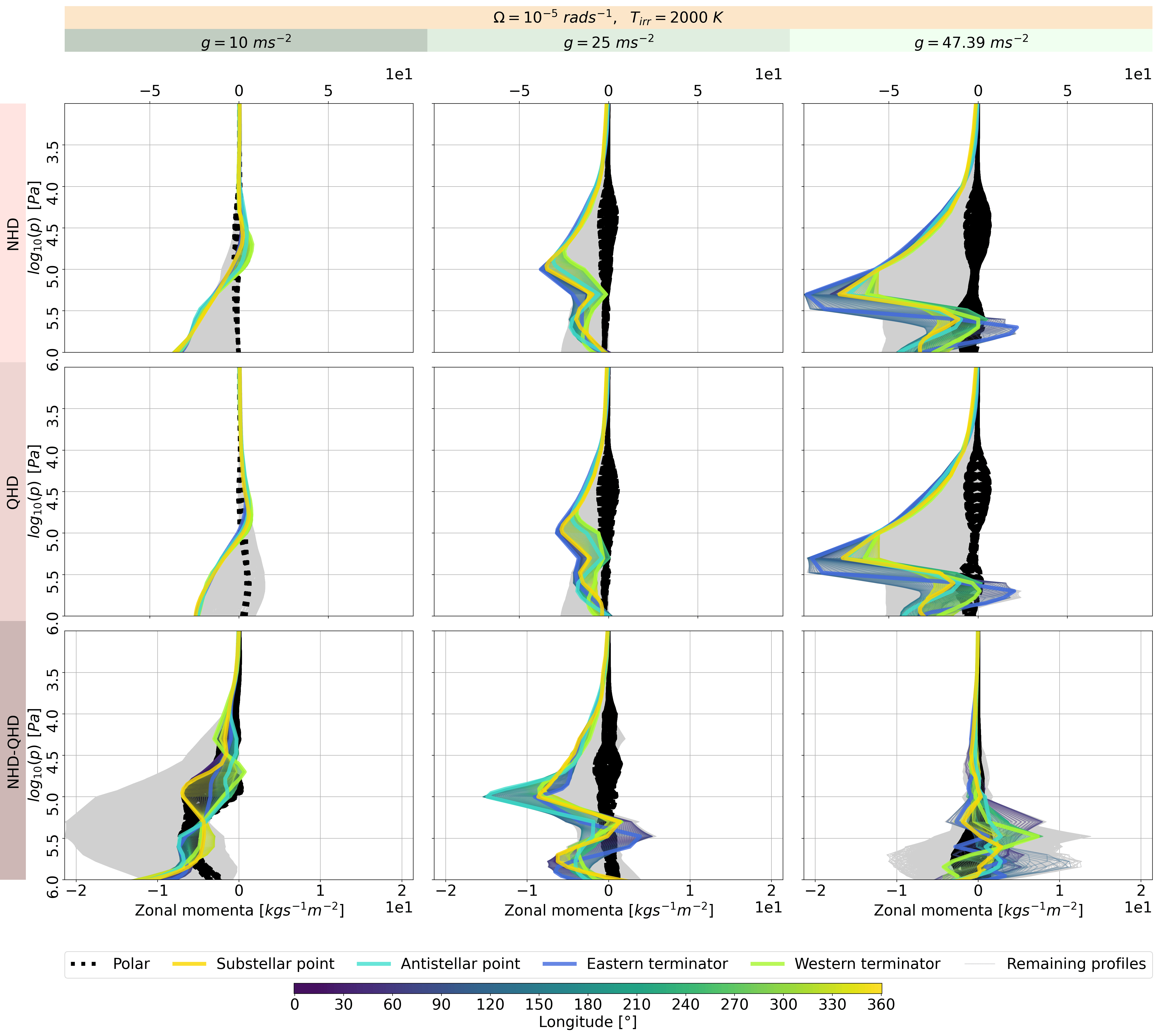}
    \caption{Zonal momenta at each grid point for NHD and QHD equation set with $\Omega = 1 \cdot 10^{-5} \:rad/s$, $T_{irr} = 2'000 \:K$ and with altering $g$. The profiles show only pressures $p\le 10^{6} \:Pa$ (without the pressure range $10^{6} \ge p \le 10^{8}$). The coloured lines indicate momenta profiles along the equator and its coordinates by the colourbar. The dotted black thin line shows momenta profiles at the latitudes 87°N and 87°S. The bold coloured lines represent momenta profiles at the western, eastern terminators, sub- and antistellar point. The grey lines represents all the other momenta profiles.}
    \label{fig:density_u_wind_composit_1e-5_2000_varying_G_upper_atmosphere}
\end{figure*}

Figure \ref{fig:density_u_wind_composit_1e-5_2000_varying_G_upper_atmosphere} shows the zonal momenta $[kg/m^3 m/s]$ along vertical profiles at each grid point for NHD and QHD equation set with $\Omega = 1 \cdot 10^{-5} \:rad/s$, $T_{irr} = 2'000 \:K$ and with altering $g$ (without the deep atmosphere). The zonal momenta along the vertical profiles in the simulations with NHD and QHD equation set become more similar the higher the gravity becomes. Furthermore, higher gravity leads to a change to an easterly jet (retrograde flow) in both cases. Another effect of higher gravity is the strengthening of the jet at pressures $p <10^{5} \:Pa$ in both cases. The jet reaches higher pressure with higher gravity in both cases. The highest momenta are found where the jet is the coldest regardless the gravity. Around the substellar point, the momenta remains still high, but the air masses get decelerated in a zone with a lot of upwelling.

\begin{figure*}
	\includegraphics[width=\textwidth]{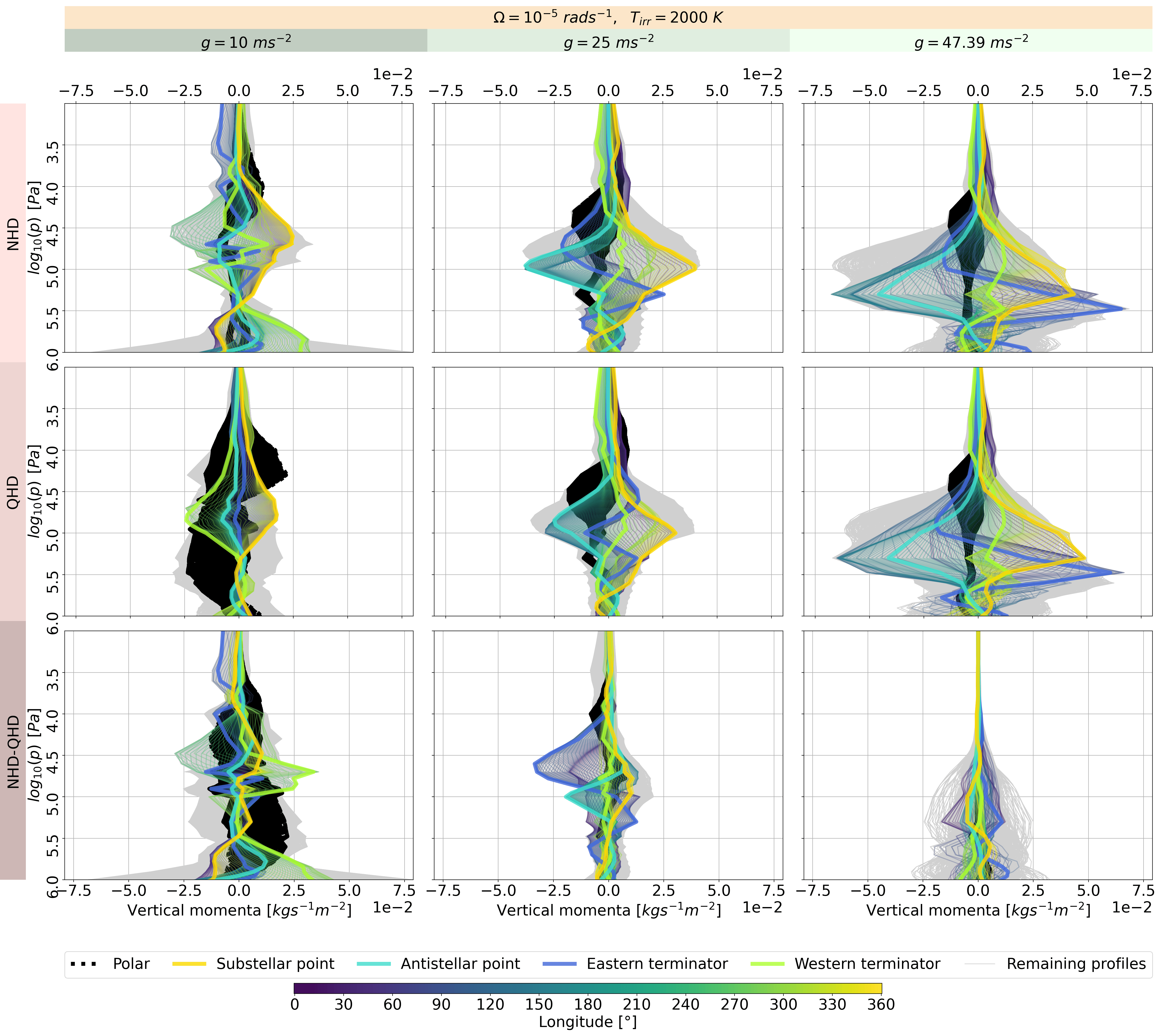}
    \caption{Vertical momenta at each grid point for NHD and QHD equation sets with $\Omega = 1 \cdot 10^{-5} \:rad/s$, $T_{irr} = 2'000 \:K$ and with altering $g$. The profiles show only pressures $p\le 10^{6} \:Pa$ (without the pressure range $10^{6} \ge p \le 10^{8}$). The coloured lines indicate momenta profiles along the equator and its coordinates by the colourbar. The dotted black thin line shows momenta profiles at the latitudes 87°N and 87°S. The bold coloured lines represent momenta profiles at the western, eastern terminators, sub- and antistellar point. The grey lines represents all the other momenta profiles.}
    \label{fig:density_w_wind_composit_1e-5_2000_varying_G_upper_atmosphere}
\end{figure*}

Figure \ref{fig:density_w_wind_composit_1e-5_2000_varying_G_upper_atmosphere} shows the vertical momenta $[kg/m^3 m/s]$ along vertical profiles at each grid point for the NHD and QHD equation sets with $\Omega = 1 \cdot 10^{-5} \:rad/s$, $T_{irr} = 2'000 \:K$ and with altering $g$ (without the deep atmosphere). Looking at the effects of increasing gravity, we see a wider range of vertical momenta when the gravity gets higher in both cases at pressures $p <10^{5} \:Pa$.

\begin{figure*}
	\includegraphics[width=\textwidth]{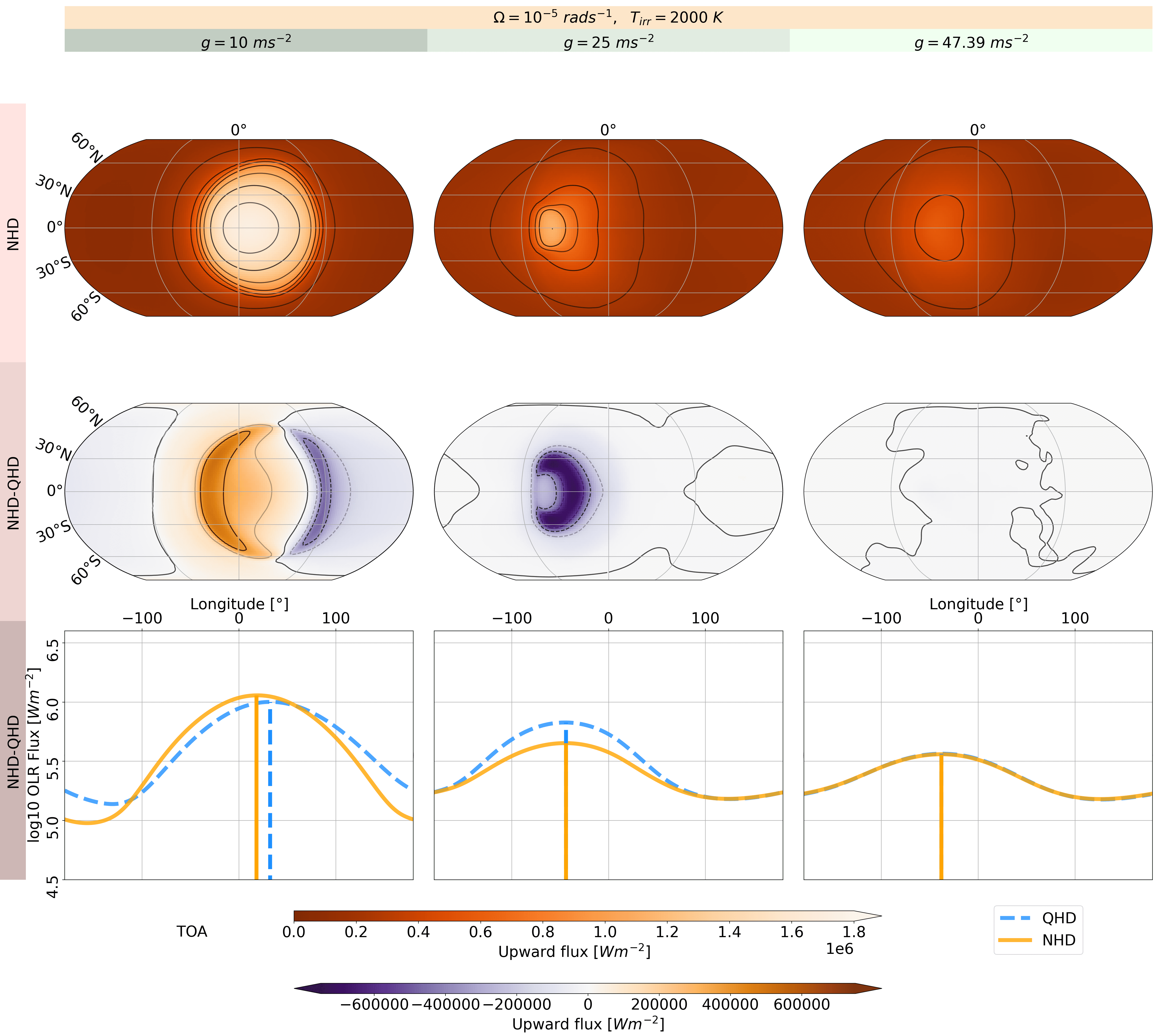}
    \caption{OLR fluxes at the top of the atmosphere for NHD and QHD equation sets with $\Omega = 1 \cdot 10^{-5} \:rad/s$, $T_{irr} = 2'000 \:K$ and with altering $g$. Third row: OLR phase curves.}
    \label{fig:flux_composit_1e-5_2000_varying_G}
\end{figure*}

Figure \ref{fig:flux_composit_1e-5_2000_varying_G} shows the OLR fluxes at the top of the atmosphere for the NHD and QHD equation set with $\Omega = 1 \cdot 10^{-5} \:rad/s$, $T_{irr} = 2'000 \:K$ and with altering $g$. Looking at the OLR phase curve, the maxima decrease with higher gravity in the NHD and QHD cases, although the QHD case stays much higher above the NHD case when gravity is moderate. When gravity gains strength, the minima switches to the western terminator. Furthermore, we see a westward shifted hotspot together with a retrograde flow like in \citet{carone2020}, but the retrograde flow extends to higher latitudes. At both terminators, small wave patterns occur in both cases with moderate and high gravity. When the rotational wind re-enters the daylight zone, the OLR phase curve to rises from the minimum at moderate and high gravity. Moreover, the slope of the OLR phase curves fall less on the upstream side of the maxima in both cases with higher gravity.

\begin{figure*}
	\includegraphics[width=\textwidth]{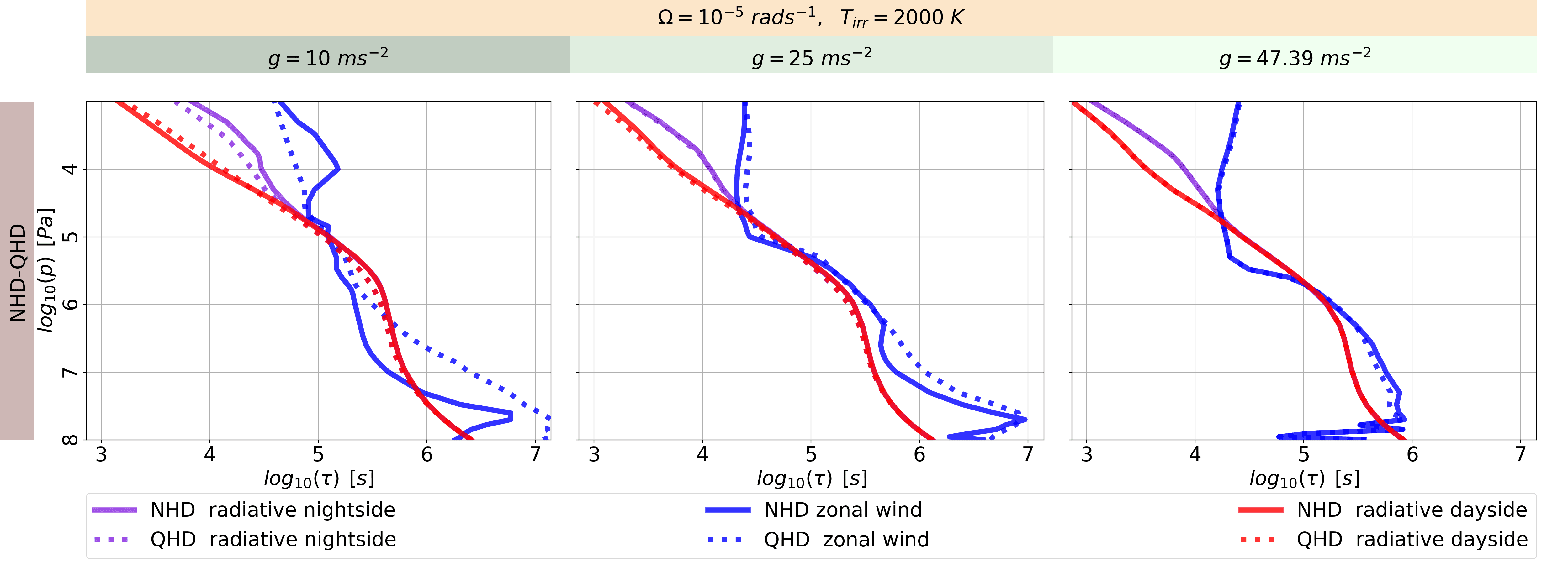}
    \caption{Radiative and zonal wind timescales for the NHD and QHD equation sets with $\Omega = 1 \cdot 10^{-5} \:rad/s$, $T_{irr} = 2'000 \: K$ and with altering $g$.}
    \label{fig:timescale_composit_1e-5_2000_varying_G}
\end{figure*}

Figure \ref{fig:timescale_composit_1e-5_2000_varying_G} shows the radiative and zonal wind timescales for the NHD and QHD equation sets with $\Omega = 1 \cdot 10^{-5} \:rad/s$, $T_{irr} = 2'000 \:K$ and with altering $g$. The timescale of the zonal wind shrinks at many heights the higher the gravity becomes. Similarly, the radiative timescales get shorter when gravity increases.

\subsection{Altering Irradiation Temperature}

\begin{figure*}
	\includegraphics[width=\textwidth]{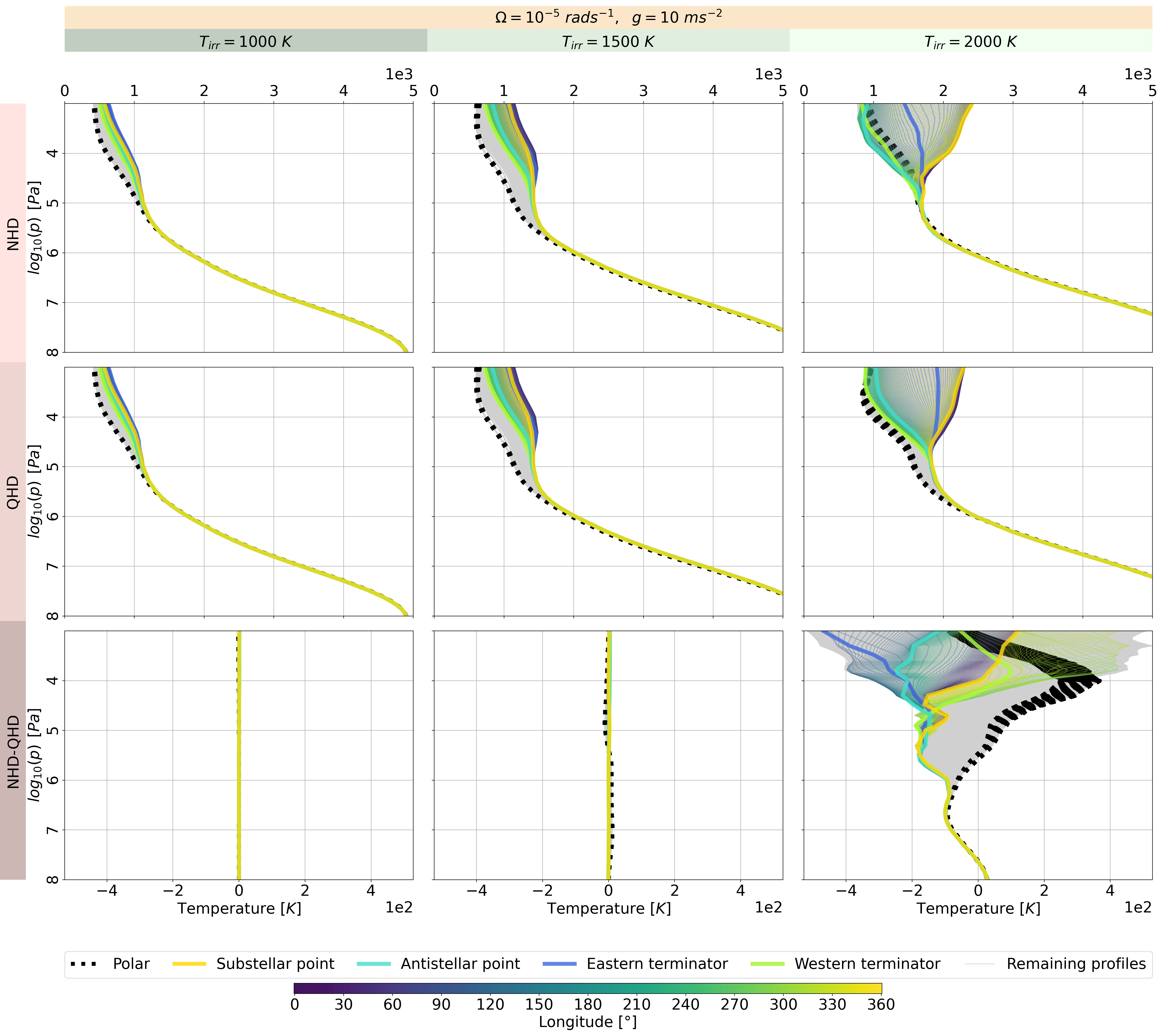}
    \caption{T-p profiles of covering entire planet for the NHD and QHD equation sets with $\Omega = 1 \cdot 10^{-5} \:rad/s$, $g = 10 \:ms^{-2}$ and with altering $T_{irr}$. The coloured lines indicate T-p profiles along the equator and its coordinates by the colourbar. The dotted black thin line shows T-p profiles at the latitudes 87°N and 87°S. The bold coloured lines represent T-p profiles at the western, eastern terminators, sub- and antistellar point. The grey lines represents all the other T-p profiles.}
    \label{fig:TP_profile_composit_10G_1e-5_varying_T}
\end{figure*}

Figure \ref{fig:TP_profile_composit_10G_1e-5_varying_T} shows the T-p profiles for the NHD and QHD equation sets with $\Omega = 1 \cdot 10^{-5} \:rad/s$, $g = 10 \:ms^{-2}$ and with altering $T_{irr}$. The range of temperatures at pressures $p \leq 10^{5} \:Pa$ decreases when the irradiation temperature decreases in both cases. Regarding differences between the NHD and QHD cases, they get smaller by a magnitude with each 500 K step in temperature. Furthermore, the temperatures at the poles get the coldest when the irradiation temperatures are equal or less than 1’500 K. Inversions start to disappear when the irradiation temperature lowers. The deep atmosphere has cooled down more the lower the irradiation temperature is set.

\begin{figure*}
	\includegraphics[width=\textwidth]{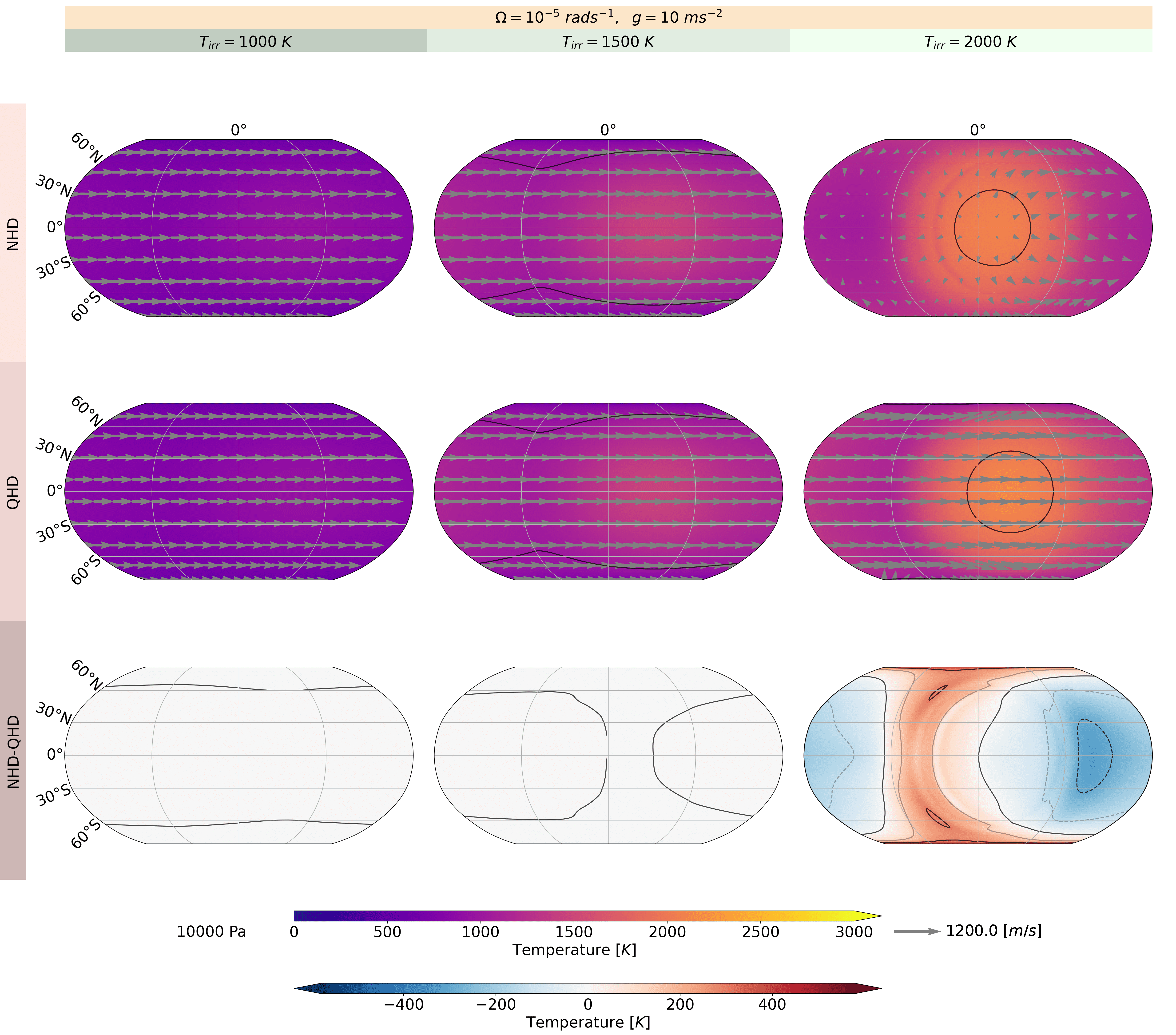}
    \caption{Temperature and wind speed at $10^{4} \:Pa$ for the NHD and QHD equation sets with $g = 10 \:ms^{-2}$, $\Omega = 1 \cdot 10^{-5}\:rad/s$ and with altering $T_{irr}$.}
    \label{fig:overview_composit_10G_1e-5_varying_T}
\end{figure*}
Figure \ref{fig:overview_composit_10G_1e-5_varying_T} shows the temperature and horizontal wind at $10^{4} \:Pa$ for the NHD and QHD equation sets with $\Omega = 1 \cdot 10^{-5} \:rad/s$, $g = 10 \:ms^{-2}$ and with altering $T_{irr}$. Lower $T_{irr}$ leads to a change from the 3 prograde jet system to a 1 prograde jet system. The jet is stronger in the 1 prograde jet system and ranges from pole to pole. The offset of the hotspot is higher at moderate $T_{irr}$. But the hotspot starts to vanish at low $T_{irr}$. The differences become minor at low $T_{irr}$.

\begin{figure*}
	\includegraphics[width=\textwidth]{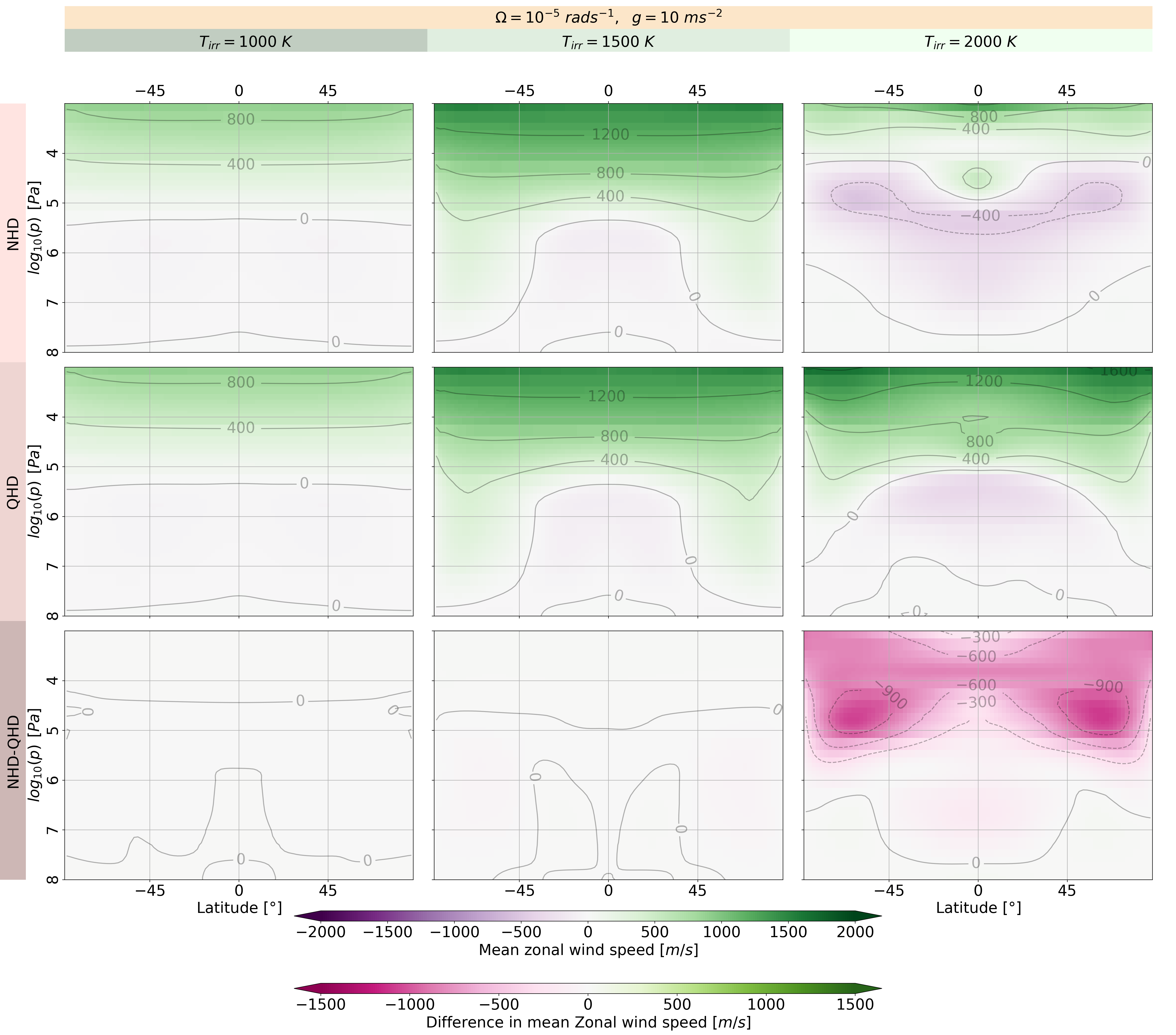}
    \caption{Zonal mean wind at each grid point for the NHD and QHD equation sets with $g = 10 \:ms^{-2}$, $\Omega = 1 \cdot 10^{-5}\:rad/s$ and with altering $T_{irr}$.}
    \label{fig:zonal_wind_composit_10G_1e-5_varying_T}
\end{figure*}
Figure \ref{fig:zonal_wind_composit_10G_1e-5_varying_T} shows the zonal mean wind for the NHD and QHD equation sets with $\Omega = 1 \cdot 10^{-5} \:rad/s$, $g = 10 \:ms^{-2}$ and with altering $T_{irr}$. At lower $T_{irr}$, the jet gets shallower and diffences between NHD and QHD case become minor. A peak of jet speeds are reached at $T_{irr}\sim 1'500 \:K$.

\begin{figure*}
	\includegraphics[width=\textwidth]{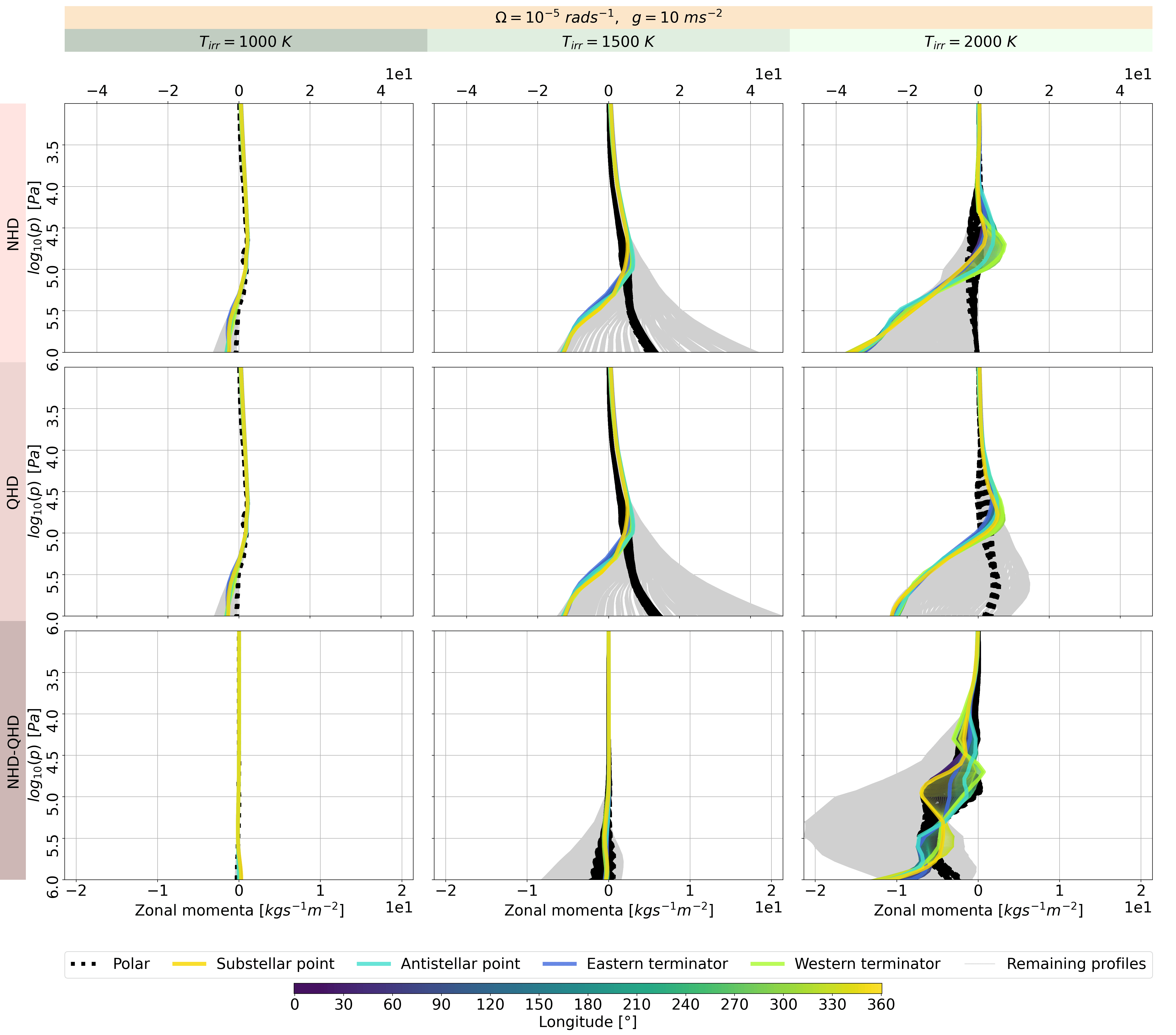}
    \caption{Zonal momenta at each grid point for the NHD and QHD equation sets with $\Omega = 1 \cdot 10^{-5} \:rad/s$, $g = 10 \:ms^{-2}$ and with altering $T_{irr}$. The profiles show only pressures $p\le 10^{6} \:Pa$ (without the pressure range $10^{6} \ge p \le 10^{8}$). The coloured lines indicate momenta profiles along the equator and its coordinates by the colourbar. The dotted black thin line shows momenta profiles at the latitudes 87°N and 87°S. The bold coloured lines represent momenta profiles at the western, eastern terminators, sub- and antistellar point. The grey lines represents all the other momenta profiles.}
    \label{fig:density_u_wind_composit_10G_1e-5_varying_T_upper_atmosphere}
\end{figure*}

Figure \ref{fig:density_u_wind_composit_10G_1e-5_varying_T_upper_atmosphere} shows the zonal momenta $[kg/m^3 m/s]$ along vertical profiles at each grid point for the NHD and QHD equation sets with $\Omega = 1 \cdot 10^{-5} \:rad/s$, $g = 10 \:ms^{-2}$ and with altering $T_{irr}$ (without the deep atmosphere). When the irradiation temperature lowers, all zonal wind components become positive at pressures $p \leq 10^{5} \:Pa$ in the NHD and QHD cases. We see an increase of zonal momenta, when the irradiation temperature decreases from $2'000$ to $1'500 \:K$. Additionally, the divergent component decreases and zonal component becomes stronger if the irradiation temperature lowers (see Helmholtz  decomposition in the \textbf{supplementary file}). Regarding differences between the NHD and QHD cases, they become a magnitude smaller at each 500 K step in temperature. 

\begin{figure*}
	\includegraphics[width=\textwidth]{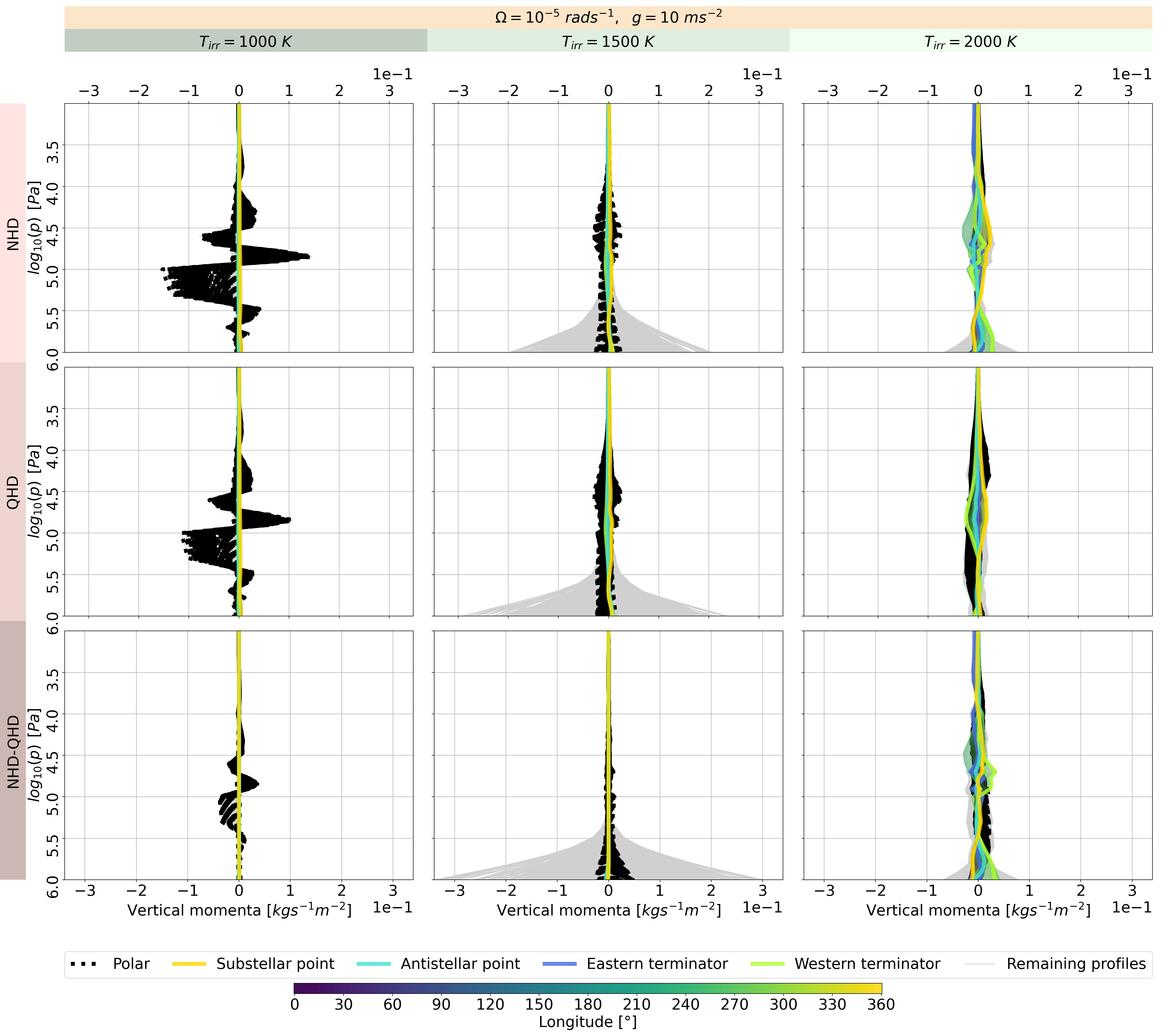}
    \caption{Vertical momenta at each grid point for the NHD and QHD equation sets with $\Omega = 1 \cdot 10^{-5} \:rad/s$, $g = 10 \:ms^{-2}$ and with altering $T_{irr}$. The profiles show only pressures $p\le 10^{6} \:Pa$ (without the pressure range $10^{6} \ge p \le 10^{8}$). The coloured lines indicate momenta profiles along the equator and its coordinates by the colourbar. The dotted black thin line shows momenta profiles at the latitudes 87°N and 87°S. The bold coloured lines represent momenta profiles at the western, eastern terminators, sub- and antistellar point. The grey lines represents all the other momenta profiles.}
    \label{fig:density_w_wind_composit_10G_1e-5_varying_T_upper_atmosphere}
\end{figure*}

Figure \ref{fig:density_w_wind_composit_10G_1e-5_varying_T_upper_atmosphere} shows the vertical momenta $[kg/m^3 m/s]$ along vertical profiles at each grid point for the NHD and QHD equation sets with $\Omega = 1 \cdot 10^{-5} \:rad/s$, $g = 10 \:ms^{-2}$ and with altering $T_{irr}$ (without the deep atmosphere).

\begin{figure*}
	\includegraphics[width=\textwidth]{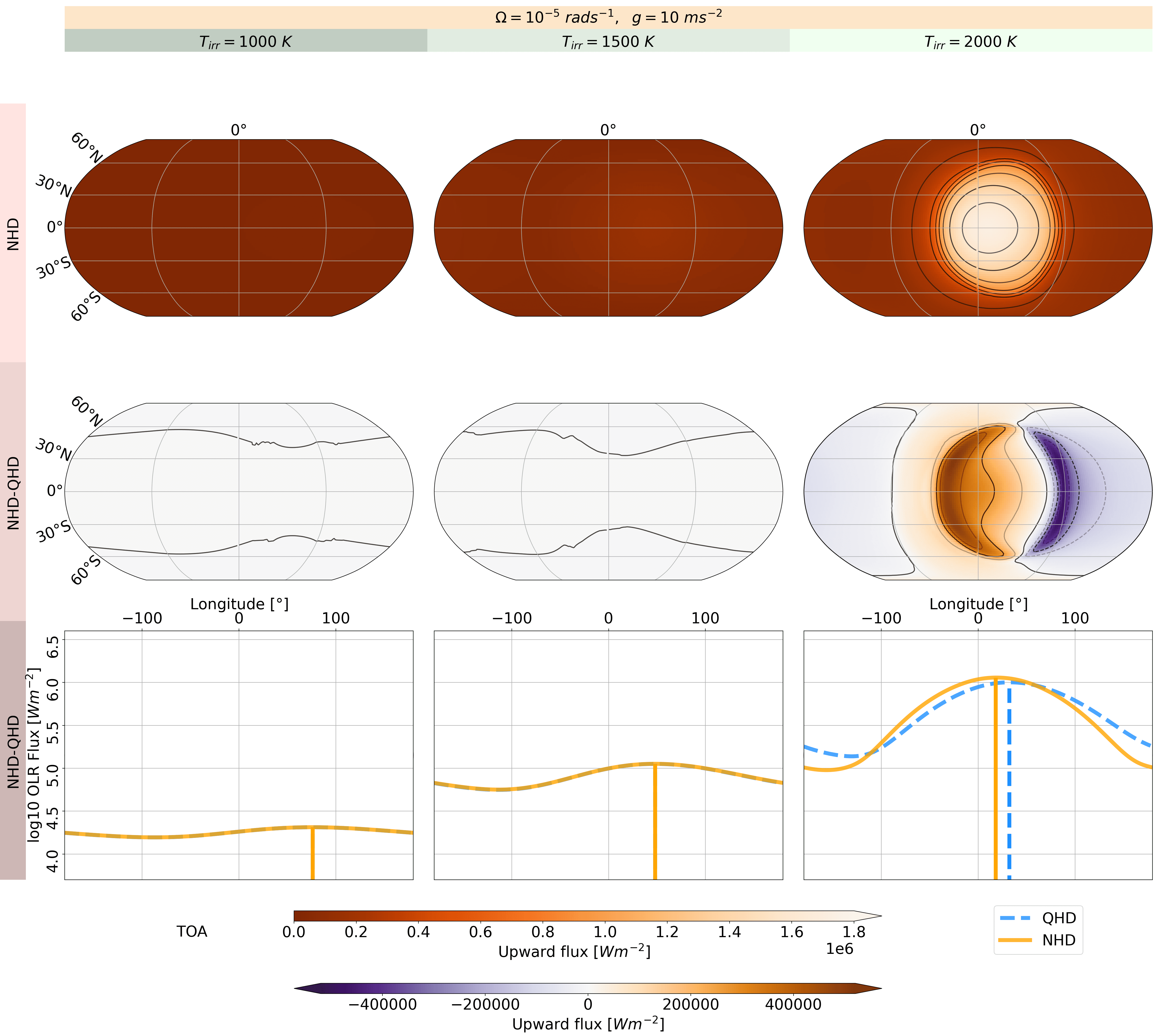}
    \caption{OLR fluxes at the top of the atmosphere for NHD and QHD equation sets with $\Omega = 1 \cdot 10^{-5} \:rad/s$, $g = 10 \:ms^{-2}$ and with altering $T_{irr}$. Third row: OLR phase curves.}
    \label{fig:flux_composit_10G_1e-5_varying_T}
\end{figure*}

Figure \ref{fig:flux_composit_10G_1e-5_varying_T} shows the OLR fluxes at the top of the atmosphere for the NHD and QHD equation sets with $\Omega = 1 \cdot 10^{-5} \:rad/s$, $g = 10 \:ms^{-2}$ and with altering $T_{irr}$. We see an increasing shift of the OLR to the East the lower the irradiation temperature is set lower. Similarly, the minima of the OLR phase curve occur around the western terminator in both simulations with lowered irradiation temperature. Furthermore, the differences between NHD and QHD equation sets in the OLR decrease when we set the irradiation temperature lower.

\begin{figure*}
	\includegraphics[width=\textwidth]{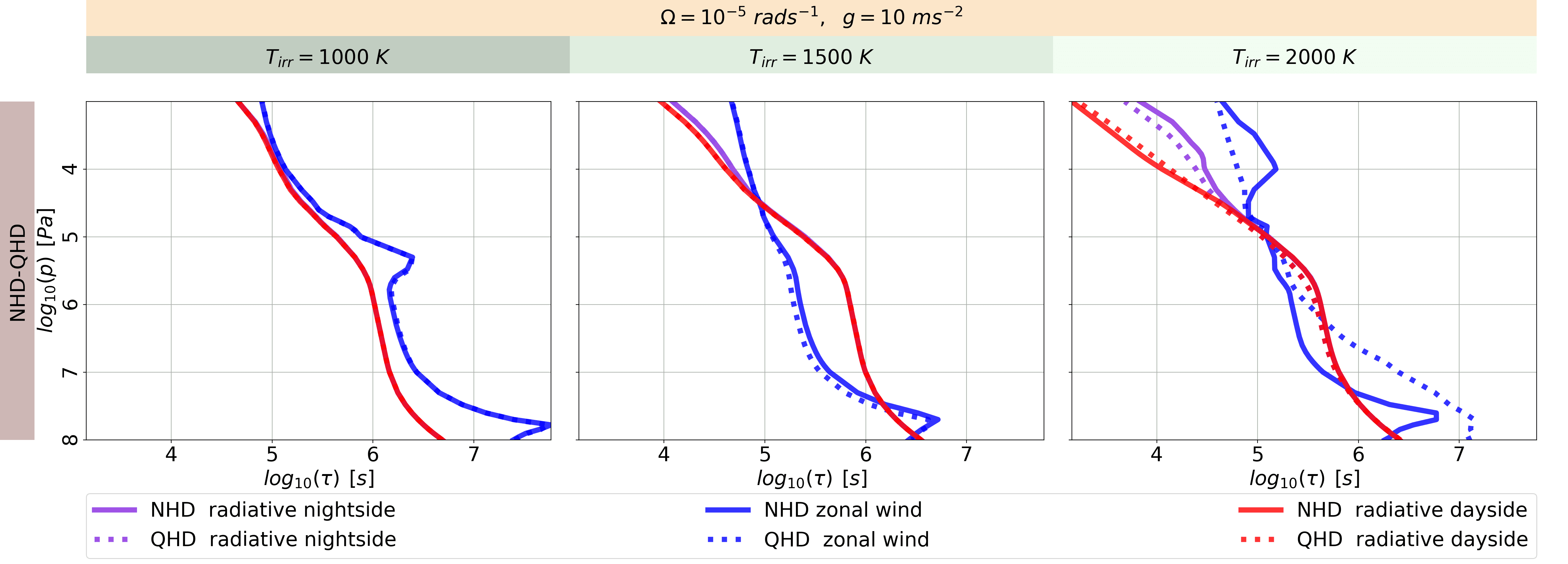}
    \caption{Radiative and zonal wind timescales for the NHD and QHD equation sets with $\Omega = 1 \cdot 10^{-5} \:rad/s$, $g = 10 \:ms^{-2}$ and with altering $T_{irr}$.}
    \label{fig:timescale_composit_10G_1e-5_varying_T}
\end{figure*}

Figure \ref{fig:timescale_composit_10G_1e-5_varying_T} shows the radiative and zonal wind timescales for the NHD and QHD equation sets with $\Omega = 1 \cdot 10^{-5} \:rad/s$, $g = 10 \:ms^{-2}$ and with altering $T_{irr}$. At a irradiation temperature of 1'500 K, the timescale of the zonal wind stays much shorter than the radiative timescales at pressures $10^{4.5} \leq p \leq 10^{7.5} \:Pa$. We see a more efficient advection by the zonal wind when the divergent component weakens. But the radiative timescales get shorter than the timescale of the zonal wind when the irradiation temperature is lowed to $1'000\: K$. The zonal wind gets weaker and therefore less efficient of advection.

\section{Discussion}

The difference between the NHD and QHD equation sets in THOR lies in the $Dv_{r}/Dt$, the Lagrangian derivative of the vertical velocity, $\mathcal{F}_{r}$, the hyperdiffusive flux and $\mathcal{A}_{r}$, the vertical component of the advection terms. These terms lead to deviations in the vertical momenta in the simulations with QHD. The altered vertical momenta affects the horizontal momenta and the temperature structure indirectly. Those changes caused by a different dynamical equations set even lead to different climate states in the simulated time period.

As first approach, we can compare to the analytic solutions of \cite{showman2011superrotation} which applied  the linearised shallow water equations. They designed the equation set to be the simplest as possible to cleanly identify specific dynamical processes, therefore, a two-layer model was implemented. Those analytic solutions were calculated with the zonal wavenumber $k=0.5$ and a rotation period of 3 Earth days. The rotation periods in our study are $7.27$, $2.3$ and $0.73$ Earth days. Therefore, all analytic solutions lie between our simulations with $\Omega = 1 \cdot 10^{-5} \:rad/s$ and $\Omega = 1 \cdot 10^{-4.5} \:rad/s$. The closest parameterisation between our results and those analytic solutions is $\tau_{rad}=1\: d$ and  $\tau_{drag}=1\: d$ which corresponds to the top left plot in Figure 3 in \cite{showman2011superrotation}. In our simulations, the radiative timescales reach values between $\tau_{rad}\sim0.12\: d$ for dayside and $\tau_{rad}\sim1.11\: d$ at the height of the jet. Furthermore, $\tau_{diff}$ becomes $1.69\cdot 10^{-4}\: d$ when $D_{h\!y\!p,h}=0.0025$ is used in the following equation (according to \cite{hammond2022})
 \begin{equation}
\tau_{d\!i\!f\!f} \sim \frac{\Delta t}{2^{2n+1}D_{h\!y\!p,h}}.
\label{eq:tau_diff}
\end{equation}
 Rossby-wave gyres do not appear in the analytic solutions \citep{showman2011superrotation} with $\tau_{rad}=1\: d$ and  $\tau_{drag}=\:1 d$. When $\tau_{rad}$ and  $\tau_{drag}$ become higher, cyclones and anticyclones become visible in the analytic solutions. In our results, we do see Rossby-wave gyres pumping zonal momenta from higher latitudes to lower latitudes, when gravity or the rotation rate gets more intense (e.g. see figure with high $g$ and high $\Omega$ in the \textbf{supplementary file} and figure \ref{fig:overview_composit_1e-5_2000_varying_G}). However, $\tau_{rad}$ is smaller in our composits with altering $\Omega$ than in the analytic solutions of the linearized shallow-water equations. The equilibrated solutions in \citet{showman2011superrotation} lead to a single maxima and minima of the geopotential $gh$ for $\tau_{rad}=\tau_{drag}=0.1\: d$ . When $gh$ for $\tau_{rad}$ or $\tau_{drag}$ become higher, 2 minima and 2 maxima evolve. In our results, we see 1 maxima and a chevron respectively 2 minima. That pattern evolves likely due to the different $\tau_{rad}$ on the day- and nightside compared to the uniform timescales in \citet{komacek2019}.
 \citet{komacek2019} did run 36 experiments with a comparable setting ($c_{P}=13'000 \:J kg^{-1}K^{-1}$, $R=3'700 \:Jkg^{-1}K^{-1}$, $a=9.437\cdot 10^{7}\:m$, $g=9.36 ms^{-2}$ and $\Omega=2.078 \cdot 10^{-5} \:s^{-1}$). They show the vertical and horizontal wind for different $\tau_{drag}$ and $T_{eq}$ at $10^{2} \:Pa$. Their simulations with $\tau_{drag} \leq 10^{6}\:s$ led to no superrotating jet and a more divergent flow, whereas simulations with $\tau_{drag} \geq 10^{6} \:s$ show a superrotating jet. Our comparable simulation with $g=10 ms^{-2}$, $\Omega=1 \cdot 10^{-5}\:rad\:s^{-1}$ and $T_{irr}= 2'000 \:K$ falls with $\tau_{diff}=1.69\cdot 10^{-4} \:d$ below the threshold of $\tau_{drag} \leq 10^{6}\: s$ and produces a similar horizontal and vertical flow pattern, although we see the similarity at $10^{4} \: Pa$ instead of at $10^{2} \:Pa$. The differences to \citet{komacek2019} probably occur because of different dynamical cores (dynamical equation sets) and spatially different radiative timescales.

\subsection{Examination of climate states}

\begin{figure*}
	\includegraphics[width=\textwidth]{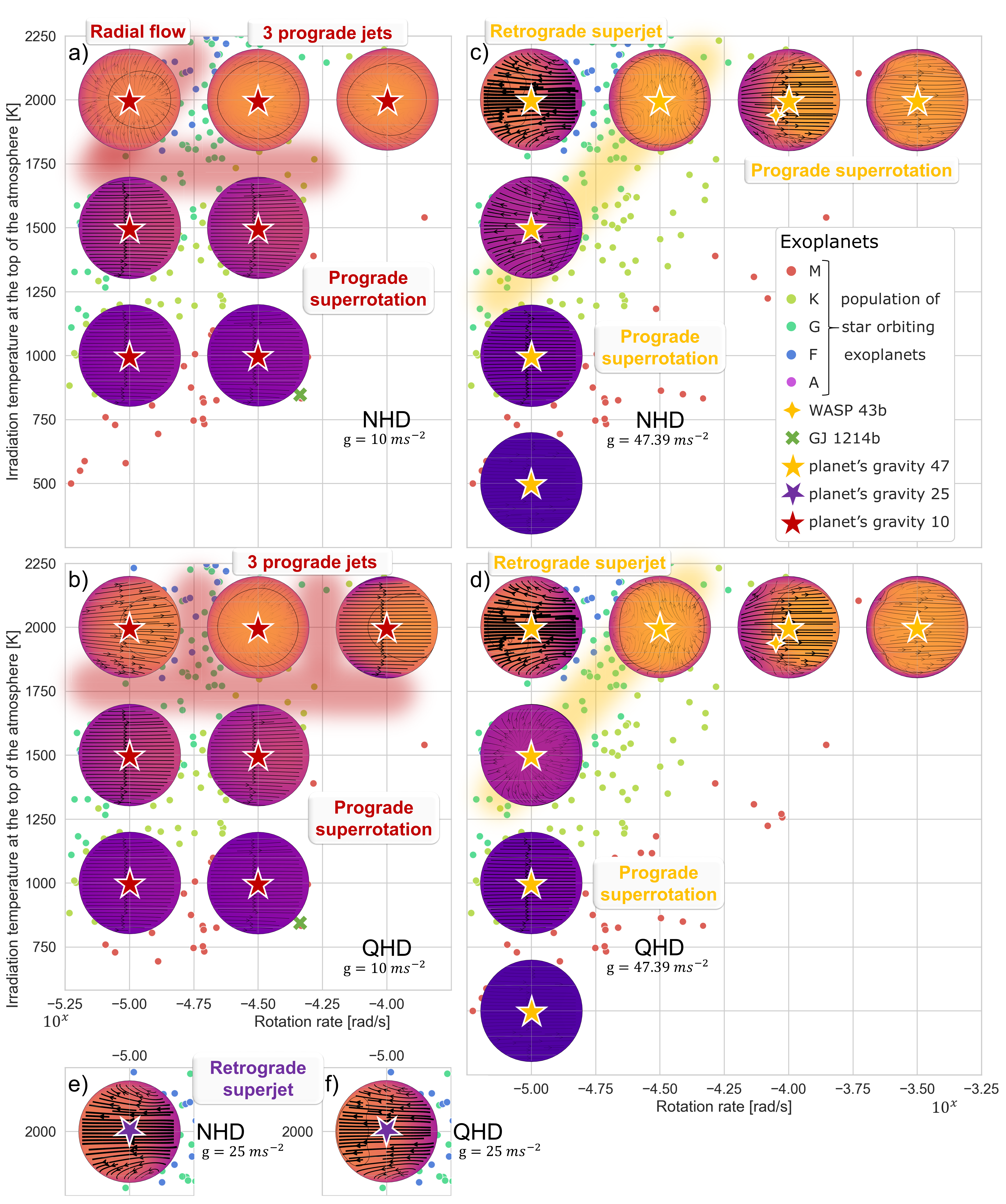}
    \caption{Classification of the simulations to circulation and climate states in the parameter grid of known exoplanets. Orthographic projections shows the temperature and horizontal wind flow on the dayside at $10^{4}\: Pa$. The shaded lines in red and yellow show assumptions for possible classifiers. Subplots a) and b) show the classification of simulations for the NHD respectively QHD case with $g = 10 \:ms^{-2}$, altering $\Omega$ and with altering $T_{irr}$. Subplots c) and d) show the classification of simulations for the NHD respectively QHD case with $g = 47.39 \:ms^{-2}$, altering $\Omega$ and with altering $T_{irr}$. Subplots e) and f) show the classification of simulations for the NHD respectively QHD case with $g = 25 \:ms^{-2}$, $\Omega = 1 \cdot 10^{-5} \:rad/s$ and $T_{irr}= 2'000\:K$.}
    \label{fig:classification}
\end{figure*}

We classify the NHD simulation outputs into climate states according to jet behaviours and manifestations of the components of the Helmholtz decomposition. The stated climate states are presented hereafter and illustrated in the Figure \ref{fig:classification}. We consider this classification as a first assumption to figure out parameters where the QHD case (and maybe GCMs with HPEs) perform not as accurately. So, it should not be seen as a definitive classification scheme.

Moreover, we computed large-scale flow quantities and other characteristic values and scales in Table \ref{tab:large-scale_flow_quantities}. We discuss those indicators in relation to the climate states in the next section.

\begin{table*}
	\centering
	\caption{ Characteristic values and scales in comparison to the circulation and climate states in the NHD case. The characteristics include scale height $H$, Rossby number $Ro$, Rossby deformation radius $L_{D}$, Rhines scale and the Brunt–Väisälä frequency $N$. of the all simulations using a similar setting as in \citep{parmentier2014PhD} and \citet{lee2020} (for details of the calculations see the appendix \ref{large_scale_flow_quantities}). We set $L=R_{p}$ and the characteristic velocity to a range between $100$ and $4'500 \:ms^{-1}$. The Rossby number and Rossby deformation radius is evaluated for the mid-latitudes. For the Rossby deformation, we set $D=H$. For the Rhines scale, we use the equatorial value for the $\beta$-Term  and use the range between $100$ and $4'500 \:ms~{-1}$ for the wind speeds. Rossby deformation radius and Rhines scale are calculated as ratios to $R_{p}$.} 
	\label{tab:large-scale_flow_quantities}
	\begin{tabular}{cccccccccll}
		\hline \hline
		$\Omega$ & $g$ & $T_{irr}$ & $T_{eq}$ & $H$ & $Ro$ & $L_{D}$ & $L_{Rh}$ & $N$ & Circulation state & Features\\
        $[rads^{-1}]$ & $[ms^{-1}]$ & $[K]$ & $[K]$ & $[km]$ & - & $[R_{p}]$ & $[R_{p}]$ & $[s^{-1}]$ & - & -\\
		\hline \hline
        \\[-1em]
		$10^{-5}$ & $10$ & $2'000$ & $1'414.21$ & $525.24$ & $0.098-4.39$ & $4.19$ & $0.83-5.54$ & $0.00816$ & 3 prograde jets & 3 deeper \\[0.5ex]
        & &  &  &  &  & &  & &and radial flow & retrograde jets\\[0.5ex]
        \\[-1em]
        \hline
        \\[-1em]
		$10^{-4.5}$ & $10$ & $2'000$ & $1'414.21$ & $525.24$ & $0.031-1.39$ & $1.32$ & $0.46-3.11$ & $0.00816$ & 3 prograde jets & 3 deeper retrograde jets\\[0.5ex]
        \\[-1em]
        $10^{-4}$ & $10$ & $2'000$ & $1'414.21$ & $525.24$ & $0.0098-0.44$ & $0.42$ & $0.26-1.75$ & $0.00816$ & 3 prograde jets & shallow equatorial\\[0.5ex]
        \\[-1em]
         & &  &  &  &  & &  & & &retrograde flow\\[0.5ex]
        \\[-1em]
        \hline
        \\[-1em]
		$10^{-5}$ & $25$ & $2'000$ & $1'414.21$ & $210.1$ & $0.098-4.39$ & $4.19$ & $0.83-5.54$ & $0.0204$ & retrograde superjet & jet width from pole to pole\\[0.5ex]
        \\[-1em]
		$10^{-5}$ & $47.39$ & $2'000$ & $1'414.21$ & $110.83$ & $0.098-4.39$ & $4.19$ & $0.83-5.54$ & $0.039$ & retrograde superjet & jet width from pole to pole\\[0.5ex]
        \\[-1em]
        \hline
        \\[-1em]
		$10^{-5}$ & $10$ & $1'000$ & $707.11$ & $262.62$ & $0.098-4.39$ & $2.96$ & $0.83-5.54$ & $0.011$ & prograde superrotation & jet width from pole to pole\\[0.5ex]
        \\[-1em]
		$10^{-5}$ & $10$ & $1'500$ & $1'060.66$ & $393.93$ & $0.098-4.39$ & $3.63$ & $0.83-5.54$ & $0.0094$  & massive prograde & jet width from\\[0.5ex]
        \\[-1em]
        & &  &  &  &  & &  & & superrotation & pole to pole\\[0.5ex]
        \\[-1em]
		\hline
        \\[-1em]
        $10^{-4.5}$ & $47.39$ & $2'000$ & $1'414.21$ & $110.83$ & $0.031-1.39$ & $1.32$ & $0.46-3.11$ & $0.039$ & interrupted prograde & 2 interrupted retrograde\\[0.5ex]
        \\[-1em]
        & &  &  &  &  & &  & &  superrotation & high-latitude jets\\[0.5ex]
        \\[-1em]
        $10^{-4}$ & $47.39$ & $2'000$ & $1'414.21$ & $110.83$ & $0.0098-0.44$ & $0.42$ & $0.26-1.75$ & $0.039$ & prograde superrotation & -\\[0.5ex]
        \\[-1em]
        \hline
        \\[-1em]
        $10^{-5}$ & $47.39$ & $1'000$ & $707.11$ & $55.42$ & $0.098-4.39$ & $2.96$ & $0.83-5.54$ & $0.055$ & retrograde superjet & jet width from pole to pole\\[0.5ex]
        \\[-1em]
        $10^{-5}$ & $47.39$ & $1'500$ & $1'060.66$ & $83.12$ & $0.098-4.39$ & $3.63$ & $0.83-5.54$ & $0.045$ & prograde superrotation & jet width from pole to pole\\[0.5ex]
        \\[-1em]
        \hline
	\end{tabular}
\end{table*}

\subsubsection{3 prograde jets}

When we alter the planetary rotation rate $\Omega$ at irradiation temperatures $T_{irr} = 2'000 \:K$ and gravity $g=10 \:ms^{-2}$, we see a transition from a climate state with a dominate divergent component to a climate state with higher Coriolis forces. A dominate large "extra-tropical" zone expands near to the equator with higher rotation rates \citep{parmentier2014PhD}. In that zone, the advection term becomes small or even negligible and the force balance is mainly made up among the Coriolis term and the pressure gradient. The Rossby number $Ro$ for $\Omega= 10^{-4.5}$ and $\Omega= 10^{-4} \: rads^{-1}$ are in the range of $0.031$ to $1.39$ respectively in the range of $0.0098$ to $0.44$. For the maximum horizontal wind speeds in our simulations, we get $Ro=0.19$ and $Ro=0.15$ for the NHD and QHD case for $\Omega= 10^{-4.5} \: rads^{-1}$. For $\Omega= 10^{-4} \: rads^{-1}$, we get $Ro=0.052$ and $Ro=0.14$ for the NHD and QHD case for the horizontal winds in our simulations. The too high wind speed in the simulation with the QHD equation set prevents the Coriolis force to act on the jet structures at $\Omega= 10^{-4} \: rads^{-1}$. At $\Omega= 10^{-4.5} \: rads^{-1}$, the horizontal wind in QHD case is more moderate than at lower $\Omega$ and therefore the balancing regarding the Coriolis force is more similar to the NHD case.

Looking at the Helmholtz decomposition at $10^{4} \: Pa$, all components are weaker than at lower rotation rates (see the plots of the Helmholtz decomposition in the \textbf{supplementary file}). The divergent component is still dominant compared to simulations with higher $g$ or lower $T_{irr}$. The rotational eddy and rotational jet components evolved moderate weakly. 

The scale height $H$ and the Brunt-V\"ais\"al\"a frequency are $525.24 \:km$ and $N = 0.00816 \:s^{-1}$ for the 3 altered $\Omega$ at $T_{irr} = 2'000 \:K$ and $g=10 \:ms^{-2}$. The Rossby deformation radius $L_{D}$ are $1.32$ and $0.42  \:R_{p}$ for $\Omega= 10^{-4.5}$ and $\Omega= 10^{-4} \: rads^{-1}$. So, we expect smaller eddy sizes at higher $\Omega$. The Rhines scales $L_{RH}$ vary between $0.46$ and $3.11\:R_{p}$. For our maximum wind speeds in our simulations, $L_{RH}$ becomes $1.14$ respectively $0.06 \:R_{p}$, for $\Omega= 10^{-4.5}$ and $\Omega= 10^{-4} \: rads^{-1}$. Such small scales let small scale vortices boost the larger atmospheric flow with their energies \citep{parmentier2014PhD}. The values for $L_{RH}$ increases with higher latitude and the likelihood for the appearance of Rossby waves. At higher latitudes, we do see planetary waves at $10^{4} \:Pa$.

The NHD case shows the emergence of high-latitude prograde jets in addition to the deeper, prograde and primary superrotating equatorial jet. \citet{showman2009} and \citet{rauscher2014} observed the 3 jet structure in their GCM simulations as well, but for both HD 189733b respectively HD 209458b with non-synchronous rotation rates.

We see differences in our simulations between the NHD and QHD equation sets growing with increased rotation rate. Wind speeds and momenta in the QHD simulations underlay those in the NHD simulations at slow rotation. But the zonal momenta in the QHD case exceed by about 5 times those in the NHD case at high rotation rates. The differences in the momenta lead to significant differences in the advection and the temperature structure at pressures $p \leq 10^{6} \:Pa$ at slow and high rotation rates. The differences in the temperature range grow from about $600 \:K$ at the slow rotation to $1'200 \:K$ at fast rotation rate. The difference between the NHD and QHD equation sets do not behave linearly and include dynamical regime  and climate changes in the QHD case. We see even very similar regimes and climates at moderate rotation rate at pressures $p \leq 10^{4} \:Pa$, but the dynamical equations lead to totally different regimes in the deep atmosphere. We noticed two dynamical regime and climate state changes by altering the rotation rate in the QHD case at pressures $p \leq 10^{4} \:Pa$. The QHD case changes from a 2 jet system with superrotation to a 3 jet system with weak extra-tropical conditions and then back to the state with 2 jets and superrotation when we alter the rotation rate. There might further dynamical regime changes and multiple stable climate states at different parameters which we did not simulate. Considering deeper atmosphere layers with pressures $p> 10^{5} \:Pa$, the range of the zonal momenta is lower in the QHD case than the NHD case at low rotation rate, but larger at high rotation rate.Furthermore, we see a slow down of overturning circulation in the standard and tidally locked coordinates with increasing $\Omega$ (see the plots of the overturning circulation in the \textbf{supplementary file}). The overturning circulations of NHD cases differs quantitatively and qualitatively from circulations in the QHD case.

In the QHD case, the terms $\frac{Dv_{r}}{Dt}$, $\mathcal{F}_{r}=0$ and $\mathcal{A}_{r}$ lead to different vertical and indirectly to higher horizontal momenta. Therefore,the QHD case implies that GCMs with HPEs simulate too high zonal velocities at these parameterisations. The higher zonal wind speeds encounter the Coriolis forces. We expect a range of critical wind speed at a given rotation rate at which the climate switches to another climate state when the extra-tropical zone is relatively large. Higher wind speeds in combination with the smaller Rossby deformation radius $L_{D}$, moderate Coriolis forces may cause totally different climate states at certain parameters. Consequently, the models show different shifts of hotspot in simulations with different hydrodynamic equation sets depending on the parameters.

 The faster rotation rates cause deviations as well with other approximations; As \citet{tort2015} has already proved for terrestrial regimes, the traditional approximation gets increasingly less valid, when the rotation becomes faster. Regarding another Coriolis term, $-2\Omega\omega\cos{\phi}$ can be neglected if $2\Omega H\cos{(\phi)U^{-1}\ll 1}$ as \citet{white1995} did show. For our simulation at low $g$ and at $\Omega= 10^{-4} \: rads^{-1}$, $2\Omega H\cos{(\phi)U^{-1}}$ is about $0.21$ and $0.11$ for a wind speed of $500 \:m^{-1}$ at the equator respectively for the mid-latitudes. Therefore, the term $-2\Omega\omega\cos{\phi}$ gets more relevant in this climate state with extra-tropical conditions and GCMs with the traditional approximation in their dynamical equation sets may predict incorrectly. \citet{mayne2019} has shown that increased rotation rate leads to significant differences in the flow and the flow becomes dominated by the Coriolis forces. Furthermore, a higher rotation rate result in a net warming on the dayside and a net cooling on the nighside in their simulation although the more complete equations manifest less those warming and cooling effects. At higher pressures, they noticed only temperature changes by a few degrees. \citet{mayne2019} suggested to analyse and compare different dynamical equation sets with a full radiative transfer solution as used in \citet{amundsen2016}. Similarly as in \citet{mayne2019}, we see a net warming on the dayside and a net cooling on the nightside at pressures $p \leq 10^{5} \:Pa$. But in the deep atmosphere, temperatures start to vary increasingly by increased rotation rates in our simulations. That difference among both studies in the deep atmosphere may arise from different type of planet: hot, fast rotating Jupiters may respond differently than on slowly rotating and warm Neptunes. Furthermore, we simulated a much larger fraction of the deep atmosphere than \citet{mayne2019}. Regarding the radiative transfer, we expect effects on the dynamics and temperature structure due to different radiative transfer implementations.  Additionally, we expect some differences in the GCM implementations which leads to varying results when comparing to other studies.
 
\subsubsection{Radial flow}

This idealised climate state has a radial and divergent flow on the dayside as well as a convergent flow on the nightside in the upper atmosphere, analogous to a global Hadley or Walker cell. Vica versa for some deeper layers. The Helmholtz decomposition would show a dominant divergent component. That climate state is an idealised and needs higher ratio of $T_{irr}$ to $\Omega$ which is likely unrealistic compared to the observed exoplanets so far. At lower rotation rates $\Omega$, at $T_{irr} = 2'000 \:K$ and $g=10 \:ms^{-2}$, we see a transition to a climate state with a dominant divergent component, a moderate weak rotational eddy component and weak rotational jet component (see the plots of the Helmholtz decomposition in the \textbf{supplementary file} and Figure \ref{fig:classification}). The 3 jet system is still present in this transitional phase. As the Coriolis forces get weaker, winds get less deflected and can flow more direct from the dayside to the nightside. We see wind flows deflected less and crossing more directly over the poles to the nightside (e.g. at $10^{4} Pa$). There are certainly more simulations in this parameter space needed to characterise that area in the parameter grid. The circulation state may change at lower $\Omega$. It cannot excluded if there is a retrograde superjet at lower $\Omega$ and if radial flow is evolved due to a balance between prograde and retrograde tendencies (similar to the simulations with higher $g$). A similar radial flow pattern was found by \citet{carone2018} for tidally locked ExoEarths (TRAPPIST 1b, TRAPPIST 1d, Proxima Centauri b and GJ 667 C f) at relativly low $\Omega$.

The gradual transition to the climate state is seen at $T_{eq}=1'414.21\:K$ and $H=525.24\:km$. The Rossby number $Ro$ varies between $0.098$ and $4.39$ from winds of $100$ to $4'500 \:ms^{-1}$. The Rossby deformation radius and Rhines scale are $4.19$ and $0.83$ - $5.54 $. The Brunt-V\"ais\"al\"a frequency remains the same as at higher $\Omega$, $N = 0.00816 \:s^{-1}$.

\subsubsection{Prograde superrotation}
This circulation and climate state occurs on the one side at high $g$ and high $T_{irr}$, on the other side at low and high $g$, at relatively high $T_{int}$ compared to the $T_{irr}$.

The $T_{int}$ lies above the value computed according to the expression in \citet{thorngren2019}, $300 \: K$ respectively $400 \: K$. The high $T_{int}$ is debatable; High $T_{int}$ might be the reality as strong magnetic fields have been detected by \citep{yadav2017,cauley2019}. The magnetic field strength determines the $T_{int}$ substantially \citep{christensen2009}. \citet{thorngren2019} excluded $T_{int}= 100 \: K$ for planets with clouds because of the cold trap, especially for $ Teq \sim 1100$ to $1600 \:K$ \citep{lines2018b}. Nevertheless, higher $T_{int}$ can be realistic because of a significant higher entropy which cause a higher internal heat flux \citep{thorngren2019}. Regarding the cooling rate of hot (ultra) Jupiters, \citet{showman&guillot2002}, \citet{guillot2002}, \citet{youdin2010} and \citet{tremblin2017} predicted a downward heat transport by the atmosphere. As a theoretical proof, \citet{mendonca2020} found heat transport from the upper into the deeper atmosphere by the atmospheric circulation. Similarly, \citet{komacek2022} saw the coupling of internal evolution and atmospheric structure with the atmospheric dynamics in their simulations.

The Rossby number $Ro$ lies between $0.098$ to $ 4.39$. For the maximum wind speeds in our simulation, we get $Ro < 1.18$. The climate state have Rossby deformation radii $L_{D} \leq 3.63 \:R_{p}$. The Rhines scales $L_{RH}$ vary between $0.83$ to $5.54 \:R_{p}$, and smaller than $2.9\:R_{p}$ for the maximum wind speeds in our simulation. The Brunt-V\"ais\"al\"a frequency is $N > 0.01 \:s^{-1}$. The scale height is $H=55.42 \:km$ for $g=47.39 \:ms^{-2}$ and  $H\leq393.93 \:km$ for $g=10 \:ms^{-2}$.

Differences between simulations outputs from NHD and QHD equation sets are quantitatively relatively small and negligible at $T_{irr}= 1'500 \: K$ and $g=47.39 \:ms^{-2}$ respectively $g=10 \:ms^{-2}$. Qualitatively, the differences are more pronounced in the circulation pattern at $10^{4} \:Pa$. In this transitional phase, the QHD performs not so well compared to clear distinguishable circulation and climate states.

A dominant rotational jet component, a dominant rotational eddy component and a weaker divergent component characterise that climate state (see the plots of the Helmholtz decomposition in the \textbf{supplementary file}). Comparable simulations were computed by \citet{kataria2015} and \citet{schneider2022} for WASP 43b with the hydrostatic primitive equations (HPEs), although our parametrisation differs by slightly higher $T_{irr}$ and slightly higher $\Omega$. Our results with the parametrisation $\Omega = 10^{-4} \:rads{-1}$, $T_{irr}= 2'000 \: K$ and $g=47.39 \:ms^{-2}$ partially agree to those of \citet{kataria2015} and \citet{schneider2022}. We see a prograde jet and Rossby gyres which transport zonal momenta to low latitudes as well as retrograde flow at high latitudes like in their study (e.g. see Figure with simulation computed with $g=47.39 \:ms^{-2}$ $T_{irr}= 2'000 \: K$ and altering $\Omega$ in the \textbf{supplementary file}). But the speed of the jet remains with $\sim 1'800 \:ms^{-1}$ for the NHD and QHD case much lower than the wind speeds of $5'500 \:ms^{-1}$ in studies of \citet{kataria2015} and \citet{schneider2022}. At this parametrisation, it seems the HPEs predict too high wind speeds compared to the NHD and QHD equation sets.
The differences between the NHD and QHD case are less than $100 \:K$ and minor compared to the low gravity.

\subsubsection{Retrograde superjet}
In this circulation and climate state, a retrograde superjet leads to a westward offset of the hotspot. This climate states occurs at high gravity and low rotation rate ($T_{irr}= 2'000 \: K$, $\Omega= 10^{-5}rads{-1}$ and $g=25 \:ms^{-2}$ respectively $g=47.39 \:ms^{-2}$; see Figure \ref{fig:classification}). This climate state has a dominant rotational jet component, a weak divergent component and a weak rotational eddy component (see the plots of the Helmholtz decomposition in the \textbf{supplementary file}). We see a transition from retrograde to prograde superrotation in the simulations with $T_{irr}= 2'000 \: K$, $\Omega= 10^{-4.5}rads{-1}$ and $g=47.39 \:ms^{-2}$ respectively partially in the simulation with $T_{irr}= 1'500 \: K$, $\Omega= 10^{-5}rads{-1}$ and $g=47.39 \:ms^{-2}$ (see Figure \ref{fig:classification}).

The equilibrium temperature lies around $T_{eq}=1'414.21 \:K$. The scale height is $110.83$ respectively $210.1 \:km$. The Rossby number is around $0.098 - 4.39$. The high winds in our simulations imply encountered Coriolis forces, partially tropical conditions. The Rossby deformation radius is $4.19$, while the Rhines scales varies in a range of $0.83-5.54$.

Differences between simulations outputs from NHD and QHD equation sets are less than $200$ and less than $100 \:K$ at $T_{irr}= 2'000 \: K$, $\Omega= 10^{-5}rads{-1}$ and $g=25 \:ms^{-2}$ respectively $g=47.39 \:ms^{-2}$ for pressures larger than $10^{4} \:Pa$. The smaller temperature differences come along with a stronger retrograde superjet.

\subsection{Implication for the superrotation}

We see a complete shift in the climate regime in our simulations towards a retrograde jet spreading to high latitudes at pressures $p \leq 10^{5.5} \:Pa$ and at low $\Omega$ when gravity increases. Many studies (e.g. \citet{showman&guillot2002, showman2009, dobbs_dixon2010, tsai2014,kataria2015,amundsen2016,zhang2017, mendonca2018}) have shown that tidally locked hot Jupiters produce an equatorial eastward wind jet in 3D simulations. The equatorial eastward jet transports heat to the nightside and shifts the hotspot to the east \citep{knutson2007}. Nevertheless, there are several exceptions among hot Jupiters; \citet{dang2018} observed a westward shifted hotspot in CoRoT 2b. Similarly, \citet{may2022} made observations of a westward shift for WASP 140b. Several factors can counter superrotation \citep{carone2020}; clouds  \citep{helling2016, parmentier2016,mendonca2018}, including variability in the cloud coverage \citet[possible for CoRoT 2b and HAT P7b]{armstrong2016,dang2018}, higher metallicity in the planet's atmosphere \citep{kataria2015,drummond2018a} and magnetic fields \citep{rogers2014,kataria2015, arcangeli2019, hindle2019} may affect the circulation significantly. Moreover, planets may evolve retrograde flow because of non-synchronous planetary rotation \citep{rauscher2014}. We suggest that the choice of the dynamical equation set may counter superrotation as well as lead to different jet systems and climate states. Furthermore, we assume additional physical schemes may alter the balances for the evolution of jet systems and climate states.

Many of the previous studies used simplified Newtonian cooling or grey RT solutions, \citep{lee2021} showed the improvements for more realistic RT solution We consider more realistic RT solution in GCMs and other schemes in addtion to the dynamical cores as a key consideration when investigating differences between dynamical equation sets.
 WASP 43b orbits its host star with  $0.8315$ days relatively quickly \citep{hellier2011} and is unusually dense. \citet{carone2020} simulated WASP 43b and got varying results compared to \citet{kataria2015}, \citet{mendonca2018} and \citet{schneider2022}; The simulations of WASP 43b in \citet{carone2020} show westward (retrograde) flow in the upper thermal photosphere ($p\le 8'000 \:Pa$) as soon as the model simulates deep wind jets. They found a strong tendency of an equatorial westward flow in the eddy-mean-flow analysis for $p < 10^{4} \: Pa$ for WASP 43b. \citet{carone2020} concluded that the deep atmosphere may significantly influence the atmospheric flow in the observable middle and upper atmosphere of hot Juptiers. \citet{deitrick2020} stated as well a retrograde flow at $10^{6} \: Pa$ in the simulations of HD 189733b. Investigating eddy transport, \citet{mayne2017} noticed a deceleration of the superrotating jet due to the evolution of the deep atmosphere (the model did not reach steady state after $10'000$ Earth days). In their study, air masses sink over the poles and rise over the equator. The horizontal temperature gradient at greater depths ($p > 10^{6} \: Pa$) powers the deep circulation. 

Retrograde flow has been noted in simulation in few cases; \citet{showman2015} performed simulations for HD 189733b altering irradiation (warm and cool Jupiters) and rotation periods ($0.55$, $2.2$, and $8.8$ Earth days). Their simulations with fast rotation or low irradiation show retrograde flow in the zonal-mean wind. More retrograde flow patterns were found for tidally locked exo-Earths with fast rotation  \citep[less than $3$ Earth days,][]{carone2015}. \citet{carone2020} showed that retrograde flow over the equator can appear on dense and hot Jupiters. \citet{mayne2017} highlighted that vertical angular momentum in balance of horizontal interactions plays a crucial role for the evolution of superrotation. \citet{carone2020} identified unusually deep wind jets \citep[already predicted by][]{thrastarson2011} accompanied by deeper convective layers. Those deep wind jets may impact the upper atmosphere ( $p< 10^{6} \: Pa$) by zonal momentum transport at depths ( $p> 10^{6} \: Pa$) that supposed to increases with faster rotation. More studies are required to understand the exact mechanisms and regimes that can produce retrograde flow.

\citet{mayne2019} analyzed indirectly the effect of gravity on the dynamic equation set via temperature contrast and the scale height. They concluded that the maximum variation appears between varying and constant $g$, when the temperature contrast is altered, and their view when $g$ is supposedly altered as well (scale height). The deep (equation) case varies roughly 30 \% to the full (equation) case at the top of the atmosphere. \citet{mayne2019} stated that the resulting flows in the simulations with the primitive and deep equation set respond independently of the treatment of $g$. Our results support the idea of independence of $g$ partially; At high gravity differences among NHD and QHD nearly vanish. It can be explained by the growing dominance of the gravity term over other terms. But at low gravity, the other terms in the NHD and QHD equation set reveal their effects and the related differences which cannot anymore be encountered by the gravity term. We cannot comment how the full equation responses in comparison to other equation sets, since the THOR model does not provide the option for varying $g$ yet. Only an extensive study on the effects of the gravity term with different dynamical equation sets can provide a full answer. The combination of high gravity in the deep atmosphere with decreasing gravity in the upper atomsphere may even lead to total different climate states than presented in here.

We have to note that the Bond albedo changes with $g$ with altered gravity and with that the incoming shortwave radiation. Therefore, we see effects of $g$ combined with radiative effects on the dynamics.

\citet{mayne2019} showed that increased planetary temperature contrast lead to an accelerated zonal flow while comparing the primitive with the full equation set. They see significant changes in the thermal structure. As a consequence the regime becomes advectively dominated. The changes in the zonal flow and advection end in changed temperature structure \citep{mayne2019}.
We see growing differences between the NHD and QHD equation set in the zonal momenta in our simulations, when we increase the irradiation temperature. At lower irradiation temperatures, the differences nearly vanish and a superrotation is evolved. The deviations in the temperature remain much smaller at lower temperatures. That is not surprising, since the temperature is not included directly in the altered terms in the QHD case, $\frac{Dv_{r}}{Dt}$, $\mathcal{F}_{r}$ and $\mathcal{A}_{r}$. Therefore, the deviations have to rise from the changed dynamics which alters the temperature advection and therefore the temperature structure at higher irradiation temperature more significantly. At higher irradiation temperatures, the spread in the T-p profiles (day - night contrast) increases with higher irradiation temperatures. Hence the temperature advection gets a more decisive role in the temperature structure of planets. In the comparable study of \citet{deitrick2020}, only minor differences appear in the temperature among simulations with NHD and QHD equation sets in the simulations of HD 189733b. They stated slightly higher velocities in the NHD case and differences of jet velocity of roughly 5 \%. THOR produces a superrotation as well in their simulation.

\citet{may2022} compared Spitzer phase curves and showed evidence for a trend of increasing phase offset with increasing orbital period at $4.5 \mu m$ (for $T_{eq} \equiv 1'300 \:K$), as already shown in \citet{parmentier2018}. Our results show larger offsets with larger orbital periods for the NHD case when gravity is low (for $T_{eq} \equiv 1'414.21 \:K$). This comes along with a weaker overturning circulation with increasing $\Omega$ (see the plots of the overturning circulation in the \textbf{supplementary file}). The QHD case does not show a trend in this regard and the offset changes more due to climate state changes at low $g$. At higher gravity, the offset switches direction due to climate state changes. We see a decrease of the eastward offset of the hotspot when superrotation is prograde, $g$ is high and $\Omega$ increases.

Moreover, \citet{zhang2017} suggested that only the radiative and advective timescales affect the hotspot offset. So, the radiative timescale should not be changed by the rotation rate. Consequently the rotation rate should change the wind speed in tidally locked hot Jupiter when the rotation rate is altered because of the trend of the offset. Therefore, faster rotation rates should lead to weaker equatorial jets. In our simulations, we see the radiative timescales changing in the NHD and QHD case when the rotation rate is altered, due to temperature advection. Moreover, the radiative timescales on the nightside vary much more than those on the dayside when the rotation rate alters. Nevertheless, we see a weakening of the equatorial jets with higher rotation rates in the NHD case. The QHD case does not show weakening, much more a strengthening with higher rotation rates. Looking at the entire parameter grid we simulated, the offset changes, when we altered $g$, $T_{\text{irr}}$ and $\Omega$. We see the offset changes due to several parameters. Similarly, \citet{hammond2018} showed dependence of the offset on a nondimensional parameter, which is related to the radius, scale height, gravity and rotation rate. \citet{may2022} observed the dependence of the offset is not only bound to the rotation rate as in hot Jupiters, but also to gravity for cooler Jupiters with consistent nightside temperature near $\sim 1'000\:K$. The different jet structures and offsets of the hotspots in our simulated parameter grid imply a dependence on multiple parameters as \citet{hammond2018} and \citet{may2022} suggested.

Comparable simulations to ours in the studies of \citet{kataria2015} (SPARC/MITgcm) and \citet{schneider2022} (expeRT/MITgcm), but computed with hydrostatic primitive equations (HPEs), show 3 times higher wind speeds for WASP 43b than our results. Unfortunately, the lower wind speeds in our simulations are mostly due to the limit imposed by the model top. Moreover, our parametrisation differs by slightly higher $T_{irr}$ and slightly higher $\Omega$. The GCM with HPE in \citet{kataria2015} predicts a superrotation with high wind speeds up to $4'800 \:ms^{-1}$. The wind speeds in our simulations lie around $\sim1'000$ and $\sim 500 \:ms^{-1}$ for our QHD case with $T_{irr}= 2'000 \: K$, $g=10 \:ms^{-2}$ and $\Omega = 1 \cdot 10^{-5}$ respectively $\Omega = 1 \cdot 10^{-4.5}$. The QHD case already predicts too high wind speeds compared to the NHD case depending on the parametrisation. The HPEs seem to predict even much higher wind speeds at this parametrisation, but it needs to be studied more extensively. Furthermore, the simulation for HD 209458b in study of \citet{kataria2015} can be classifed in a transitional state between our \textit{3 prograde jets} and the \textit{radial flow}. Therefore, we expect elements of a 3 prograde jets combined with a dominant divergent component if computed with the NHD equation set.

On the other hand, if we compare simulations for HD 189733b, the THOR model (with the double-grey dual band radiative transfer scheme) produces a prograde superrotation in the study of \citet{deitrick2020} with wind speeds up to $\sim 5'600 \:ms^{-1}$ in the NHD case, even higher than in the QHD case. \citet{kataria2016} simulated HD 189733b as well. The zonal mean wind speed goes a bit beyond $3'200 \:ms^{-1}$, but it remains lower than in study of \citet{deitrick2020}. Although \citet{deitrick2020} and \citet{kataria2016} predict a superrotation, the jet maxima is found at 2 magnitude higher pressures in \citet{kataria2016}. We consider different physical scheme as well combination with different dynamical equation sets have an effect on the jet structure and the climate state, but is has to be investigated further. In a comparison of radiative schemes, \citet{lee2021} showed different radiative transfer schemes can lead to different wind speed and temperature structures.

Hot Jupiter climates are often associated with a equatorial prograde superrotating jet (see \citet{showman2020} for full review). That concept is often supported by GCM simulations which show a prograde superrotation. Comparing jet systems in different studies, most simulations for hot Jupiters (e.g.\citet{kataria2015}, \citet{kataria2016}, \citet{amundsen2016} and \citet{schneider2022}) show only prograde superrotation. So far, only \citet{carone2020} predicts a retrograde flow for WASP 43b, embedded in a strong superrotation, with the GCM MITgcm with HPEs. Like in \citet{carone2020}, we see retrograde flow in similar cases depending on the parametrisation, but we did not explicitly simulate WASP 43b. Nevertheless, we predict even a retrograde superjet in one of the 4 different circulation states. The evolution of climate states and the jet structures depend on the parametrisation and choice of the dynamical equation set. Zonal momentum transport may play a crucial role for the evolution of retro-, prograde and cross-the-poles wind flow. Such association with the momentum transport was found by \citet{carone2020}. They associate the upwards zonal momentum transport to a deep jet which leads to the retrograde flow in the upper atmosphere. Such momenta transport can be missed by HPEs, since they ignore several terms of the full equation set related to momenta transport such as $2\Omega\cos{(\phi)}$, $\frac{-uw}{r}$ and $\frac{-uv}{r}$ \citep[see the full review on dynamical equation sets in][]{mayne2014a}. Even the NHD case does represent the full equation set, since $g$ does not altered with the altitude. We illustrated some effects of $g$ on the different dynamics and outcomes by altering $g$. Therefore, our simulation outcome may change drastically depending on the parametrisation when the full equation set is implemented in THOR.

Regarding the evolution of different jet system, \citet{serveev2022} demonstrated an interesting case of climate bistability in TRAPPIST-1e. They found 2 distinct jet systems for a $10^{5} \:Pa$ nitrogen-dominated atmosphere. They characterised 1 strong equatorial prograde jet (with strong day-night contrast) and 2 mid-latitude prograde jets (with weak day-night contrast). In their numerical experiments, the bistability was highly sensitive to the model setup, such as initial conditions, surface boundary conditions, physical parameterisation of convection and cloud radiative effects. They found a balance between the zonally asymmetric heating, mean overturning circulation, and mid-latitude baroclinic instability. As not the only study, \citet{edson2011}, \citet{noda2017} and \citet{carone2018} discovered transitional states between well defined jet systems and climate states similarly to our study. Some rocky exoplanets seem to be sensitive not only to GCM setup \citet{serveev2022}, but as well to the GCM choice as shown by \citet{sergeev2022b} and \citet{turbet2022}. As an addition to these studies, our study shows that choice of the dynamical equation set within a GCM leads to evolution of different climate state.

The discussion about the dynamics on hot Jupiters \citep[e.g.][]{kataria2015, mendonca2018, mayne2019,carone2020,deitrick2020,schneider2022} together with our results have no reached a consensus yet. Further studies of the dynamics of hot Jupiters with GCMs with the full equation set are needed.

Many simulations uses HPEs which come along with shortcomings due to their approximations made for Earth. We demonstrate with our comparison that such approximations can lead to complete changes in the jet structure and climate state that just arise from the choice of the dynamical equation set. In several parameter settings, we see prograde superrotation, but as well deviations from prograde superrotation, such as retrograde superjet, disrupted superrotation and a 3 jet instead of 1 jet structure. Nonetheless, we should be careful since climate state and observational features may change over long integration times (e.g. $50'000 -250'000$ Earth days) as \citet{wang2020} has shown. They saw the evolution of 2 prograde off-equatorial jets to a single prograde equatorial jet ranging up to the poles. Also, they found the hotspot shift becomes eastward after long integration times. Regarding the reason of the long convergence, they hint to the long radiative timescales in the deep atmosphere. They run simulations for the warm sub-Neptune GJ 1214b with the GCM LMDZ with HPEs and with two-stream grey gas RT scheme. Our comparable simulations have too high $T_{int}$ in comparison to GJ 1214b and might be in a different climate state.

We assume the climate states on hot Jupiters are more diverse than the simple superrotation. \citet{armstrong2016} found a westward shift of the hotpsot and brightness peak with Kepler measurements of HAT-P-7b. Similarly, \citet{may2022} observed a westward offset of the hotspot for WASP 140b. Moreover, \citet{dang2018} presented thermal phase observations of the hot Jupiter CoRoT 2b obtained with the Infrared Array Camera (IRAC) on the Spitzer Space Telescope. They detected a westward offset of the hotspot of $23 \pm 4^{\circ}$. The large westward offset in \citet{dang2018} might be another evidence of retrograde flow or even retrograde superjet in hot Jupiter atmospheres. Simulations including magnetohydrodynamics (MHD) predicted a westward flow \citep{rogers2014}. A more recent study \citep{hindle2019} showed simulations with MHD which led to westward shifts of the hotspot for HAT P-7b and CoRoT 2b. For these reasons, we conclude hot Jupiter atmospheres might be more diverse than so far assumed.

\subsection{Limitations and future improvements}

The GCM THOR can encounter numerical instability when the gradient between the nightside and dayside temperatures is too large \citep{deitrick2020}, most problematic when modelling ultra hot Jupiters. As a consequence, we could not simulate pressures lower than $\sim 7 \cdot 10^{2}$ or $ 10^{3} \:Pa$ (depending on the parametrisation) which affects the dynamics and temperature structure to some degree. Future updates to the THOR GCM will address the issue of large day-night temperature gradients.
 
\citet{mayne2019} performed simulations for warm, tidally locked and slowly rotating Neptunes and super Earths with a duration of $1'000$ Earth days. They saw the evolution of the maximum zonal wind speed and structure ceased in their simulations at lower pressures (pseudo-steady). The deep, high pressure atmosphere still evolve slowly in their simulations after $1'000$ Earth days. The slow evolution of the deep atmosphere does not appear to have a significant effect on the dynamics of the upper, low pressure atmosphere for hot Jupiters \citep{mayne2017}. Contrary, \citet{carone2020} suggest advection of zonal momenta upwards from the deeper atmosphere.

In this study, we run the simulations for $5'000$ Earth days and for a certain number of Earth days and did not set the duration according to a convergence condition. The computation time would take too long for 2 dozens of simulation cases to finish the study in a meaningful time. We simulated the deep atmosphere to $10^{8}\: Pa$ which needs significantly more time to converge \citep{mayne2017}. However, we simulated the deep atmosphere to stabilize THOR, especially for the first few hundreds days. Regarding sufficient time periods for convergence to steady state, \citet{wang2020} simulated GJ 1214b for $50'000$ days to observe the transition from 2 equatorial jet into 1 jet. Such long integration times are beyond our current computational resources for parameter grid we computed. \citet{christie2022} set the simulation time on basis of evolved features which different models create early on. Important feature such as the equatorial jet can be evolved in $7'800$ days \citep{menou2012} for GJ 1214b. A shorter run time was used in \citet{komacek2022}, but with a shallower atmosphere and a surface pressure of $10^{6}\: Pa$. A more detailed analysis on the convergent time for deep atmosphere is done by \citet{schneider2022}. They did run simulations with a surface pressure of $10^{8}\: Pa$ for WASP 43b and HD 209458b for $12'000$ Earth days. While HD 209458b did converge within the $12'000$ Earth days, WASP 43b did evolve steadily during the full simulation time. The temperature change rate drops from $\sim 1.5$ to $\sim 0.05 \:Kd^{-1}$ at the end of the simulation. Regarding the final state of the deeper atmosphere, \citet{schneider2022} confirmed the independence of the initial conditions for WASP 43b. As \citet{serveev2022} showed the high sensitivity of the model setup in relation to the evolution of the distinct climate states, more studies are needed to examine bistability and even multistability of exoplanets.

At lower resolution, THOR approaches steady state around $2'500-3'000$ Earth days for simulations of HD 189733 b, while high-resolution simulations converge after $10'000$ Earth days (\citet{deitrick2020}, indicated by the superrotation index according to \citet{mendonca2020}). The zonal flow undergoes a quick development and changes only very little after ~$2'000$ Earth days \citep{deitrick2020} in the simulations of HD 189733b. They showed as well that the upper atmosphere reached steady state, although the lower atmosphere did not reach it in their simulations with $g_{level}=5$ (around $2^\circ$). For hot Jupiters, we expect even shorter convergence times due to higher temperatures so that $5'000$ days are sufficient to observe differences between NHD and QHD equation sets at pressures $p \leq 10^{6} \:Pa$.
 
Higher resolutions conserve mass better as \citet{deitrick2020} noted that THOR conserves mass at $g_{level}=5$ (around $2^\circ$) slightly less well than at $g_{level}=6$ (around $1^\circ$), although the output looks qualitatively very similar. Moreover, terms such as the $\cos{\phi}$ become relevant for the mesoscale motion \citep{draghici1989} on Earth. Furthermore, more complex atmospheric motions may appear if the model resolution increases like on Jupiter \citep{schneider2009, gastine2021,heimpel2022}. On exoplanets, a higher resolution may lead to larger differences among simulations with different dynamical equation sets. Furthermore, mass, energy, numerical dissipation and integration errors lead to gradual changes of the total axial momentum \citep[see more details in][]{mendonca2020,deitrick2020}.

Regarding the gravity, THOR has a constant value throughout the atmosphere. A decreasing gravity with height would change the simulation outputs and their realism. We expect further implications for the QHD equation set and other approximations, especially at higher altitudes respectively at lower pressures ($p<10^{5} \:Pa$), since we find the largest differences at low gravity.

\section{Summary and conclusions}

For exoplanet atmosphere GCMs, several hydrodnymaic equation sets are used across the literature. However, only a few studies have compared the differences between equation sets and their effects on the atmospheric dynamical properties \citep{mayne2019,deitrick2020}.
This will be important to consider as spectral phase curve data is produced by JWST.

In this study, we compared the NHD and QHD equation sets \citep[following the nomenclature and definitions in][]{deitrick2020} in the GCM THOR. We simulated atmospheres across a parameter grid to reveal the validity of the equation sets for a wide range of the exoplanet population. Additionally, we implemented a two-stream non-grey "picket-fence" scheme to THOR which increases the realism of the radiative transfer in the model.

Our results show significant differences between the NHD and QHD equation sets in the GCM THOR for fast rotation rates, lower gravity and higher irradiation temperatures. The NHD and QHD equation sets in THOR differ only in the terms $Dv_{r}/Dt$, the Lagrangian derivative of the vertical velocity, $\mathcal{F}_{r}$, the hyperdiffusive flux and $\mathcal{A}_{r}$, the vertical component of the advection term. But those terms cause significantly different results in the dynamics and the vertical temperature structure in several regimes. Depending on the parameters, the NHD and QHD equation sets even evolve to different dynamics, radiative regime and climate state.

Overall, our study shows the evolution of different climate states which arise just due to different selection of Navier-Stokes equations and approximations. We show the implications of approximations made for Earth, but used for non Earth-like planets. Our results agree qualitatively to comparable studies of \citet{mayne2019} and \citet{deitrick2020}. \citet{mayne2019} made a similar comparison, but with the Met Office Unified Model. They compared simulations of slow-rotating, small Neptune-sized planets with the primitive and deep equation set. \citet{deitrick2020} used THOR in a similar comparison of the NHD and QHD equation sets and showed already significant differences in the dynamics in two regimes (Earth like case and HD 189733b). We showed that differences between the NHD and QHD equation sets can vary depending on the parametrisation and choice of the dynamical equation set. Finally, our results show the relevance in the use of different dynamical equation sets depending on planetary and system properties.

Future investigations may extend this study by comparing the full equation set, NHD equation set and hydrostatic, shallow approximations in GCMs. Additionally, \citet{mayne2019} suggested to implement chemical equilibrium \citep{drummond2018b,drummond2018c} and a cloud scheme like in \citet{lines2018a}. A more sophisticated spectral RT scheme like \citet{deitrick2022} may also alter our findings. Longer simulations times, similar to \citet{wang2020}, and GCMs with the full equation set may reveal new circulation and climate states as well as multistabilities.

\section*{Acknowledgements}

P.A. Noti and E.K.H. Lee are supported by the SNSF Ambizione Fellowship grant (\#193448).
Financial support to R.D. was provided by the Natural Sciences and Engineering Research Council of Canada (NSERC; Discovery Grant RGPIN-2018-05929), the Canadian Space Agency (Grant 18FAVICB21), and the European Research Council (ERC; Consolidator Grant 771620).
M.H. gratefully acknowledges funding from Christ Church, Oxford.
Data and plots were processed and produced using PYTHON version 3.9 \citep{van1995python} and the community open-source PYTHON packages \emph{Bokeh} \citep{bokeh}, \emph{Matplotlib} \citep{hunter2007}, \emph{cartopy} \citep{cartopy1,cartopy2}, \emph{jupyter} \citep{kluyver2016}, \emph{NumPy} \citep{harris2020}, \emph{pandas} \citep{reback2020pandas}, \emph{SciPy} \citep{jones2001}, \emph{seaborn} \citep{waskom2021}, \emph{windspharm} \citep{dawson2016} and \emph{xarray} \citet{hoyer2017}. Calculations were performed on UBELIX (\url{http://www.id.unibe.ch/hpc}), the HPC cluster at the University of Bern. We thank the IT Service Office (Ubelix cluster), the Physikalisches Institut and the Center for Space and Habitability at the University of Bern for their services.

\section*{Data Availability}

 We used the development version of the GCM THOR (available on \url{https://github.com/exoclime/THOR}). The code for the picket-fence scheme and the new mode for the initial conditions were uploaded on the lead author’s GitHub: \url{https://github.com/PA-NOTI/THOR_picket_fence_scheme}. The code on the lead author’s GitHub was used to run the GCM THOR simulations. The added features got integrated in the main Github of the GCM THOR. The input and output files of the GCM THOR are available on Zenodo, \href{https://doi.org/10.5281/zenodo.7620774}{DOI: 10.5281/zenodo.7620774} and \href{https://zenodo.org/record/8014271}{DOI: 10.5281/zenodo.8014271}. All other data and code are available from the authors on a collaborative basis.



\bibliographystyle{mnras}
\bibliography{literature} 




\appendix

\section{Tidally locked coordinates and velocities}
\label{tidally}

For analysis of symmetries in the atmosphere of a tidally locked
planet, we make use the ‘tidally locked coordinate system’ suggested by \citet{Koll2015}. In the transformation, the traditional latitude-longitude system ($\vartheta,\lambda$) get replaced by the ‘tidally locked coordinate system’ ($\vartheta', \lambda'$) as the following:

The coordinates are effectively a rotation of regular latitude-longitude
coordinates, so that the polar axis runs from the substellar
point to the antistellar point. They define the tidally locked latitude
$\vartheta$0 to be the angle to the terminator, and the tidally locked longitude
to be the angle about the substellar-antistellar axis. That rotation of the coordinate system results into the tidally locked coordinates according to \citet{Koll2015} as
\begin{subequations}
\begin{align}
\vartheta' &= sin^{-1}(\cos{\vartheta}\cos{\lambda}) , \\
\lambda' &= tan^{-1} \Bigg(\frac{\sin{\lambda}}{\tan(\vartheta)} \Bigg), 
\end{align}
	\label{eq:tidallycoordinates}
\end{subequations}
where $\vartheta'$ is the tidally locked latitudes, $\lambda'$ the tidally locked longitude, $\vartheta$ the original latitude and  $\lambda'$ the orginal longitude.
The tidally locked wind velocities consist of fractions of the original zonal and meridional wind components. The fractions change depending on the coordinates. According to \citet{Koll2015}, the tidally-locked zonal and meridional wind $u'$ and $v'$ are defined as
\begin{subequations}
\begin{align}
u' &= \cos{\vartheta} \Bigg(\frac{\partial \lambda '}{\partial \lambda} \frac{u}{\cos{\vartheta}} + \frac{\partial \lambda '}{\partial \vartheta}v \Bigg) , \\
v' &= \frac{\partial \vartheta'}{\partial \lambda}\frac{u}{\cos{\vartheta}} + \frac{\partial \vartheta'}{\partial \vartheta} v, 
\end{align}
	\label{eq:tidallywind}
\end{subequations}
where $u$ and $v$ are the zonal and meridional wind components of the original coordinate system.

\section{Streamfunction and tidally-locked streamfunction}
\label{streamfuction}

For analyzing the mass flow, we performed the tidally-locked streamfunction $\Psi '$ and the Eulerian mean meridional streamfunction $\Psi$ in the same fashion as \citet{hammond2021} as
\begin{subequations}
\begin{align}
\Psi &= \frac{2 \pi a \cos{\vartheta}}{g}  \int\limits_{0}^{p}[v]_{\lambda} dp , \\
\Psi ' &= \frac{2 \pi a \cos{\vartheta '}}{g}  \int\limits_{0}^{p}[v']_{\lambda '} dp, 
\end{align}
	\label{eq:streamfunction}
\end{subequations}
where $g$ declares the gravity, $a$ the equatorial radius, $[v]_{\lambda}$ averaged wind over longitude and the $[v']_{\lambda '}$ an averaged wind over tidally-locked longitude.

\section{Helmholtz decomposition}
\label{helmholtz}

We performed a Helmholtz decomposition according to \citet{hammond2021} to analyse changes due to altered parameters in our grid which might be discovered in the components of the total circulation such as the overturning circulation, stationary waves, and superrotating jet. In the Helmholtz decomposition, the total circulation is split up into the divergent and rotational components $u_{d}$ and $u_{r}$ \citep{dutton2002}:
\begin{subequations}
\begin{align}
u &= u_{d}+u_{r}= \\
 &= \bigtriangledown \chi + k \times \bigtriangledown \psi, 
\end{align}
	\label{eq:helmholtz_components}
\end{subequations}
where $\chi$ stands for the velocity potential function and $\psi$ for a streamfunction which are defined as:
\begin{subequations}
\begin{align}
 \bigtriangledown ^{2} \chi &=  \delta, \\
 \bigtriangledown ^{2} \psi &=  \zeta, 
\end{align}
	\label{eq:divergent_vorticity}
\end{subequations}
where $\delta$ is the divergence and $\zeta$ the vorticity.

\section{OLR phase curve}
\label{OLR}

\citet{cowan2008} formulated the phase curve as
\begin{equation}
F=\int_{\lambda_{1}}^{\lambda_{2}}\int_{\vartheta_{1}}^{\vartheta_{2}}\int_{-\pi/2}^{\pi/2} R^{2}\frac{F_{T\!O\!A}}{\pi} \cos^{2}{(\theta)}\cos{(\vartheta-\alpha)} \,d\phi \,d\vartheta \,d\lambda,
 \label{eq:phase_curve}
\end{equation}
where $F_{T\!O\!A}$ is the flux at the top of the atmosphere coming from the each atmospheric column of the GCM at a given wavelength $\lambda$, $\phi$ and $\vartheta$ declare the latitude and longitude and $\alpha$ the orbital phase angle.
\citet{deitrick2022} introduced a formalism to calculate the $F$ on an icosahedral grid as
\begin{equation}
F= \sum_{i=1}^{N_{g\!r\!i\!d}} \frac{F_{T\!O\!A,i}}{\pi} \mu_{i}\frac{A_{i}}{R^{2}_{p}},
 \label{eq:phase_curve_icosahedral}
\end{equation}
where $A_{i}$ declares the area of each control volume at the top of the atmosphere $R_{p}$, the radius of the planet and 
\begin{equation}
\mu_{i}=
\begin{cases}
\cos{(\phi)}\cos{(\vartheta - \alpha)},  &\alpha - \frac{\pi}{2} < \vartheta < \alpha + \frac{\pi}{2},\\
0 , & \vartheta > \alpha + \frac{\pi}{2} \:or\: \vartheta < \alpha - \frac{\pi}{2}.
\end{cases}
 \label{eq:phase_curve_mu}
\end{equation}
We take the approach of \citet{deitrick2022} adapt it to a longitude-latitude grid and limit it to long-wave radiation.
\citet{kelly2021} defined the surface area of a grid-cell in a longitude-latitude grid on the sphere as
\begin{equation}
A_{S}=\int_{\phi{1}}^{\phi{2}}\int_{\vartheta{1}}^{\vartheta{2}} R_{p}^{2} \cos{(\vartheta)} \,d\phi \,d\vartheta = R_{p}^{2}(\vartheta{2}-\vartheta{1}) (\sin{(\phi_{2})} -\sin{(\phi_{1}})).
 \label{eq:area_grid-cell}
\end{equation}
By switching to longitude-latitude grid, we modify the Equation \ref{eq:phase_curve_icosahedral}  with Equation \ref{eq:area_grid-cell} and reformulate as OLR phase curve as
\begin{equation}
F_{OLR}= \sum_{i=1}^{N_{g\!r\!i\!d}} \frac{F_{O\!L\!R, T\!O\!A, i}}{\pi} \mu_{i} (\Delta \vartheta) (\sin{(\phi+\Delta\phi)} -\sin{(\phi-\Delta\phi)}),
 \label{eq:phase_curve_lonlat}
\end{equation}
where $\Delta \vartheta$ is the longitudinal width of a grid cell, $\Delta\phi$ the latitudinal width and we defined $\mu_{i}$ as
\begin{equation}
\mu_{i}=
\begin{cases}
\cos{(\phi)}\cos{(\vartheta - \alpha)},  &\cos{(\phi)}\cos{(\vartheta - \alpha)} \geq 0,\\
0 , & \cos{(\phi)}\cos{(\vartheta - \alpha)} < 0.
\end{cases}
 \label{eq:phase_curve_mu_new}
\end{equation}

\section{Radiative and zonal timescales}
\label{timescales}
We computed the radiative timescale as in \citet{showman&guillot2002} as
\begin{equation}
 \tau_{r\!a\!d} \sim \frac{P}{g}\frac{c_{p}}{4\sigma_{B} T^{3}},
 \label{eq:radiative_timescales}
\end{equation}
where $P \:[Pa]$ declares the pressure, $g \:[ms^{-2}]$ the gravity, $\sigma_{B}$ the Stefan-Boltzmann constant, $c_{p} \:[J kg^{-1}K^{-1}]$ the heat capacity at constant pressure and $T \:[K]$ the temperature.
The zonal timescales was calculated as well like in \citet{showman&guillot2002} as
\begin{equation}
 \tau_{z\!o\!n\!a\!l} \gtrsim \frac{R}{u_{m\!a\!x}},
 \label{zonal_timescales}
\end{equation}
where $R$ is the planetary radius and $u_{m\!a\!x}$ the maximum of the zonal wind speed.
We computed the radiative and zonal timescale for each layer with the related values.

\section{Large-scale flow quantities}
\label{large_scale_flow_quantities}
For the analytics, we used several quantities for the large-scale flow characteristics. The scale height $H$ is defined in \citet{parmentier2014PhD} and we reformulated it as

\begin{equation}
 H =\frac{k_{B}T}{mg}=\frac{R_{d}T}{g},
 \label{eq:scale_height}
\end{equation}
 where $k_{B}\: [m^{2} kg s^{-2} K^{-1}]$ is the Bolzmann constant, $T \:[K]$ the temperature of the gas, $g \:[ms^{-2}]$ the gravity, $m\: [kg]$ the mass of the gas and $R_{d}\: [J kg^{-1} K^{-1}]$ the specific gas constant.

 The Rossy number indicates the balance in the momentum equation between the Coriolis and the advection term \citep{parmentier2014PhD,kataria2016}:
 \begin{equation}
 Ro \equiv \frac{U}{fL},
 \label{eq:rossby_number}
\end{equation}
where $L \: [m]$ is the typical horizontal scale, $U \:[m/s]$ the typical wind speed and $f \:[rads^{-1}]$ the Coriolis parameter as
 \begin{equation}
 f=2\Omega\sin{(\vartheta)},
 \label{eq:coriolis_parameter}
\end{equation}
where $\Omega \:[rad]s^{-1}$ represents the rotation rate of the planet and $\vartheta [rad]$ the latitude. The typical horizontal scale is typically calculated as the Rossby deformation radius $L_{D} \:[m]$ (see hereafter). The coriolis force becomes negligible and the advection, pressure gradient and dissipation remain the terms relevant in the force balance, when the Rossby number is much larger than one \citep{parmentier2014PhD}. On the other hand, a much smaller Rossby number indicates a force balance among pressure gradient and Coriolis force.

Pressure gradients may equalised by gravity waves, unless the gravity waves are not deflected by Coriolis force. The Rossby deformation radius $L_{D} \:[m]$ defines the distance at which the gravity waves get deflected by the Coriolis force \citep{parmentier2014PhD}:
 \begin{equation}
 L_{D} = \frac{ND}{f},
 \label{eq:rossby_deformation_radius}
\end{equation}
where $N \: [s^{-1}]$ is the Brunt-V\"ais\"al\"a frequency (actually, the oscillation frequency of gravity waves) and $D \: [m]$ the vertical length scale of the atmosphere. The vertical length scale of the atmosphere is  calculated at the order of one scale height, so $D=H$. The the Brunt-V\"ais\"al\"a frequency is defined in an isothermal atmosphere as \citep{parmentier2014PhD}:
 \begin{equation}
 N = \sqrt{\frac{c_{p}g}{R_{d}H}},
 \label{eq:brunt_vaisala_frequency}
\end{equation}
where $c_{p} [JK^{-1}]$ represents the heat capacity (we corrected a typing mistake in \citet{parmentier2014PhD}).
The Rhines scale $L_{Rh}$ indicates the scale at which the transition from dominant linear advection to the appearance of an inverse cascade occurs. The inverse cascade is the energy injection of small scales vortices into larger atmospheric flow. The Rhines scale is also known as an indicator for flow reorganization into the bands of alternating zonal jets, often called zonation \citep{sukoriansky2007}. In unsteady flow regimes, the Rhine scale might e associated with the moving energy front propagating towards the decreasing wavenumbers. The Rhines scale is defined as \citep{parmentier2014PhD}:
 \begin{equation}
 L_{Rh} = \pi \sqrt{\frac{U}{\beta}},
 \label{eq:rhines_scale}
\end{equation}
where $\beta$ corresponds to the meridional gradient of the Coriolis force, also known as the "$\beta$-effect, and defined as  \citep{parmentier2014PhD,kataria2016}: 
 \begin{equation}
 \beta = \frac{2\Omega\cos{(\vartheta)}}{R_{p}},
 \label{eq:beta_term}
\end{equation}
where $R_{p}$ is the radius of the planet.


\bsp	
\label{lastpage}
\end{document}